\documentclass[11pt,nofootinbib,preprint]{revtex4-1}
\pdfoutput=1
\usepackage[T1]{fontenc}
\usepackage[latin9]{inputenc}
\usepackage{times}
\usepackage{graphicx}
\usepackage{tabularx}
\usepackage{amsmath}
\usepackage{amstext}
\usepackage{amssymb}
\usepackage{xfrac}
\usepackage[colorlinks,citecolor=green]{hyperref}
\usepackage{graphicx}

\usepackage{amsmath}
\usepackage{amstext}
\usepackage{amssymb}
\usepackage{amsfonts}
\usepackage{longtable,booktabs}
\usepackage{hyperref}
\usepackage{url}
\usepackage{subfigure}
\usepackage{dsfont}
\usepackage{booktabs}
\usepackage{amsbsy}
\usepackage{amsthm}
\usepackage{bm}
\usepackage{esint}
\usepackage{multirow}
\usepackage{hyperref}
\usepackage{cleveref}
\usepackage{mathrsfs}
\usepackage{amsfonts}
\usepackage{amsbsy}
\usepackage{dcolumn}
\usepackage{bm}
\usepackage{multirow}
\usepackage{color}
\usepackage{extarrows}
\usepackage{datetime}
\usepackage{comment}
\usepackage[super]{nth}
\usepackage{tikz-cd}
\usepackage{float}

\makeatletter

\providecommand{\tabularnewline}{\\}

\def\Z{\mathbb{Z}}

\makeatother

\begin{document}
\title{ Topological Orders, Braiding Statistics, and Mixture of Two Types of Twisted $BF$ Theories in Five Dimensions}
\author{Zhi-Feng Zhang}
\author{Peng Ye}
\email{yepeng5@mail.sysu.edu.cn}
\affiliation{School of Physics, Sun Yat-sen University, Guangzhou, 510275, China}
\date{\today}
\begin{abstract}
Topological orders are a prominent paradigm for describing quantum many-body systems without symmetry-breaking orders. We present a topological quantum field theoretical (TQFT) study on topological orders in five-dimensional spacetime ($5$D) in which \textit{topological excitations} include not only point-like \textit{particles}, but also two types of spatially extended objects: closed string-like \textit{loops} and two-dimensional closed \textit{membranes}. Especially, membranes have been rarely explored in the literature of topological orders. By introducing higher-form gauge fields, we construct exotic TQFT actions that include mixture of two distinct types of $BF$ topological terms and many twisted topological terms. The gauge transformations are properly defined and utilized to compute level quantization and classification of TQFTs.  Among all TQFTs, some are not in Dijkgraaf-Witten cohomological classification. To characterize topological orders, we concretely construct all braiding processes among topological excitations, which leads to very exotic links formed by closed spacetime trajectories of particles, loops, and membranes. For each braiding process, we  construct  gauge-invariant Wilson operators and calculate the associated braiding statistical phases.  As a result, we obtain expressions of link invariants all of which have manifest geometric interpretation. Following Wen's definition, the boundary theory of a topological order exhibits gravitational anomaly. We expect that the characterization and classification of 5D topological orders in this paper encode information of 4D gravitational anomaly. Further consideration, e.g., putting TQFTs on 5D manifolds with boundaries, is left to future work.
 \end{abstract}
\maketitle
\tableofcontents{}



\section{Introduction}
As the most notable examples of beyond-symmetry-breaking orders, topological orders have attracted a lot of attentions for decades \cite{Wen2019,wen_stacking,string1.5,wenZootopoRMP}. One of most prominent features of   the research in topological orders is  joint efforts from  inter-disciplinary fields. Triggered by the condensed matter side, e.g., the fractional quantum Hall effect, topological orders have also shed light on frontiers of  high-energy physics, mathematical physics, and quantum information science \cite{hartnoll2021quantum,KitaevTopoEE,Levin_Wen_TEE,kong2020mathematical,kong2021mathematical}. In condensed matter physics, topological orders are gapped phases of matter and lack of any local order parameters that characterize symmetry-breaking patterns, which calls for a new revolution on the traditional solid-state physics and statistical physics of phases and phase transitions.  The low-energy effective field theories of topologically ordered phases of matter are usually topological quantum field theories (TQFTs) \cite{sarma_08_TQC,Witten1989}. The $2$D boundary of $3$D topological orders is governed by conformal field theory (CFT) \cite{wen_edge1995} and   gravitational anomaly in a more general sense \cite{Wen_grav_prd2013,string8}.\footnote{\emph{Important convention:} In this paper, when we mention $2$D, $3$D, etc., we refer to the dimension of spacetime. If the $3$-dimensional space, instead of spacetime, is considered, we would emphasize it as $3$D space, etc. }  The algebraic theory of topological orders in $3$D belongs to a subclass of tensor category \cite{string1}, which play a very fundamental role similar to group theory in symmetry-breaking orders. As a generalization of fermionic and bosonic statistics, anyonic  statistics in $3$D turns out to be described by  the mathematics of braid group, which was previously discovered through the path-integral of indistinguishable particles \cite{wu84}. The field of topological orders has also borrowed many exciting ideas from quantum information science, which leads to rapid developments we have been witnesses to, such as the long-range entanglement nature of topological orders and stabilizer codes as exactly solvable lattice models that admit topological orders \cite{CGW2010LRE,Kitaev2009,Kitaev2006,2003AnPhy.303....2K}.

  \textit{Topological excitations} are a central concept of topological orders.  Above the topologically ordered ground states,  {topological excitations} look very odd and behave like  fractionalized degrees of freedom. For example, anyons, a kind of exotic particles in $3$D world, carry fractionalized electron charge and fractionalized statistics, which can be regarded as a consequence of electron fractionalization in the fractional quantum Hall systems.
In addition to point-like excitations, \emph{spatially extended excitations}, e.g., string-like loop excitations and two-dimensional closed membrane excitations have been constructed in topological orders of $4$D and higher dimensions. Their exotic entanglement, symmetry enrichment, braiding statistics, TQFTs, and higher-category theory have been studied intensively \cite{lantian3dto1,lantian3dto2,yp18prl,ypdw,wang_levin1,bti2,jian_qi_14,string5,PhysRevX.6.021015,string6,ye16a,YeGu2015,corbodism3,YW13a,ye16_set,2018arXiv180101638N,2016arXiv161008645Y,string4,PhysRevLett.114.031601,3loop_ryu,string10,2016arXiv161209298P,Ye:2017aa,Tiwari:2016aa,2012FrPhy...7..150W,zhang2021compatible}.   Recently, the so-called ``fracton physics'' of spatially extended excitations is also discussed, where both mobility and deformability of  spatially extended excitations are restricted either partially or completely \cite{ye19a,2021LiYeFracton,pretko18string}.

  \textit{Braiding statistics} among topological excitations is served as  a topological order parameter to classify and characterize topological orders \cite{wen_stacking}.    In $3$D, braiding statistics among anyons is known to form $S$ and $T$ matrices.  In $4$D, particles cannot be anyonic but the presence of loop excitations makes braiding statistics even more bizarre. First, we can consider a discrete gauge group $G=\prod_i \Z_{N_i}$. All particles carry and thus are labeled by periodic gauge charges. Likewise, all loops carry and thus are labeled by    periodic gauge fluxes. Then, braiding statistics can be
   particle-loop braiding \cite{hansson2004superconductors,abeffect,PRESKILL199050,PhysRevLett.62.1071,PhysRevLett.62.1221,ALFORD1992251}, multi-loop braiding \cite{wang_levin1}, and particle-loop-loop braiding (i.e., Borromean-Rings braiding) \cite{yp18prl}. Recently, ref.~\cite{zhang2021compatible} exhausted all possible combinations of braiding processes in order to obtain a complete list of topological orders in $4$D.  The basic guiding principle there is to find TQFTs that have well-defined gauge transformations. All TQFTs for describing these braiding processes have structures of a multi-component $BF$ term \cite{Horowitz1990}  in the presence of some twists. The $BF$ term in $4$D has a form of $BdA$ where $B$ and $A$ are respectively $2$- and $1$-form gauge fields. Twists consist of  $AAdA$, $AAAA$ \cite{2016arXiv161209298P}, and $AAB$ \cite{yp18prl}.

Although topological orders in $4$D and $3$D are already interesting and directly relevant to experimentally realizable systems of condensed matter physics,  in this paper, we move forward to investigate topological orders and TQFTs in $5$D, which turns out to exhibit  highly unexplored features of both physics and mathematics.
\emph{Firstly}, one of the most attractive features in $5$D topological orders is the existence of membrane excitations, which are geometrically two-dimensional compact manifolds and form three-dimensional world-volumes. Membrane excitations can participate in nontrivial braiding processes that are expected to  go beyond braidings in $3$D and $4$D.
 \emph{Secondly}, understanding $5$D topological orders is also useful for understanding gravitational anomalies in $4$D according to Wen's definition \cite{Wen_grav_prd2013,string8}. In general, the boundary of a topologically ordered state on an open manifold has gravitational anomaly, which strictly forbids  the consistent existence of the boundary theory \text{alone}  by removing the bulk topological ordered state. This fact is true and robust even though all global symmetries are broken. Therefore, we are allowed to  investigate gravitational anomalies in $4$D by means of topological orders in $5$D.
 \emph{Thirdly},  when gauge group is still $G=\prod_{i=1}^n\Z_{N_i}$,  TQFTs in $5$D may contain two types of $BF$ terms, i.e., $CdA$ and $\tilde{B}dB$, where $C$ is $3$-form, $B$ and $\tilde{B}$ are two different $2$-form, $A$ is $1$-form.  Therefore, for each $\Z_{N_i}$ subgroup, there are two choices for assignment of gauge charges and corresponding $BF$ terms. In such a mixed $BF$ theory, if we further add twisted terms (i.e., twists), e.g., $AAAAA$, $AAAdA$, $AdAdA$, $AAC$, $AAAB$, $ABB$, $BAdA$, $AAdB$, the resulting gauge theories are expected to be very complex and host exotic topological orders.   Given $BF$ terms, the additional twists are not always compatible with each other, as some of combinations of twists unavoidably  violate gauge invariance of either \textit{usual} type or \textit{large} type.  This phenomenon is similar to but much more intricate than {that in} $4$D studied in ref.~\cite{zhang2021compatible} due to the presence of two types of $BF$ terms.

In this paper, for understanding low energy physics of  $5$D topological orders, we construct  gauge-invariant TQFT actions and define gauge transformations for all gauge fields. More specifically,  TQFT actions consist of  two distinct types of $BF$ terms with twisted topological terms,
dubbed $BF$ theories.  By means of  gauge transformations, we obtain the quantization and periods of the coefficients of all topological terms, which leads to TQFT classification. Some of TQFTs are beyond Dijkgraaf-Witten cohomological classification $H^5\left(G, \mathbb{R}/\mathbb{Z}\right)$. Furthermore, we construct gauge-invariant observables (generalized Wilson operators) of TQFT actions.  In $5$D topological orders, a braiding process results in  {a nontrivial link formed by the closed world-lines of particles, world-sheets of loops, and/or world-volumes of membranes}.    It is natural to expect a linking number or link invariant in $5$D to characterize such a braiding process. For our $5$D $BF$ theories,    we relate the expectation value of a Wilson operator  to counting intersections of sub-manifolds in $5$D. It should be noted that the link invariants are obtained from our physical theory via the principle of  gauge invariance. In this sense, our physical theory provides an alternative route to understand link theory of higher-dimensional compact manifolds. The latter manifolds are physically realized as closed spacetime trajectories of topological excitations in our 5D condensed matter systems.

This paper is organized as follows. In section~\ref{sec_topo_excitation_braid}, we define topological excitations in the hydrodynamical approach, and then introduce two types of $BF$ terms. We also add twisted terms to $BF$ terms and study the resulting TQFT. Typically, we calculate the expectation values of Wilson operators
and reveal their geometric interpretation. Next, in section~\ref{sec_tqft_classification}, we exhaust all possible TQFTs that are gauge invariant such that all braiding processes encoded in a given TQFT are mutually compatible.  Classification of these TQFTs is discussed for different gauge groups, collected in table~\ref{tab_classi_1-3Zn} ($\mathbb{Z}_{N_1}$, $\mathbb{Z}_{N_1}\times \mathbb{Z}_{N_2}$, and $\prod_{i=1}^{3}\mathbb{Z}_{N_i}$). For $G=\prod_{i=1}^{n}\mathbb{Z}_{N_i}$ with $n\geq 4$,  details are presented in section~\ref{subsec_class_4Zn}. Section~\ref{section_concl_main} is devoted to a conclusion and outlook. Some technical details are collected in appendices.

\section{Topological excitations and braiding statistics in $5$-dimensional
topological orders}\label{sec_topo_excitation_braid}
Besides the robust ground state degeneracy, topological excitations
and their braiding statistics also serve as part of the definition
of topological orders. In section~\ref{subsec_excitation_$5$D}, we intuitively explain what topological excitations in $5$D look like and how to understand the braiding processes among
them after reviewing the cases
in $3$D and $4$D. Then,  in section~\ref{subsec_2_types_BF}, we introduce two types
of $BF$ terms that describe braiding processes of two topological
excitations. Last, as shown in section~\ref{subsec_type_1_BF_a_twist} and~\ref{subsec_mixed_BF_a_twist}, by adding a twisted term to $BF$ terms, we construct
TQFT actions that describe braiding processes involving multiple topological
excitations. Especially, for each braiding process, we find a gauge-invariant observable (generalized Wilson operator) whose vacuum expectation
value is expressed by the intersection of some sub-manifolds in $5$D.
We find that such   gauge-invariant observables  are closely related with linking numbers of spatially extended topological excitations in $5$D topological orders.
\subsection{Topological excitations and their braiding processes in $3$D, $4$D, and $5$D spacetime}\label{subsec_excitation_$5$D}
It is beneficial for us to review topological excitations and their
braiding processes in $3$D and $4$D before we move into the $5$-dimensional
spacetime. Anyon in the fractional quantum Hall effect (FQHE),
a typical $3$D topological order, may be the most well-known example
of topological excitations in $3$D.  The self-statistics of anyons is captured by the phase difference
after exchanging two anyons in the $2$D space. If viewed in $3$D,
the world-lines of two anyons form a braid during the exchange. Geometrically,
anyons are particle excitations, the only possible excitation in $3$D topological orders.\footnote{In $3$D topological orders, loop excitations are regarded equivalent to particle excitations. More concretely, because a loop excitation is impenetrable in $2$D space, it actually behaves like a particle excitation. {Therefore, we do not consider loop excitations in $3$D topological orders.} } In $4$D topological orders, topological excitation spectrum is composed by particle
excitations and loop excitations \cite{lantian3dto1,lantian3dto2}. Braiding processes in
$4$D topological orders can be divided into three classes~\cite{zhang2021compatible}: particle-loop
braiding~\cite{Witten1989,horowitz89,Horowitz1990,hansson2004superconductors,Baez2011,bti6}, multi-loop (three or four) braiding~\cite{wang_levin1,2016arXiv161209298P}, and particle-loop-loop
braiding~\cite{yp18prl}. In particle-loop braiding, a particle excitation moves around
a loop excitation such that its spatial trajectory and the loop excitation
form a Hopf link. This Hopf link can also be identified from the intersection
of the world-line of the particle and the world-sheet of the loop.\footnote{For simplicity, when we mention particle, loop, or membrane, we refer to topological excitations
in topological orders. } In multi-loop braiding, three or
four loops are linked in a special manner and their intersecting world-sheets
indicate the topological invariant characterizing such braiding. In
particle-loop-loop braiding, one particle moves around two unlinked loops such
that its spatial trajectory and these two loops form Borromean rings~\cite{yp18prl,RN874}, or in
general a Brunnian link.

In $5$D, besides particles and loops, there is a kind of
exotic topological excitations, dubbed as \emph{membranes}, which look like closed 2-dimensional surfaces\footnote{We only consider membrane excitation that has a shape of a $2$-sphere in this paper. Membrane excitations with other shapes, e.g., $2$-torus, may have more fascinating properties.}. The membrane excitations in $3$D space have not been considered because they are
impenetrable. However, in $4$D space, the interior of a membrane excitation becomes
accessible due to the extra dimension. Therefore, nontrivial braiding
processes involving particles, loops and membranes are possible in
$5$D.

\begin{figure}[htb]
  \centering
  \includegraphics[width=0.8\textwidth]{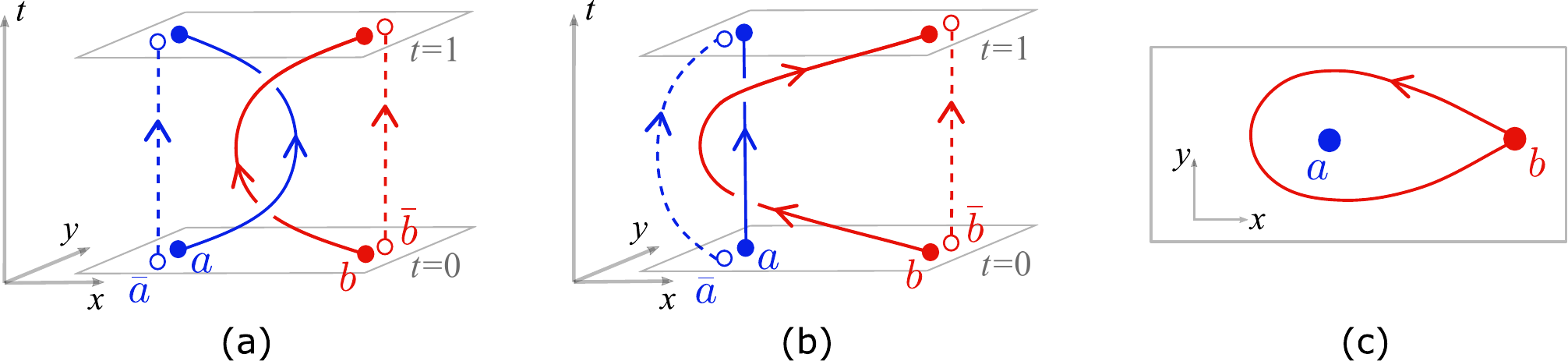}
  \caption{Braiding of two particles (e.g., anyons) in $3$D spacetime, world-lines of particles and their projection in $2$D space. (a) At $t=0$, two pairs of particles (solid circles) and antiparticle (empty circles) are created. Two particles ($a$ and $b$) move in $3$D spacetime such that their world-lines (solid lines) cross each other. At $t=1$, two particles meets their own antiparticles ($\bar{a}$ and $\bar{b}$). The $a$-$\bar{a}$ and $b$-$\bar{b}$ pairs are annihilated. Since the world-line of an antiparticle (dash line) can be understood as that of a particle with a reverse direction, the particle-antiparticle pair's world-lines can be viewed as a single closed one of the particle. For a braiding process of two particles in $3$D, the two closed world-lines are linked. (b) One of the two particles (e.g., $a$) can be assumed static in the $xy$-plane. By slightly modifying the world-line of $b$ and keeping the two closed world-lines linked, the braiding process contributes the same phase shift. (c) The world-lines of $a$ and $b$ illustrated in (b) are projected to the $xy$-plane. After that, $b$'s world-line is appeared as a closed spatial trajectory encircling the static particle $a$. In FQHE, an anyon moving around another one is exactly the braiding of them.}
  \label{fig_anyon}
\end{figure}

\begin{figure}[htb]
  \centering
  \includegraphics[width=0.7\textwidth]{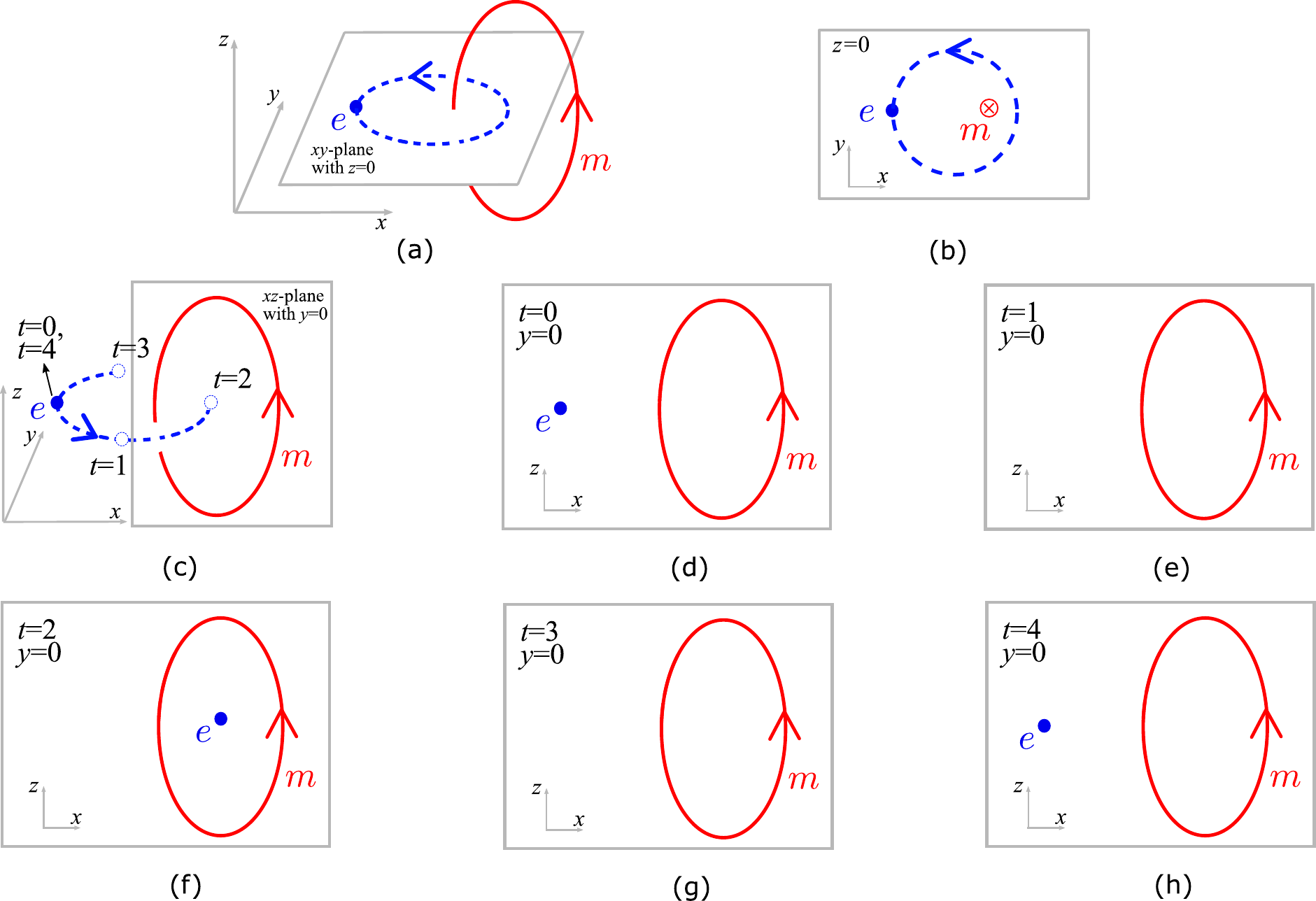}
  \caption{A particle-loop braiding in $3$D space and its projection on $2$D planes. (a) Particle-loop braiding is one important braiding process in $4$D topological order. In this braiding process, if we assume the loop $m$ to be static, what we see is that the particle $e$ encircles the loop such that its spatial trajectory and loop $m$ form a Hopf link. (b) For an observer living on the $xy$-plane with $z=0$, the particle-loop braiding in $3$D appears as a particle $e$ encircling a flux that penetrates this plane. (c) The same particle-loop braiding in $3$D space as that in (a). The particle's positions at different moments are labeled. (d)-(e) Snapshots of particle-loop braiding observed in $xz$-plane with $y=0$ at different moments. In this plane, the loop excitation appears as its complete form while the spatial trajectory of the particle is not continuous any longer. An observer living on this plane can only see the particle at $t=0,2,4$. At $t=1$ or $t=3$, this particle is located at $xz$-planes with different $y$-coordinates. To this planar observer's knowledge, a particle cannot cross a loop excitation to reach its interior area. What illustrated in (d)-(h) happens due to an extra dimension (the particle is able to move in $y$-direction) that cannot be seen by this observer. On the other hand, an observer in $2$D plane can detect a \nth{3} dimension by such an ``anomalous'' phenomenon. This inspire us that we can learn about braiding process in $4$D space ($5$D spacetime) by observing it in different $3$D spaces with a fixed \nth{4} coordinate.}
  \label{fig_particle_loop}
\end{figure}

In a braiding process of $5$D topological order, excitations move in $5$D spacetime in
a particular manner, resulting in their $1$D world-line, $2$D world-sheet,
and/or $3$D world-volume intersecting in a corresponding fashion.
The $5$D TQFT can tell us about how these manifolds intersect in $5$D
though which is not easy to imagine for us living in $3$D. To visualize
braiding processes in $5$D topological order, we need to present
them in lower dimensions. For this purpose, we can perform the following
procedures. First, noticed that only the relative motion of excitations
matters in a braiding process, we can assume one of the excitations
to be static in $4$D space, then we project the world-lines/world-sheets/world-volumes
from $5$D spacetime to $4$D space. For the static excitation, it
is just fixed in $4$D space; for other excitations doing relative
motion, their spatial trajectories are left in $4$D space after projection. An intuitive example is the anyon braiding in $3$D, shown in figure~\ref{fig_anyon}, in which the world-lines of anyons can be projected to spatial trajectories
in $2$D plane.
Yet, the static excitation and spatial trajectories
in $4$D space are still strange to observer in $3$D space. Our strategy
is to furthermore project these spatial trajectories to slices of
$4$D space, i.e., $3$D space. As an example, figure~\ref{fig_particle_loop} shows the projection of spatial trajectory of particle-loop braiding from $3$D space to $2$D plane.
This example shows how an observer in $2$D space can
understand braiding process in $4$D spacetime. In summary, by projecting
the general world-lines in $5$D spacetime to spatial trajectories
in $4$D space, then viewing them in $3$D slices, we can observe the braiding processes in $5$D topological order. This method will be explained by several pictures when discussing braiding and $5$D TQFT in the following main text.

\subsection{Two types of $BF$ terms}\label{subsec_2_types_BF}
As the effective theory of topological order, TQFT describes braiding processes by revealing how the world-lines (in general meaning) of excitations intersect in spacetime. A braiding process is nontrivial only if the world-lines of excitations form a link that is homotopic invariant. By constructing gauge-invariant observables, we know how to count the intersections of world-lines to obtain a result that is invariant under homotopic mapping of world-lines. In this and following sections, we use TQFT actions and gauge-invariant observables to study braiding processes in $5$D topological order.

The study of $BF$ term has a long history \cite{Horowitz1990}. Discovered from quantum gravity and high energy physics, now $BF$ theories have been introduced into frontiers of condensed matter physics. In
an $n$-dimensional spacetime manifold $M$, the $BF$ term is the wedge product
of a $p$-form $\mathscr{B}$ and an $\left(n-p\right)$-form $\mathscr{F}=d\mathscr{A}$
where $\mathscr{A}$ is an $\left(n-p-1\right)$-form: $\mathscr{B}\wedge d\mathscr{A}$.
$BF$ term in $3$D and $4$D reads $\tilde{A}dA$ and $BdA$ respectively,
where $\tilde{A}$ and $A$ are $1$-form, $B$ is $2$-form. Theories with $BF$ terms
have been applied to many different physical systems~\cite{hansson2004superconductors,YeGu2015,bti2,bti6,bti1,Ye:2017aa,ye16a,PhysRevB.99.205120}.
It is easy to verify that $\tilde{A}dA$ ($BdA$) is the only $BF$ term in
$3$D ($4$D). However, in $5$D, by taking $p=2$ or $p=3$, there
are two types of $BF$ terms, i.e., $CdA$ and $\widetilde{B}dB$ which are respectively called type-I and type-II in this paper.
Here, $C$ is a $3$-form, $\widetilde{B}$
and $B$ are two different $2$-form's. These two types of $BF$ terms
in $5$D are the cornerstone of this paper. With the action $S\sim \int_{M}\mathscr{B}d\mathscr{A}$,
one can calculate the linking number of a $p$ and $\left(n-p-1\right)$-dimensional
sub-manifolds~\cite{Horowitz1990,2016arXiv161209298P}. In this section,
we study the action with type-I and type-II $BF$ term in details,
revealing the connection between braiding processes in $5$D topological
orders and $BF$ terms.

\begin{figure}[H]
  \centering
  \includegraphics[width=0.7\textwidth]{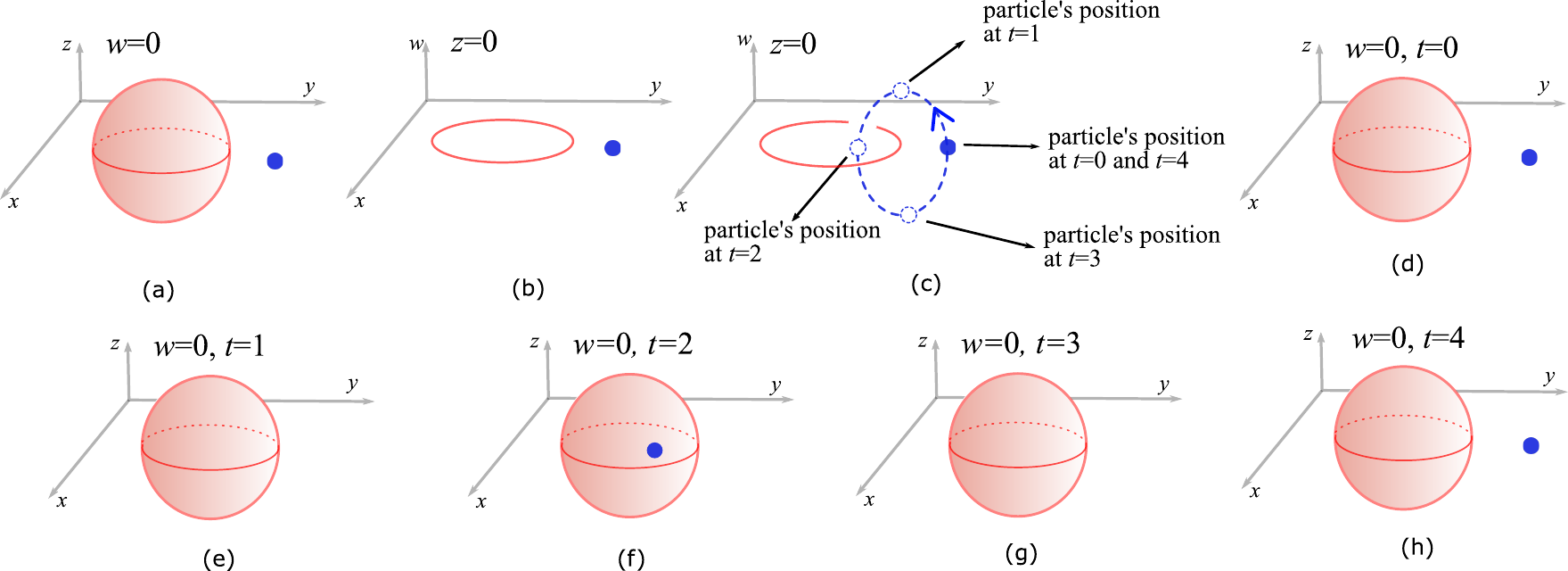}
  \caption{Observing the particle-membrane braiding described by $CdA$ term in $3$D space. (a) A membrane (red) with the shape of  a $2$-sphere in $xyz$-space with $w=0$ and a particle (blue) are illustrated. This membrane is assumed to be static in space. (b) If observed in $xyw$-space with $z=0$, this membrane appears as a ``loop''. (c) In $xyw$-space with $z=0$, the particle-membrane braiding in $4$D space looks like a particle-loop braiding. Starting at $t=0$, the particle moves in the $xyw$-space with $z=0$ and returns to its initial position at $t=4$. The particle's spatial trajectory forms a closed $1$D curve that is linked with the ``loop'' (projection of membrane in this $3$D space). The $w$-coordinate of particle, $w_p$, varies in the following way: at $t=0$, $w_p =0$; for $0 <t< 2$, $w_p > 0$; at $t=2$, $w_p = 0 $ again; for $2 <t< 4$, $w_p <0$; finally, the particle returns to its initial position thus $w_p = 0$ at $t=4$. (d)-(h) Snapshots of particle-membrane braiding at different moments in $xyz$-space with $w=0$. At $t=0$, the particle is located outside the membrane. At $t=2$, the particle appears inside the membrane. At $t=4$, this particle returns to its starting position; in $xyz$-space with $0$, it appears outside the membrane again. For other moments, the particle is located in $xyz$-space with $w\neq 0$. In $4$D space, $\Xi_1$ in eq.~(\ref{eq_vev_CdA}) is projected to a $3$D Seifert hypersurface and $\gamma_1$ in that appears as a closed $1$D spatial trajectory. In $xyw$-space with $w=0$ shown in (c), the aforementioned $3$D Seifert hypersurface appears as a $2$D one whose boundary is exactly the ``loop''. The spatial trajectory and the $2$D Seifert surface intersect at one point at $t=2$ implying that $\left|\#\left(\Xi_{1}\cap\gamma_{1}\right)\right|=1$ with a sign determined by orientation. In $xyz$-space with $w=0$ shown in (d)-(h), the spatial trajectory is the union of particle's positions at different moments which intersects with the $3$D Seifert hypersurface at $t=2$. This also indicates $\left|\#\left(\Xi_{1}\cap\gamma_{1}\right)\right|=1$.}
  \label{fig_CdA}
\end{figure}

We first look at the action with type-I $BF$ term:
\begin{equation}
S=\int\frac{N_{1}}{2\pi}C^{1}dA^{1},
\label{eq_action_CdA}
\end{equation}
which is invariant up to boundary term under gauge transformations
\begin{equation}
A^{1}\rightarrow A^{1}+d\chi^{1},\ C^{1}\rightarrow C^{1}+dT^{1},
\end{equation}
where 0-form $\chi^{1}$ and 2-form $T^{1}$ are $\mathbb{U}\left(1\right)$
gauge parameters with $\oint d\chi^{1}\in2\pi\mathbb{Z}$ and $\oint dT^{1}\in2\pi\mathbb{Z}$.
We consider the following gauge invariant observable (Wilson operator)
\begin{align}
\mathcal{W}\left(\omega_{1},\gamma_{1}\right)= & \exp\left({\rm i}m_{1}\int_{\omega_{1}}C^{1}\right)\exp\left({\rm i}e_{1}\int_{\gamma_{1}}A^{1}\right)\nonumber\\
= & \exp\left[{\rm i}m_{1}\int C^{1}\wedge\delta^{\perp}\left(\omega_{1}\right)+{\rm i}e_{1}\int A^{1}\wedge\delta^{\perp}\left(\gamma_{1}\right)\right],
\end{align}
where $\omega_{1}$ is a closed $3$D volume and $\gamma_{1}$ is
a closed $1$D curve; charges $m_{1},e_{1}\in\mathbb{Z}$.\footnote{Subscripts ($1,2,\cdots$, or general $i,j$) are used to distinguish manifolds that labeled by the same Greek letter. In this paper, $\gamma_i$ labels  different $1$D closed curves; $\sigma_{i}$ and $\widetilde{\sigma}_{i}$ label different $2$D closed surfaces; $\omega_i$ stands for different $3$D closed volumes. $\Sigma_i$, $\Omega_i$, and $\Xi_i$ are the Seifert (hyper)surfaces: $\partial \Sigma_{i}=\gamma_i$, $\partial \Omega_{i}=\sigma_i$, and $\partial \Xi_{i}=\omega_i$. $\mu_{i} |_X$, $\nu_{i} |_X$, and $\lambda_{i} |_X$ respectively stand for general $1$D open curve, $2$D open surface, and $3$D open volume on manifold $X$.\label{footnote_symbol_notation}} The expectation
value is given by $\left\langle \mathcal{W}\left(\omega_{1},\gamma_{1}\right)\right\rangle =\frac{1}{\mathcal{Z}}\int\mathscr{D}A\mathscr{D}C\mathcal{W}\left(\omega_{1},\gamma_{1}\right)\exp\left({\rm i}S\right)$ where
the partition function is defined as $\mathcal{Z}=\int \mathscr{D}C\mathscr{D}A\exp\left({\rm i}S\right)$.
By integrating out $C^{1}$ we get $dA^{1}=-\frac{2\pi m_{1}}{N_{1}}\delta^{\perp}\left(\omega_{1}\right)$
which can be solved by $A^{1}=-\frac{2\pi m_{1}}{N_{1}}\delta^{\perp}\left(\Xi_{1}\right)$
with $\partial\Xi_{1}=\omega_{1}$, i.e., $\Xi_{1}$ is a Seifert
hypersurface bounded by $\omega_{1}$. $\delta^{\perp}\left(\omega_{1}\right)$
is the $2$-form valued delta function distribution supported on $\omega_{1}$
and $\delta^{\perp}\left(\Xi_{1}\right)$ is {similarly defined}. Plugging the solution back
to $\left\langle \mathcal{W}\left(\omega_{1},\gamma_{1}\right)\right\rangle $, we get
\begin{align}
\left\langle \mathcal{W}\left(\omega_{1},\gamma_{1}\right)\right\rangle = & \exp\left[-\frac{{\rm i}2\pi e_{1}m_{1}}{N_{1}}\int\delta^{\perp}\left(\Xi_{1}\right)\wedge\delta^{\perp}\left(\gamma_{1}\right)\right]\nonumber\\
= & \exp\left[-\frac{{\rm i}2\pi e_{1}m_{1}}{N_{1}}\#\left(\Xi_{1}\cap\gamma_{1}\right)\right],
\label{eq_vev_CdA}
\end{align}
i.e., $\left\langle \mathcal{W}\left(\omega_{1},\gamma_{1}\right)\right\rangle $ is determined
by counting the intersection of $\Xi_{1}$ and $\gamma_{1}$. In $3$D, the linking number of two closed curves $\gamma_{i}$ and
$\gamma_{j}$ is defined by the intersection number $\#\left(\Sigma_{i}\cap\gamma_{j}\right)$
where $\Sigma_{i}$ is a Seifert surface of $\gamma_{i}$. Analogous
to this, $\#\left(\Xi_{1}\cap\gamma_{1}\right)$
defines the linking number of $\gamma_1$ and $\omega_1$ (whose Seifert hypersurface is $\Xi_1$) in $5$D. In Sec~\ref{subsec_excitation_$5$D}, we mention that $5$D TQFT describes braiding process via intersection pattern of world-lines of excitations in $5$D. The action~(\ref{eq_action_CdA}) and $\left\langle\mathcal{W}\left(\omega_{1},\gamma_{1}\right)\right\rangle$ describe the braiding process of a particle and a membrane once we interpret $\gamma_1$ ($\omega_1$) as the closed world-line (world-volume) of particle (membrane). Figure~\ref{fig_CdA} illustrates this particle-membrane braiding by projecting it in different $3$D space.

\begin{figure}[H]
  \centering
  \includegraphics[width=0.8\textwidth]{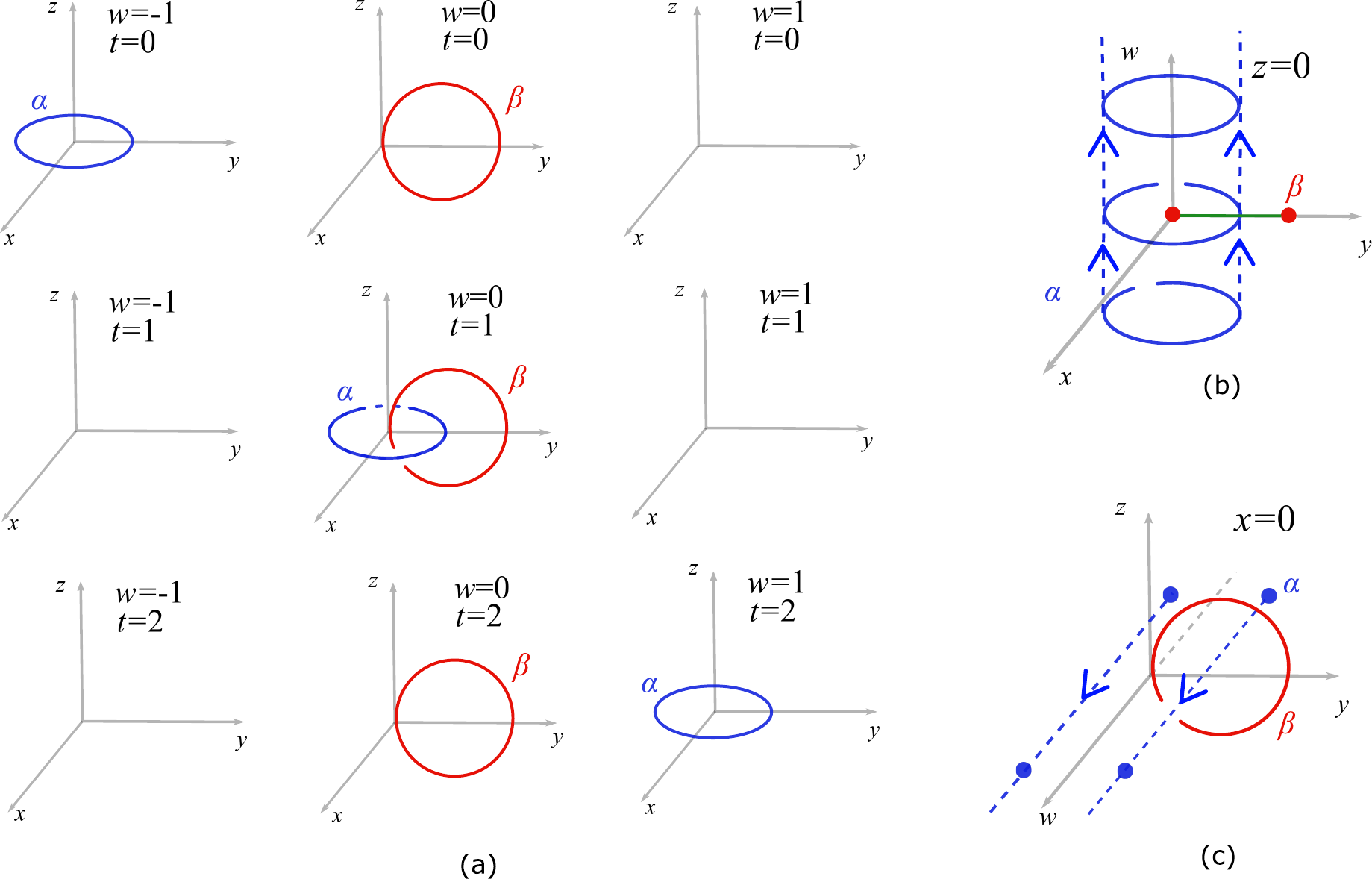}
  \caption{Diagrammatic representation of loop-loop braiding described by $\widetilde{B}dB$ term. (a) Loop-loop braiding process viewed in $xyz$-space with different $w$- and $t$-coordinates. Loop $\alpha$ (blue) lives on $xy$-plane with $z_{\alpha}=0$ and $w_{\alpha}=-1$ at $t=0$. Loop $\beta$ (red) lives on $yz$-plane with $x_{\beta}=0$ and $w_{\beta}=0$. Loop $\beta$ is assumed to be static in space. When $t=0$, loop $\alpha$ can only be observed in $xyz$-space  with $w=-1$ and so does loop $\beta$ in $xyz$-space with $w=0$; as for $xyz$-space with $w=1$, neither of two loops could be observed. Starting at $t=0$, loop $\alpha$ moves in $w$-direction such that at $t=1$ $w_{\alpha}=0$ and loop $\alpha$ and $\beta$ are linked in $xyz$-space with $w=0$. Next, loop $\alpha$ continues to move such that $w_{\alpha}=1$ at $t=2$. At $t=2$, loop $\alpha$ and $\beta$ are not linked obviously. Then loop $\alpha$ returns to its initial position in such a way that it would be linked with loop $\beta$ again. During this process, these two loops never intersect with each other since no points of them share a same $4$D spatial coordinate. (b) Part of the spatial trajectory of loop $\alpha$ (indicated by blue dash lines) observed in $xyw$-space with $z=0$. In this $3$D space, loop $\beta$ appears as two points (red solid circles). In $4$D space, $\widetilde{\Omega}_1$ in eq.~(\ref{eq_vev_BdB}) is projected to a $2$D Seifert surface and $\sigma_1$ becomes a $2$D closed spatial trajectory. A further projection to $xyw$-space with $z=0$ makes this $2$D Seifert surface appear as a $1$D one (the green segment) whose boundary is two points (red solid circles, exactly the projection of loop $\beta$ in this $3$D space). This $1$D Seifert surface intersect with loop $\alpha$'s spatial trajectory at one point, which implies $\left|\#\left(\widetilde{\Omega}_1 \cap \sigma_1\right)\right|=1$ with a sign determined by orientation. (c) Viewed in $wyz$-space with $x=0$, loop $\alpha$ appears as two points (blue solid circles) and part of its spatial trajectory is shown as the blue dash lines. This spatial trajectory intersect with loop $\beta$'s $2$D Seifert surface at one point, indicating that $\left|\#\left(\widetilde{\Omega}_1 \cap \sigma_1\right)\right|=1$ as well.}
\label{fig_BdB}
\end{figure}

Next we focus on the action that consists of a type-II $BF$ term:
\begin{equation}
S=\int\frac{N_{1}}{2\pi}\widetilde{B}^{1}dB^{1}
\end{equation}
which is invariant up to boundary terms under
\begin{equation}
B^{1}\rightarrow B^{1}+dV^{1},\ \widetilde{B}^{1}\rightarrow \widetilde{B}^{1}+d\widetilde{V}^{1}.
\end{equation}
Here, $1$-form $V^{1}$ and 1-form $\widetilde{V}^{1}$ are $\mathbb{U}\left(1\right)$
gauge parameters with $\oint dV^{1}\in2\pi\mathbb{Z}$ and $\oint d\widetilde{V}^{1}\in2\pi\mathbb{Z}$. We point out that $\widetilde{B}^1$ and $B^1$ are two independent  $2$-form gauge field variables. The corresponding gauge-invariant observable is
\begin{equation}
\mathcal{W}\left(\widetilde{\sigma}_{1},\sigma_{1}\right)= \exp\left({\rm i}m_{1}\int_{\widetilde{\sigma}_{1}}\widetilde{B}^{1}\right)\exp\left({\rm i}e_{1}\int_{\sigma_{1}}B^{1}\right),
\end{equation}
where $\widetilde{\sigma}_{1}$ and $\sigma_{1}$ are two closed $2$D surfaces; charges $m_{1},e_{1}\in\mathbb{Z}$.
Similar to the calculation of $\left\langle \mathcal{W}\left(\omega_{1},\gamma_{1}\right)\right\rangle $ above,
we get the solution $B^{1}=-\frac{2\pi m_{1}}{N_{1}}\delta^{\perp}\left(\widetilde{\Omega}_{1}\right)$
with $\partial\widetilde{\Omega}_{1}=\widetilde{\sigma}_{1}$,
i.e., $\widetilde{\Omega}_{1}$ is a Seifert hypersurface bounded
by $\widetilde{\sigma}_{1}$. Plugging this solution back,
we obtain
\begin{align}
\left\langle \mathcal{W}\left(\widetilde{\sigma}_{1},\sigma_{1}\right)\right\rangle = & \exp\left[-\frac{{\rm i}2\pi e_{1}m_{1}}{N_{1}}\int\delta^{\perp}\left(\widetilde{\Omega}_{1}\right)\wedge\delta^{\perp}\left(\sigma_{1}\right)\right]\nonumber\\
= & \exp\left[-\frac{{\rm i}2\pi e_{1}m_{1}}{N_{1}}\#\left(\widetilde{\Omega}_{1}\cap\sigma_{1}\right)\right],
\label{eq_vev_BdB}
\end{align}
which indicates that $\left\langle \mathcal{W}\left(\widetilde{\sigma}_{1},\sigma_{1}\right)\right\rangle $
is determined by counting the intersection of $\widetilde{\Omega}_{1}$
and $\sigma_{1}$. Similarly, $\widetilde{\sigma}_{1}$ and
$\sigma_{1}$ can be viewed as closed world-sheets of two loop excitations whose linking number is given by $\#\left(\widetilde{\Omega}_{1}\cap\sigma_{1}\right)$. Therefore, the type-II $BF$ term $\widetilde{B}^{1}dB^{1}$ describes
to the braiding process of two loops and the braiding statistical phase can be extracted from $\left\langle \mathcal{W}\left(\widetilde{\sigma_{1}},\sigma_{1}\right)\right\rangle$. Figure~\ref{fig_BdB} provides a diagrammatic representation of this loop-loop braiding by projecting the world-sheets to spatial trajectories of loops from $5$D to $4$D space.

\subsection{Braidings in type-I $BF$ theory with a twist\label{subsec_type_1_BF_a_twist}}
So far we have investigated TQFT actions that consist of single $BF$ term. The coefficient $\frac{N_1}{2\pi}$ of the $BF$ term encodes the $\mathbb{Z}_{N_{1}}$ gauge group. If a $\prod_{i=1}^{n}\mathbb{Z}_{N_{i}}$
topological order is considered, i.e., the gauge group is $G=\prod_{i=1}^{n}\mathbb{Z}_{N_{i}}$,
there are twisted terms allowed in TQFT actions besides $BF$ terms.
As examples, we present some TQFT actions that consist of $BF$ terms
and \emph{one} twisted term in $4$D $\prod_{i=1}^{n}\mathbb{Z}_{N_{i}}$
topological orders~\cite{Kapustin2014,YeGu2015,2016arXiv161209298P,yp18prl,PhysRevB.99.235137,Tiwari:2016aa}. With twisted term $AAdA$ or $AAAA$, TQFT actions
$S=\int\sum_{i}\frac{N_{i}}{2\pi}B^{i}dA^{i}+qA^{i}A^{j}dA^{k}$ and
$S=\int\sum_{i}\frac{N_{i}}{2\pi}B^{i}dA^{i}+qA^{i}A^{j}A^{k}A^{l}$
describe multi-loop braidings (three-loop and four-loop, respectively; $q$ is the proper coefficient).
With twisted term $AAB$, the action $S=\int\sum_{i}\frac{N_{i}}{2\pi}B^{i}dA^{i}+qA^{i}A^{j}B^{k}$
describes the particle-loop-loop braiding. It is natural to expect that TQFT actions of $BF$ terms and twisted terms in $5$D
are related to braiding processes of topological excitations. We divide
the TQFT actions into $3$ classes according to the type of their
$BF$ terms, namely type-I $BF$ theory, type-II $BF$ theory, and
mixed $BF$ theory. In the remaining part of this section, we consider $BF$ theory with only
\emph{one} twisted term. Consistent combination of all {twisted} terms will be explored in section~\ref{sec_tqft_classification}.

 We start with   type-I $BF$ terms with a twisted term formed by $A$ and $C$, i.e., $AAC$, $AdAdA$, $AAAdA$, and $AAAAA$.

\emph{$AAC$ twisted topological term}. By considering three $C^{i}dA^{i}$
terms, we can introduce the twisted term $A^{i}A^{j}C^{k}$.  The corresponding
TQFT action is
\begin{equation}
S=S_{BF}+S_{AAC}=\int\sum_{i=1}^{3}\frac{N_{i}}{2\pi}C^{i}dA^{i}+qA^{1}A^{2}C^{3},
\label{eq_S_CdA+AAC}
\end{equation}
where $q$ is a quantized and periodic coefficient, $q=\frac{pN_{1}N_{2}N_{3}}{\left(2\pi\right)^{2}N_{123}}$, $p\in\mathbb{Z}_{N_{123}}$, $N_{123}$ is the greatest common divisor of $N_1$, $N_2$ and $N_3$. The property of coefficient $q$ results from two requirements on the TQFT action: large gauge invariance and flux identification. In appendix~\ref{sec_quantization_of_coefficient} we derive the quantization and periodicity of $AAC$ and other twisted terms. It should be pointed out that
the indices $i$, $j$, and $k$ in $A^{i}A^{j}C^{k}$ have to be
mutually different. For example, the $A^{1}A^{2}C^{2}$ term is prohibited, because either $A^{2}$ or $C^{2}$ has to be the Lagrange multiplier, which means that they cannot simultaneously appear in the twisted term. A more
rigorous explanation is that the an action with $A^{1}A^{2}C^{2}$
term would no more be gauge-invariant, which is similar to the  {$4$D
case $S\sim\int \sum_{i=1}^{2}B^{i}dA^{i}+A^{1}A^{2}B^{2}$ studied in the Sec. IV A of ref.~\cite{zhang2021compatible}.}

This action (\ref{eq_S_CdA+AAC}) is invariant under gauge transformations:\
\begin{align}
A^{1}\rightarrow A^{1}+d\chi^{1},\  & C^{1}\rightarrow C^{1}+dT^{1}-\frac{2\pi q}{N_{1}}\left(\chi^{2}C^{3}-A^{2}T^{3}+\chi^{2}dT^{3}\right),\nonumber \\
A^{2}\rightarrow A^{2}+d\chi^{2},\  & C^{2}\rightarrow C^{2}+dT^{2}+\frac{2\pi q}{N_{2}}\left(\chi^{1}C^{3}-A^{1}T^{3}+\chi^{1}dT^{3}\right),\nonumber \\
C^{3}\rightarrow C^{3}+dT^{3},\  & A^{3}\rightarrow A^{3}+d\chi^{3}-\frac{2\pi q}{N_{3}}\left[\left(\chi^{1}A^{2}+\frac{1}{2}\chi^{1}d\chi^{2}\right)-\left(\chi^{2}A^{1}+\frac{1}{2}\chi^{2}d\chi^{1}\right)\right],
\label{eq_GT_AAC}
\end{align}
where $\chi^{i}$ and $T^{i}$ are 0-form and 2-form gauge parameters, respectively.
The gauge-invariant observable is
\begin{align}\mathcal{W}= & \exp\left\{ {\rm i}\int_{\omega_{1}}e_{1}\left[C^{1}+\frac{1}{2}\frac{2\pi q}{N_{1}}\left(d^{-1}A^{2}C^{3}-d^{-1}C^{3}A^{2}\right)\right]\right.\nonumber\\
 & +{\rm i}\int_{\omega_{2}}e_{2}\left[C^{2}+\frac{1}{2}\frac{2\pi q}{N_{2}}\left(d^{-1}C^{3}A^{1}-d^{-1}A^{1}C^{3}\right)\right]\nonumber\\
 & +\left.{\rm i}\int_{\gamma_{3}}e_{3}\left[A^{3}+\frac{1}{2}\frac{2\pi q}{N_{3}}\left(d^{-1}A^{1}A^{2}-d^{-1}A^{2}A^{1}\right)\right]\right\}
\label{eq_obs_AAC}
\end{align}
with $e_{1}$, $e_2$, and $e_3$ being integers. The derivation of gauge transformation (\ref{eq_GT_AAC}) and verification of gauge-invariance of (\ref{eq_obs_AAC}) is detailed in appendix~\ref{appendix_gauge_invariance_AAC}.
The operation $d^{-1}$   is defined as $d^{-1}C^{3}|_{\omega_{i}}\equiv\int_{\lambda_{3} |_{\omega_{i}}}C^{3}$ where $\lambda_{3} |_{\omega_{i}}$ is an open $3$D volume on $\omega_{i}$ with $i=1,2$. As a $2$-form, $d^{-1}C^{3}$ is
well-defined on $\omega_{i}$ if and only if $C^{3}$ is exact on
$\omega_{i}$, i.e., $\int_{\omega_{i}}C^{3}=0$.
The action of $d^{-1}$ on $A^{i}$ is defined as $d^{-1}A^{i}|_{\omega_{j}}\equiv\int_{\mu_{i}|_{\omega_{j}}}A^{i}$
and $d^{-1}A^{i}|_{\gamma_{3}}\equiv\int_{\mu_{i}|_{\gamma_{3}}}A^{i}$,
where $\mu_{i}$ is an open curve on $\omega_{j}$ or $\gamma_{3}$.
As a $0$-form, $d^{-1}A^{i}$ is well-defined on $\omega_{j}$ or $\gamma_{3}$ if
and only if the integral of $A^{i}$ over any $1$-dimensional closed
sub-manifolds of $\omega_{j}$ or $\gamma_{3}$ is zero. Since $A^{1}$, $A^{2}$, and $C^{3}$ are required to be exact on specific manifolds, it is straightforward
that $dA^{1}$, $dA^{2}$, and $dC^{3}$ are zero on corresponding manifolds. This in fact guarantees
the gauge invariance of (\ref{eq_obs_AAC}). From a geometric perspective, the {exactness} condition implies restrictions  {that some sub-manifolds are not linked}. In other words, the gauge invariance of observable is associated with the geometric interpretation of braiding process. The gauge fields and their fluxes are defined on specific sub-manifolds. For example, consider a gauge field $\mathscr{A}$ whose flux $d\mathscr{A}$ is defined on a closed $1$-manifold $\gamma_1$: $d\mathscr{A}=\delta^{\perp}\left(\gamma_1\right)$. Say, $\mathscr{A}$ is exact on another closed $1$-manifold $\gamma_2$, this means that $\int_{\Sigma}d\mathscr{A}=\int_{\gamma_2}\mathscr{A}=0$ with $\partial \Sigma=\gamma_2$. Since $\int_{\Sigma}d\mathscr{A}=\int\delta^{\perp}\left(\Sigma\right)\wedge\delta^{\perp}\left(\gamma_1\right)=\#\left(\Sigma\cap\gamma_1\right)$, the linking number of $\gamma_1$ and $\gamma_2$ is $0$ due to $\mathscr{A}$ is exact on $\gamma_2$. On the other hand, the gauge invariance of $\mathcal{W}$ ensures that $\left\langle\mathcal{W}\right\rangle$ is invariant under homotopic mapping of world-lines of excitations. However, sometimes $\left\langle\mathcal{W}\right\rangle$ does not clearly show how world-lines are linked . The exactness condition, required to meet gauge invariance, will give some hints since they reveal which sub-manifolds are not linked.

\begin{figure}
  \centering
  \includegraphics[width=0.8\textwidth]{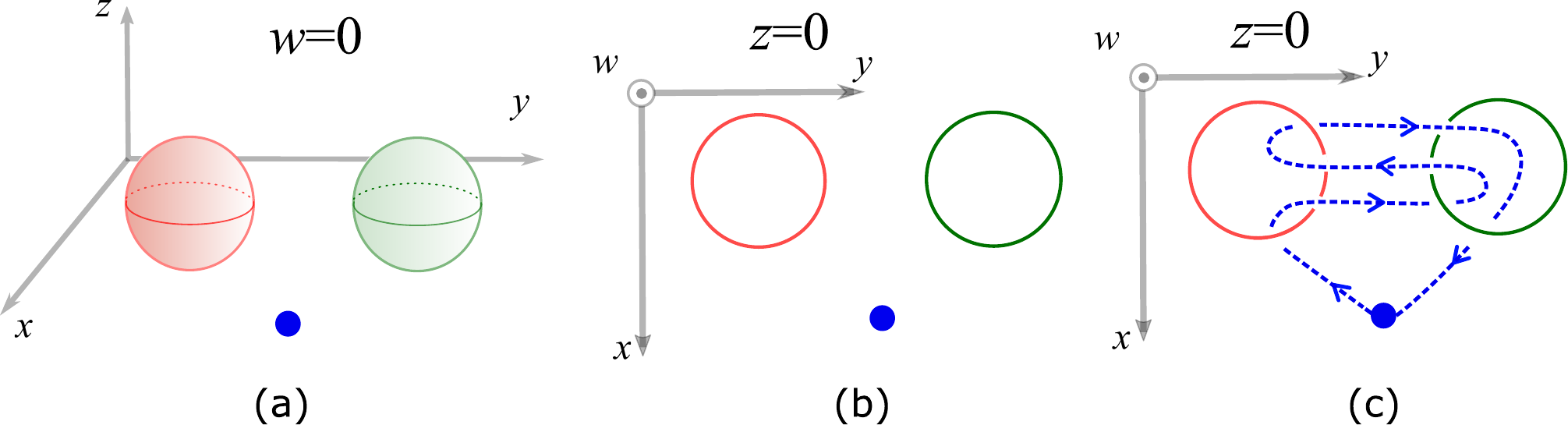}
  \caption{Particle-membrane-membrane braiding described by $AAC$ term. (a) Two membranes and one particle viewed in $xyz$-space with $w=0$. These two membranes are assumed to be static in space. (b) These two membrane excitations appears as two loops in $xyw$-space with $z=0$. (c) If viewed in $xyw$-space with $z=0$, the spatial trajectory of particle and two loops (projections of two membranes) form Borromean rings. In Borromean rings, any two of the two loops and particle's trajectory are not linked. This means that the particle's spatial trajectory is not linked with any of two membranes and  neither the two membranes, which matches the exactness conditions that imply specific sub-manifolds are not linked. This particle-membrane-membrane braiding can be considered as the $5$D analog of particle-loop-loop braiding described by $AAB$ term in $4$D \cite{yp18prl}.}
  \label{fig_AAC}
\end{figure}

The expectation value is given by $\left\langle \mathcal{W}\right\rangle =\frac{1}{\mathcal{Z}}\int\mathscr{D}A\mathscr{D}C\mathcal{W}\exp\left({\rm i}S\right)$.
Integrating out $C^{1}$, $C^{2}$, and $A^{3}$ implies $A^{1,2}=-\frac{2\pi e_{1,2}}{N_{1,2}}\delta^{\perp}\left(\Xi_{1,2}\right)$ and $C^{3}=-\frac{2\pi e_{3}}{N_{3}}\delta^{\perp}\left(\Sigma_{3}\right)$
with $\partial\Xi_{1,2}=\omega_{1,2}$ and $\partial\Sigma_{3}=\gamma_{3}$.  Here, {$\omega$ stands for $3$D closed volume, $\gamma$ stands for $1$D closed curve, etc, as explained in footnote \ref{footnote_symbol_notation}.}
Putting these solutions back in $\left\langle \mathcal{W}\right\rangle $,
we have
\begin{align}
\int qA^{1}A^{2}C^{3}= & -\frac{\left(2\pi\right)^{3}qe_{1}e_{2}e_{3}}{N_{1}N_{2}N_{3}}\int\delta^{\perp}\left(\Xi_{1}\right)\wedge\delta^{\perp}\left(\Xi_{2}\right)\delta^{\perp}\left(\Sigma_{3}\right),\nonumber\\
= & -\frac{pN_{1}N_{2}N_{3}}{\left(2\pi\right)^{2}N_{123}}\frac{\left(2\pi\right)^{3}e_{1}e_{2}e_{3}}{N_{1}N_{2}N_{3}}\#\left(\Xi_{1}\cap\Xi_{2}\cap\Sigma_{3}\right)\nonumber\\
= & -\frac{2\pi pe_{1}e_{2}e_{3}}{N_{123}}\#\left(\Xi_{1}\cap\Xi_{2}\cap\Sigma_{3}\right);
\end{align}
\begin{align}
\int_{\omega_{1}}e_{1}\left[\frac{q}{2}\left(d^{-1}A^{2}C^{3}-d^{-1}C^{3}A^{2}\right)\right]= & \frac{\pi pe_{1}e_{2}e_{3}}{N_{123}}\left[\int\delta^{\perp}\left(\omega_{1}\right)\wedge\delta^{\perp}\left(\Sigma_{3}\right)\wedge d^{-1}\delta^{\perp}\left(\Xi_{2}\right)|_{\omega_{1}}\right.\nonumber\\
 & -\left.\int\delta^{\perp}\left(\omega_{1}\right)\wedge\delta^{\perp}\left(\Xi_{2}\right)\wedge d^{-1}\delta^{\perp}\left(\Sigma_{3}\right)|_{\omega_{1}}\right].
\end{align}
By definition, $d^{-1}\delta^{\perp}\left(\Xi_{2}\right)|_{\omega_{1}}=\int_{\mu_{2}|_{\omega_{1}}}\delta^{\perp}\left(\Xi_{2}\right)=\int \delta^{\perp}\left(\mu_{2}|_{\omega_{1}} \right) \wedge \delta^{\perp}\left(\Xi_{2}\right)$, etc. Thus
\begin{align}
\int_{\omega_{1}}e_{1}\left[\frac{q}{2}\left(d^{-1}A^{2}C^{3}-d^{-1}C^{3}A^{2}\right)\right]= & \frac{\pi pe_{1}e_{2}e_{3}}{N_{123}}\left[\int\delta^{\perp}\left(\omega_{1}\right)\wedge\delta^{\perp}\left(\Sigma_{3}\right)\wedge\int\delta^{\perp}\left(\mu_{2}|_{\omega_{1}}\right)\wedge\delta^{\perp}\left(\Xi_{2}\right)\right.\nonumber\\
 & -\left.\int\delta^{\perp}\left(\omega_{1}\right)\wedge\delta^{\perp}\left(\Xi_{2}\right)\wedge\int\delta^{\perp}\left(\lambda_{3}|_{\omega_{1}}\right)\wedge\delta^{\perp}\left(\Sigma_{3}\right)\right]\nonumber\\
= & \frac{\pi pe_{1}e_{2}e_{3}}{N_{123}}\left[\#\left(\omega_{1}\cap\Sigma_{3}\cap\mu_{2}|_{\omega_{1}}\cap\Xi_{2}\right)\right.\nonumber\\
 & -\left.\#\left(\omega_{1}\cap\Xi_{2}\cap\lambda_{3}|_{\omega_{1}}\cap\Sigma_{3}\right)\right].
\end{align}
In a similar manner, we can calculate the remaining parts of $\left \langle\mathcal{W}\right\rangle$
and end up with
\begin{align}
\left\langle \mathcal{W}\right\rangle = & \exp\left\{ -\frac{{\rm i}2\pi pe_{1}e_{2}e_{3}}{N_{123}}\#\left(\Xi_{1}\cap\Xi_{2}\cap\Sigma_{3}\right)\right.\nonumber\\
 & +\frac{{\rm i}\pi pe_{1}e_{2}e_{3}}{N_{123}}\left[\#\left(\omega_{1}\cap\Sigma_{3}\cap\mu_{2}|_{\omega_{1}}\cap\Xi_{2}\right)-\#\left(\omega_{1}\cap\Xi_{2}\cap\lambda_{3}|_{\omega_{1}}\cap\Sigma_{3}\right)\right]\nonumber\\
 & +\frac{{\rm i}\pi pe_{1}e_{2}e_{3}}{N_{123}}\left[\#\left(\omega_{2}\cap\Xi_{1}\cap\lambda_{3}|_{\omega_{2}}\cap\Sigma_{3}\right)-\#\left(\omega_{2}\cap\Sigma_{3}\cap\mu_{1}|_{\omega_{2}}\cap\Xi_{1}\right)\right]\nonumber\\
 & +\left.\frac{{\rm i}\pi pe_{1}e_{2}e_{3}}{N_{123}}\left[\#\left(\gamma_{3}\cap\Xi_{2}\cap\mu_{1}|_{\gamma_{3}}\cap\Xi_{1}\right)-\#\left(\gamma_{3}\cap\Xi_{1}\cap\mu_{2}|_{\gamma_{3}}\cap\Xi_{2}\right)\right]\right\} .
\label{eq_exp_AAC}
\end{align}
Geometrically, $\left\langle \mathcal{W}\right\rangle $ is determined by counting the intersections of several sub-manifolds in $5$D. According to their dimension, we can interpret their relations with excitations. The $1$D $\gamma_i$ can be understood as the closed world-line of particle. The $3$D $\omega_i$ can be regarded as the closed world-volume of membrane. $\Sigma_i$ and $\Xi_i$ are Seifert (hyper)surfaces: $\partial \Sigma_i=\gamma_i$ and $\partial \Xi_i=\omega_i$. We can see that all these sub-manifolds are related to one particle and two membranes. In this sense, we believe that the $AAC$ term along with its gauge-invariant observable~(\ref{eq_obs_AAC}) describe the braiding process of one particle and two membranes. Exploited with the exactness conditions, figure~\ref{fig_AAC} illustrates this particle-membrane-membrane braiding in $3$D space.

In addition, the expectation value (\ref{eq_exp_AAC}) is expressed as the sum of several terms: $\left\langle \mathcal{W}\right\rangle =\exp\left(-\frac{{\rm i}2\pi pe_{1}e_{2}e_{3}}{N_{123}}\mathfrak{L}\right)$ with
\begin{align}
\mathfrak{L} = & \#\left(\Xi_{1}\cap\Xi_{2}\cap\Sigma_{3}\right)\nonumber\\
 & -\frac{1}{2}\left[\#\left(\omega_{1}\cap\Sigma_{3}\cap\mu_{2}|_{\omega_{1}}\cap\Xi_{2}\right)-\#\left(\omega_{1}\cap\Xi_{2}\cap\lambda_{3}|_{\omega_{1}}\cap\Sigma_{3}\right)\right]\nonumber\\
 & -\frac{1}{2}\left[\#\left(\omega_{2}\cap\Xi_{1}\cap\lambda_{3}|_{\omega_{2}}\cap\Sigma_{3}\right)-\#\left(\omega_{2}\cap\Sigma_{3}\cap\mu_{1}|_{\omega_{2}}\cap\Xi_{1}\right)\right]\nonumber\\
 & -\frac{1}{2}\left[\#\left(\gamma_{3}\cap\Xi_{2}\cap\mu_{1}|_{\gamma_{3}}\cap\Xi_{1}\right)-\#\left(\gamma_{3}\cap\Xi_{1}\cap\mu_{2}|_{\gamma_{3}}\cap\Xi_{2}\right)\right].
 \label{eq_linking_number_AAC}
\end{align}
$\#\left(\Xi_{1}\cap\Xi_{2}\cap\Sigma_{3}\right)$ counts the signed
intersections of $\Xi_{1}$, $\Xi_{2}$ and $\Sigma_{3}$ and other terms
have similar geometric meanings. Each term in $\mathfrak{L}$ (\ref{eq_linking_number_AAC})
is not homotopic invariant because it depends on the choices of Seifert
(hyper)surfaces. However, $\mathfrak{L}$ as the sum of these terms,
is homotopic invariant due to the gauge invariance of $\left\langle \mathcal{W}\right\rangle $.
We see that a mathematical invariant is obtained via physical gauge-invariant
field theory. The formula of $\mathfrak{L}$ is similar to the equation
(9) of ref.~\cite{yp18prl} which computes the Milnor's triple linking
number from a $4$D TQFT. In fact, the Milnor's triple linking number
of $3$ closed curves is also expressed as the sum of several homotopic-variant
terms \cite{milnor1954link,mellor2003geometric}. Analogously, eq.~(\ref{eq_linking_number_AAC})
can be regarded as the ``triple linking number'' of closed world-lines (in general meaning) of two membranes and one particle in $5$D. In the following main text, we will see more observables whose expectation values are expressed in a similar
manner. It is interesting to reveal the mathematical structure behind them.

\emph{$AdAdA$ twisted topological term}. Next, we consider the $AdAdA$ term, i.e., the Chern-Simons term in $5$D.\footnote{When $BF$ term is not included, the theory is studied in ref.~\cite{lapa17}.} For a gauge group $G=\prod_{i=1}^{n}\mathbb{Z}_{N_{i}}$,
the type-I $BF$ theory with $AdAdA$ term has the form of
\begin{equation}
S=S_{BF}+S_{AdAdA}=\int\sum_{i=1}^{n}\frac{N_{i}}{2\pi}C^{i}dA^{i}+qA^{i}dA^{j}dA^{k}
\end{equation}
with gauge transformations $A^{i}\rightarrow A^{i}+d\chi^{i}$ and
$C^{i}\rightarrow C^{i}+dT^{i}$. The indices of $A^{i}dA^{j}dA^{k}$, same or different, take values from $\left\{ 1,2,\cdots,n\right\} $. The simplest type-I $BF$ theory with $AdAdA$ term  is
\begin{equation}
S=\int\frac{N_{1}}{2\pi}C^{1}dA^{1}+qA^{1}dA^{1}dA^{1}\label{eq_action_3_membrane_zn1}
\end{equation}
with $q=\frac{p}{\left(2\pi\right)^{2}}$, $p\in\mathbb{Z}_{N_{1}}$. This action (\ref{eq_action_3_membrane_zn1}) describes a $\mathbb{Z}_{N_{1}}$ $5$D topological order.
The gauge-invariant observable for (\ref{eq_action_3_membrane_zn1}) is
\begin{equation}
\mathcal{W}=\exp\left({\rm i}e_{1}\int_{\omega_{1}}C^{1}\right)=\exp\left[{\rm i}e_{1}\int C^{1}\wedge\delta^{\perp}\left(\omega_{1}\right)\right]
\end{equation}
with the expectation value
\begin{equation}
\left\langle \mathcal{W}\right\rangle =\exp\left[-\frac{{\rm i}2\pi pe_{1}e_{1}e_{1}}{N_{1}N_{1}N_{1}}\#\left(\Xi_{1}\cap\omega_{1}\cap\omega_{1}\right)\right].
\end{equation}
If the gauge group is $G=\mathbb{Z}_{N_{1}}\times\mathbb{Z}_{N_{2}}$,
the $AdAdA$ twisted term can also be $A^{1}dA^{2}dA^{1}$ or $A^{2}dA^{1}dA^{2}$.
They are the only two linearly independent
terms because $d\left(A^{1}A^{2}dA^{1}\right)=dA^{1}A^{2}dA^{1}-A^{1}dA^{2}dA^{1}$
and $d\left(A^{1}A^{2}dA^{2}\right)=dA^{1}A^{2}dA^{2}-A^{1}dA^{2}dA^{2}$. The corresponding
type-I $BF$ theory and gauge-invariant observable is
\begin{align}
S= & \int\sum_{i=1}^{2}\frac{N_{i}}{2\pi}C^{i}dA^{i}+qA^{1}dA^{2}dA^{1},\\
\mathcal{W}= & \exp\left({\rm i}\sum_{i=1}^{2}e_{i}\int_{\omega_{i}}C^{i}\right)=\exp\left[{\rm i}\sum_{i=1}^{2}e_{i}\int C^{i}\wedge\delta^{\perp}\left(\omega_{i}\right)\right],\\
\left\langle \mathcal{W}\right\rangle = & \exp\left[-\frac{{\rm i}2\pi pe_{1}e_{2}e_{1}}{N_{1}N_{2}N_{1}}\#\left(\Xi_{1}\cap\omega_{2}\cap\omega_{1}\right)\right];
\end{align}
or
\begin{align}
S'= & \int\sum_{i=1}^{2}\frac{N_{i}}{2\pi}C^{i}dA^{i}+q'A^{2}dA^{1}dA^{2},\\
\mathcal{W}'= & \exp\left({\rm i}\sum_{i=1}^{2}e_{i}\int_{\omega_{i}}C^{i}\right)=\exp\left[{\rm i}\sum_{i=1}^{2}e_{i}\int C^{i}\wedge\delta^{\perp}\left(\omega_{i}\right)\right],\\
\left\langle \mathcal{W}'\right\rangle = & \exp\left[-\frac{{\rm i}2\pi pe_{2}e_{1}e_{2}}{N_{2}N_{1}N_{2}}\#\left(\Xi_{2}\cap\omega_{1}\cap\omega_{2}\right)\right].
\end{align}
The coefficients of twisted terms are $q=\frac{p}{\left(2\pi\right)^{2}}$, $p\in\mathbb{Z}_{N_{12}}$ and $q'=\frac{p'}{\left(2\pi\right)^{2}}$, $p'\in\mathbb{Z}_{N_{12}}$; $\mathbb{Z}_{N_{12}}$ is the greatest common divisor of $N_1$ and $N_2$.
The general case is that $G=\prod_{i=1}^{3}\mathbb{Z}_{N_{i}}$ in which
the twisted term is $A^{1}dA^{2}dA^{3}$. In fact, $A^{1}dA^{2}dA^{3}$, $A^{2}dA^{3}dA^{1}$, and $A^{3}dA^{1}dA^{2}$ are identical up to a boundary term because $d\left(A^{1}A^{2}dA^{3}\right)=A^{2}dA^{3}dA^{1}-A^{1}dA^{2}dA^{3}$
and $d\left(A^{1}dA^{2}A^{3}\right)=A^{3}dA^{1}dA^{2}-A^{1}dA^{2}dA^{3}$. The type-I $BF$ theory
with $A^{1}dA^{2}dA^{3}$ is
\begin{equation}
S =\int\sum_{i=1}^{3}\frac{N_{i}}{2\pi}C^{i}dA^{i}+q A^{1}dA^{2}dA^{3}\label{eq_action_3_membrane_zn1zn2zn3}
\end{equation}
with $q =\frac{p }{\left(2\pi\right)^{2}}$, $p \in\mathbb{Z}_{N_{123}}$.
For this type-I $BF$ theory~(\ref{eq_action_3_membrane_zn1zn2zn3}), the gauge-invariant observable is
\begin{equation}
\mathcal{W} =\exp\left({\rm i}\sum_{i=1}^{3}e_{i}\int_{\omega_{i}}C^{i}\right)=\exp\left[{\rm i}\sum_{i=1}^{3}e_{i}\int C^{i}\wedge\delta^{\perp}\left(\omega_{i}\right)\right],
\end{equation}
and its expectation value is
\begin{equation}
\left\langle \mathcal{W} \right\rangle =\exp\left[-\frac{{\rm i}2\pi p e_{1}e_{2}e_{3}}{N_{1}N_{2}N_{3}}\#\left(\Xi_{1}\cap\omega_{2}\cap\omega_{3}\right)\right].
\end{equation}
We conclude that the $A^{i}dA^{j}dA^{k}$ twisted term describes the braiding process of three membranes, in which membranes move in $5$D leaving their intersecting world-volumes described by $\#\left(\Xi_{i}\cap\omega_{j}\cap\omega_{k}\right)$. For the action $S\sim\int \sum BdA +A^i dA^j dA^k$, the expectation value of gauge-invariant observable takes the form of $\left\langle\mathcal{W}\right\rangle\sim\exp\left[\#\left(\Xi_i\cap \omega_j \cap \omega_k\right)\right]$. All these three sub-manifolds are related with membranes: $\omega_j$ and $\omega_k$ can be regarded as the closed world-volumes of membrane $i$ and $j$; $\Xi_i$ is the Seifert hypersurface of $\omega_i$, the closed world-volume of membrane $i$. $\left(\Xi_i\cap \omega_j \cap \omega_k\right)$ can be seen as the intersection of three membranes' world-volumes. Therefore, we believe that the $AdAdA$ term describes the braiding of three membrane whose phase shift is related to $\#\left(\Xi_i\cap \omega_j \cap \omega_k\right)$.

\emph{$AAAdA$ twisted topological term}. Then we consider the $AAAdA$ twisted term. Since $A$ is $1$-form,
the three $A$'s in $AAAdA$ have to be different, thus $G=\prod_{i=1}^{n}\mathbb{Z}_{N_{i}}$
with $n\geq3$ is required. If $G=\prod_{i=1}^{3}\mathbb{Z}_{N_{i}}$
, the twisted term can be $A^{1}A^{2}A^{3}dA^{1}$, $A^{1}A^{2}A^{3}dA^{2},$
or $A^{1}A^{2}A^{3}dA^{3}$. Take $A^{1}A^{2}A^{3}dA^{1}$ as an example,
the type-I $BF$ theory with $A^{1}A^{2}A^{3}dA^{1}$ is
\begin{equation}
S=S_{BF}+S_{AAAdA}=\int\sum_{i=1}^{3}\frac{N_{i}}{2\pi}C^{i}dA^{i}+qA^{1}A^{2}A^{3}dA^{1}
\end{equation}
with $q=\frac{pN_{1}N_{2}N_{3}}{\left(2\pi\right)^{3}N_{123}}$, $p\in\mathbb{Z}_{N_{123}}$.
The corresponding gauge transformations are
\begin{align}
A^{1}\rightarrow A^{1}+d\chi^{1},\  & C^{1}\rightarrow C^{1}+dT^{1}-\frac{2\pi q}{N_{1}}\left(A^{1}d\chi^{2}A^{3}+A^{1}A^{2}d\chi^{3}+A^{1}d\chi^{2}d\chi^{3}\right),\nonumber \\
A^{2}\rightarrow A^{2}+d\chi^{2},\  & C^{2}\rightarrow C^{2}+dT^{2}+\frac{2\pi q}{N_{2}}\left(d\chi^{1}A^{3}A^{1}+d\chi^{1}d\chi^{3}A^{1}\right),\nonumber \\
A^{3}\rightarrow A^{3}+d\chi^{3},\  & C^{3}\rightarrow C^{3}+dT^{3}+\frac{2\pi q}{N_{3}}\left(-d\chi^{1}A^{2}A^{1}-d\chi^{1}d\chi^{2}A^{1}\right).
\end{align}
If $G=\prod_{i=1}^{4}\mathbb{Z}_{N_{i}}$, possible $AAAdA$ terms
are $A^{1}A^{2}A^{3}dA^{4}$, $A^{2}A^{3}A^{4}dA^{1}$, $A^{3}A^{4}A^{1}dA^{2}$,
and $A^{4}A^{1}A^{2}dA^{3}$. Since $d\left(A^{1}A^{2}A^{3}A^{4}\right)=dA^{1}A^{2}A^{3}A^{4}-A^{1}dA^{2}A^{3}A^{4}+A^{1}A^{2}dA^{3}A^{4}-A^{1}A^{2}A^{3}dA^{4}$,
only three of them are linearly independent. We consider $A^{1}A^{2}A^{3}dA^{4}$
as an example and the corresponding type-I $BF$ theory is
\begin{equation}
S=S_{BF}+S{}_{AAAdA}=\int\sum_{i=1}^{4}\frac{N_{i}}{2\pi}C^{i}dA^{i}+qA^{1}A^{2}A^{3}dA^{4}
\end{equation}
with $q=\frac{pN_{1}N_{2}N_{3}}{\left(2\pi\right)^{3}N_{123}}$, $p\in\mathbb{Z}_{N_{1234}}$, $N_{1234}$ is the greatest common divisor of $N_1$, $N_2$, $N_3$, and $N_4$. The gauge transformations are
\begin{equation}
A^{i}\rightarrow  A^{i}+d\chi^{i},\ C^{i}\rightarrow C^{i}+dT^{i}+\frac{2\pi q}{N_{i}}\sum_{j,k}\epsilon^{ijk4}\left(A^{j}\chi^{k}dA^{4}-\frac{1}{2}\chi^{j}d\chi^{k}dA^{4}\right).
\end{equation}
The gauge-invariant observable is
\begin{equation}
\mathcal{W}=\exp\left\{ {\rm i}\sum_{i=1}^{4}\int_{\omega_{i}}e_{i}\left[C^{i}+\frac{1}{2}\frac{2\pi q}{N_{i}}\sum_{j,k}\epsilon^{ijk4}\left(d^{-1}A^{j}\right)A^{k}dA^{4}\right]\right\}
\end{equation}
and its expectation value is
\begin{align}
\left\langle \mathcal{W}\right\rangle = & \exp\left\{ \frac{{\rm i}2\pi pe_{1}e_{2}e_{3}e_{4}}{N_{4}N_{123}}\#\left(\Xi_{1}\cap\Xi_{2}\cap\Xi_{3}\cap\omega_{4}\right)\right.\nonumber\\
 & +\left.\frac{{\rm i}\pi pe_{1}e_{2}e_{3}e_{4}}{N_{4}N_{123}}\sum_{i=1}^{4}\sum_{j,k}\epsilon^{ijk4}\#\left(\omega_{i}\cap\Xi_{k}\cap\omega_{4}\cap\mu_{j}|_{\omega_{i}}\cap\Xi_{j}\right)\right\}. \label{eq_vev_AAAdA}
 \end{align}
Similar to the case of $AdAdA$ twisted term, $\omega_i$ and $\Xi_i$ can be interpreted as membrane's closed world-volume and its Seifert hypersurface. Eq.~(\ref{eq_vev_AAAdA}) calculates the intersection of world-volumes of four membranes to give the phase shift. It makes sense to consider $AAAdA$ term describing the braiding of four membranes.

\emph{$AAAAA$ twisted topological term}. The last possible twisted term in type-I $BF$ theory is $AAAAA$. The
corresponding action is
\begin{equation}
S=S_{BF}+S_{AAAAA}=\int\sum_{i=1}^{5}\frac{N_{i}}{2\pi}C^{i}dA^{i}+qA^{1}A^{2}A^{3}A^{4}A^{5},
\end{equation}
where $q=\frac{pN_{1}N_{2}N_{3}N_{4}N_{5}}{\left(2\pi\right)^{4}N_{12345}},\ p\in\mathbb{Z}_{N_{12345}}$, $\mathbb{Z}_{N_{12345}}$ is the greatest common divisor of $N_1$, $N_2$, $N_3$, $N_4$, and $N_5$.
The gauge transformations are
\begin{align}A^{i}\rightarrow & A^{i}+d\chi^{i},\nonumber\\
C^{i}\rightarrow & C^{i}+dT^{i}-\frac{2\pi q}{N_{i}}\sum_{k<l<m}\epsilon^{ijklm}\chi^{j}A^{k}A^{l}A^{m}-\frac{2\pi q}{N_{i}}\sum_{j<k,l<m}\epsilon^{ijklm}\chi^{j}d\chi^{k}A^{l}A^{m}\nonumber\\
 & -\frac{2\pi q}{N_{i}}\sum_{j<k<l}\sum_{m=1}^{5}\epsilon^{ijklm}\chi^{j}d\chi^{k}d\chi^{l}A^{m}-\frac{2\pi q}{N_{i}}\sum_{j<k<l<m}\epsilon^{ijklm}\chi^{j}d\chi^{k}d\chi^{l}d\chi^{m}.
\end{align}
The gauge-invariant observable is
\begin{equation}
\mathcal{W}=\exp\left[{\rm i}\sum_{i=1}^{5}\int_{\omega_{i}}e_{i}\left(C^{i}+\frac{1}{4}\frac{2\pi q}{N_{i}}\sum_{k<l<m}\sum_{j=1}^{5}\epsilon^{ijklm}d^{-1}A^{j}A^{k}A^{l}A^{m}\right)\right]
\end{equation}
and its expectation value is
\begin{align}
\left\langle \mathcal{W}\right\rangle = & \exp\left\{ \frac{{\rm i}2\pi p\prod_{i=1}^{5}e_{i}}{N_{12345}}\#\left(\Xi_{1}\cap\Xi_{2}\cap\Xi_{3}\cap\Xi_{4}\cap\Xi_{5}\right)\right.\nonumber\\
 & +\left.\sum_{i=1}^{5}\frac{{\rm i}2\pi p\prod_{i=1}^{5}e_{i}}{4N_{12345}}\sum_{k<l<m}\sum_{j=1}^{5}\epsilon^{ijklm}\#\left(\omega_{i}\cap\Xi_{k}\cap\Xi_{l}\cap\Xi_{m}\cap\mu_{j}|_{\omega_{i}}\cap\Xi_{j}\right)\right\}.
\label{eq_vev_AAAAA}
\end{align}
As cases discussed above, $\omega_i$ and $\Xi_i$ can be treated as membrane's closed world-volume and its Seifert hypersurface. Consider a braiding of five membranes, eq.~(\ref{eq_vev_AAAAA}) exactly gives the phase shift by counting the intersection of five membranes' world-volumes.

\subsection{Braidings in mixed $BF$ theory with a twist\label{subsec_mixed_BF_a_twist}}
In a type-II $BF$ theory whose only $BF$ term is $\widetilde{B}dB$, there is no twisted term since $2$-form $\widetilde{B}$ and $B$ cannot make up a $5$-form twisted term. Therefore the only possible type-II $BF$ theory is $S=\int \sum_{i=1}^{n}\frac{N_{i}}{2\pi}\widetilde{B}^{i}dB^{i}$ with gauge group $G=\prod_{i=1}^{n}\mathbb{Z}_{N_{i}}$. We will come back to this action in section~\ref{sec_tqft_classification}.

Besides type-I or type-II $BF$ theory,  we study \emph{mixed} $BF$ theory with a twist which consists of \emph{two types} of $BF$ terms and one twisted term. Possible twisted terms in mixed $BF$ theory include $BBA$, $BAdA$, $AAdB$, and $AAAB$.

\emph{$BAdA$ twisted topological term}. One example of mixed $BF$ theory with $BAdA$ term is
\begin{equation}
S=S_{BF}+S_{BAdA}=\int\frac{N_{1}}{2\pi}C^{1}dA^{1}+\frac{N_{2}}{2\pi}\widetilde{B}^{2}dB^{2}+qB^{2}A^{1}dA^{1},
\end{equation}
where $q=\frac{pN_{2}N_{1}}{\left(2\pi\right)^{2}N_{12}},p\in\mathbb{Z}_{N_{12}}$.
This mixed $BF$ action is invariant up to boundary terms under
\begin{align}
A^{1}\rightarrow A^{1}+d\chi^{1},\  & C^{1}\rightarrow C^{1}+dT^{1}-\frac{2\pi q}{N_{1}}dV^{2}A^{1},\nonumber \\
B^{2}\rightarrow B^{2}+dV^{2},\  & \widetilde{B}^{2}\rightarrow\widetilde{B}^{2}+d\widetilde{V}^{2}-\frac{2\pi q}{N_{2}}d\chi^{1}A^{1}.
\end{align}
The gauge-invariant observable is
\begin{equation}
\mathcal{W}=\exp\left\{ {\rm i}\left[e_{1}\left(\int_{\omega_{1}}C^{1}+\frac{2\pi q}{N_{1}}\int_{\Xi_{1}}B^{2}dA^{1}\right)+e_{2}\left(\int_{\sigma_{2}}\widetilde{B}^{2}-\frac{2\pi q}{N_{2}}\int_{\Omega_{2}}A^{1}dA^{1}\right)\right]\right\} ,
\end{equation}
where $\partial\Omega_{2}=\sigma_{2}$ and $\partial\Xi_{1}=\omega_{1}$.
The expectation value is
\begin{equation}
\left\langle \mathcal{W}\right\rangle =\exp\left[-\frac{{\rm i}2\pi pe_{2}e_{1}e_{1}}{N_{1}N_{12}}\#\left(\Omega_{2}\cap\Xi_{1}\cap\omega_{1}\right)\right].
\end{equation}
Other possible  $BAdA$ terms include $B^{3}A^{1}dA^{2}$, $B^{3}A^{2}dA^{1}$, etc. Notice that
$d\left(B^{3}A^{1}A^{2}\right)=dB^{3}A^{1}A^{2}+B^{3}dA^{1}A^{2}-B^{3}A^{1}dA^{2}$,
only two of them are linearly independent. Thus the $AAdB$ twisted term can always be expressed by two proper $BAdA$ twisted terms, up to a boundary term. For now, we consider an example of mixed $BF$ theory with $B^{3}A^{1}dA^{2}$:
\begin{equation}
S=\int\sum_{i=1}^{2}\frac{N_{i}}{2\pi}C^{i}dA^{i}+\frac{N_{3}}{2\pi}\widetilde{B}^{3}dB^{3}+qB^{3}A^{1}dA^{2}
\end{equation}
which is gauge-invariant under
\begin{align}
A^{1}\rightarrow A^{1}+d\chi^{1},\  & C^{1}\rightarrow C^{1}+dT^{1}-\frac{2\pi q }{N_{1}}dV^{3}A^{2},\nonumber \\
A^{2}\rightarrow A^{2}+d\chi^{2},\  & C^{2}\rightarrow C^{2}+dT^{2},\nonumber \\
B^{3}\rightarrow B^{3}+dV^{3},\  & \widetilde{B}^{3}\rightarrow\widetilde{B}^{3}+d\widetilde{V}^{3}-\frac{2\pi q }{N_{3}}d\chi^{1}A^{2}.
\end{align}
The coefficient $q$ is quantized and periodic: $q=\frac{pN_{3}N_{1}}{\left(2\pi\right)^{2}N_{13}},p\in\mathbb{Z}_{N_{123}}$.
For this action, the gauge-invariant observable is
\begin{equation}
\mathcal{W}=  \exp\left\{ {\rm i}\left[e_{1}\left(\int_{\omega_{1}}C^{1}+\frac{2\pi q}{N_{1}}\int_{\Xi_{1}}B^{3}dA^{2}\right)+e_{2}\int_{\omega_{2}}C^{2}+e_{3}\left(\int_{\sigma_{3}}B^{3}-\frac{2\pi q}{N_{3}}\int_{\Omega_{3}}A^{1}dA^{2}\right)\right]\right\} .
\end{equation}
with its expectation value being
\begin{equation}
\left\langle \mathcal{W}\right\rangle =\exp\left[-\frac{{\rm i}2\pi pe_{1}e_{2}e_{3}}{N_{2}N_{13}}\#\left(\Omega_{3}\cap\Xi_{1}\cap\omega_{2}\right)\right],
\label{eq_vev_B3A1dA2}
\end{equation}
where $\omega_{1,2}$ are closed $3$D volumes, $\sigma_{3}$ is closed $2$D surface; $\Xi_1$ and $\Omega_3$ are Seifert hypersurfaces: $\partial \Xi_{1}=\omega_{1}$ and $\partial \Omega_{3}=\sigma_{3}$. If we consider the $A^{1}A^{2}dB^{3}$ or $A^{2}B^{3}dA^{1}$ term, the corresponding TQFT action, gauge transformation, and observable along with its expectation value, would be slightly different but share a similar pattern.

We notice that the $BAdA$ term shares a similar form of the $AAdA$ term in $4$D TQFT which describes the $3$-loop braiding in $4$D topological order. Since $BAdA$ and $AAdA$ term both have an $AdA$ part, the Chern-Simons term in $3$D, their corresponding braiding processes may share some similarities.

\begin{figure}
  \centering
  \includegraphics[width=0.8\textwidth]{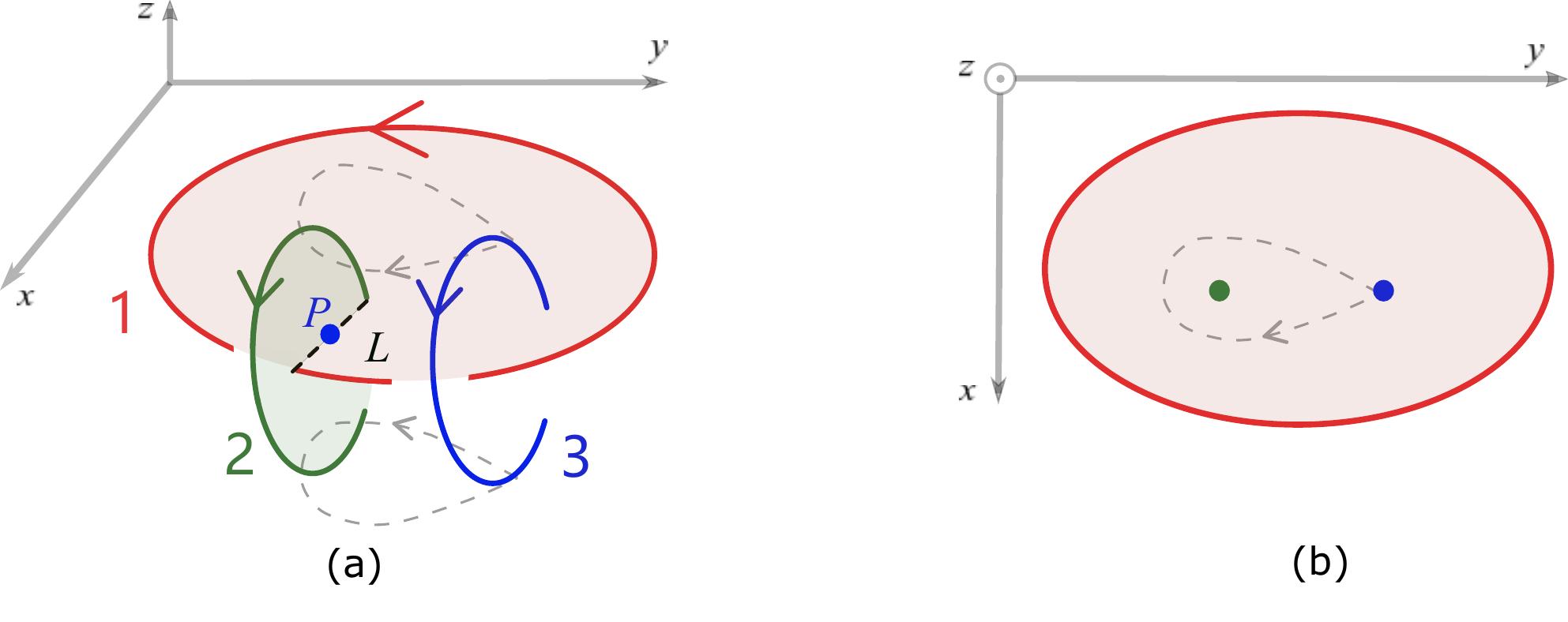}
  \caption{$3$-loop braiding described by $AAdA$ term in $3$D space. (a) The configuration of $3$-loop braiding described by $A^1 A^2 dA^3$ term. For this $3$-loop braiding, its phase shift is related to $\#\left(\Omega_{1}\cap\Omega_{2}\cap\sigma_{3}\right)$ as indicated in eq.~(\ref{eq_AAdA_ob_vev}). This linking number can be calculated via projecting $\Omega_1$, $\Omega_2$, and $\sigma_3$ from $4$D spacetime to $3$D space. After projection, $\Omega_1$ appears as the Seifert surface (red shaded area) of the static base loop (loop $1$); $\Omega_2$ appears as the Seifert surface (green shaded area) of loop $2$; $\sigma_3$ becomes the closed spatial trajectory of loop $3$. We can see that the \emph{projection} of $\left(\Omega_1 \cap \Omega_2\right)$ to $3$D space is the segment $L$ (black dash line). $\left(\Omega_1 \cap \Omega_2\right)$ is the world-sheet generated by $L$. Now we consider the spatial trajectory of loop $3$: shrinking itself, passing through loop $2$, expanding itself, finally back to its initial position, loop $3$ swaps a closed two-dimensional surface. This surface intersects with segment $L$ at a point $P$ (blue solid circle). In other words, the intersection points, $\left(\Omega_1 \cap \Omega_2 \cap \sigma_3\right)$, share the same spatial coordinates as that of point $P$. Since the spatial trajectory of loop $3$ only intersects with $L$ once, $\left(\Omega_1 \cap \Omega_2\right)$ and $\sigma_3$ intersect only at one specific moment, i.e., $\left(\Omega_1 \cap \Omega_2\cap \sigma_3\right)$'s temporal coordinate has only one value. At last, we can conclude that there is only $1$ intersection point of $\Omega_1$, $\Omega_2$, and $\sigma_3$, i.e., $\left|\#\left(\Omega_1 \cap \Omega_2 \cap \sigma_3\right)\right|=1$ with a sign determined by orientation. (b) The $3$-loop braiding can be viewed as an anyon braiding on the Seifert surface of the base loop.}
  \label{fig_AAdA}
\end{figure}

We first briefly review the $3$-loop braiding in $4$D spacetime
described by the $AAdA$ twisted term. Figure~\ref{fig_AAdA} illustrates a typical
$3$-loop braiding which can be described by the TQFT action
\begin{equation}
S=\int\sum_{i=1}^{3}\frac{N_{i}}{2\pi}B^{i}dA^{i}+\frac{pN_{1}N_{2}}{\left(2\pi\right)^{2}N_{12}}A^{1}A^{2}dA^{3}
\end{equation}
with $p\in\mathbb{Z}_{N_{123}}$. The gauge-invariant observable and
its expectation value are \cite{2016arXiv161209298P}
\begin{align}
\mathcal{W}= & \exp\left\{{\rm i}\left[\sum_{i=1}^{3}e_{i}\left(\int_{\sigma_{i}}B^{i}+\sum_{j}\frac{p\epsilon^{ij}}{2\pi N_{j}}\int_{\Omega_{i}}A^{j}dA^{3}\right)\right]\right\},\\
\left\langle \mathcal{W}\right\rangle = & \exp\left[\frac{2\pi{\rm i}pe_{1}e_{2}e_{3}}{N_{123}}\#\left(\Omega_{1}\cap\Omega_{2}\cap\sigma_{3}\right)\right].\label{eq_AAdA_ob_vev}
\end{align}
In eq. (\ref{eq_AAdA_ob_vev}), $\sigma_{i}$ is the closed $2$D
world-sheet of loop $i$ and $\Omega_{i}$ is the $3$D Seifert hypersurface
with $\partial\Omega_{i}=\sigma_{i}$. Derived from a $4$D field
theory, $\Omega_{1}\cap\Omega_{2}\cap\sigma_{3}$, intersection of
world-sheet and Seifert hypersurfaces, is a geometric object in $4$D
which is not so intuitive for an observer living in $3$D. However,
we can project it from $4$D spacetime to $3$D space so
that we can have a better understanding. More concretely, being projected
on $3$D space, the closed world-sheet of loop appears as
the closed spatial trajectory of loop, and the Seifert hypersurfaces
are observed as $2$D Seifert surfaces. Figure~\ref{fig_AAdA} illustrates how
to calculate $\#\left(\Omega_{1}\cap\Omega_{2}\cap\sigma_{3}\right)$
from a $3$-loop braiding.

\begin{figure}
  \centering
  \includegraphics[width=0.8\textwidth]{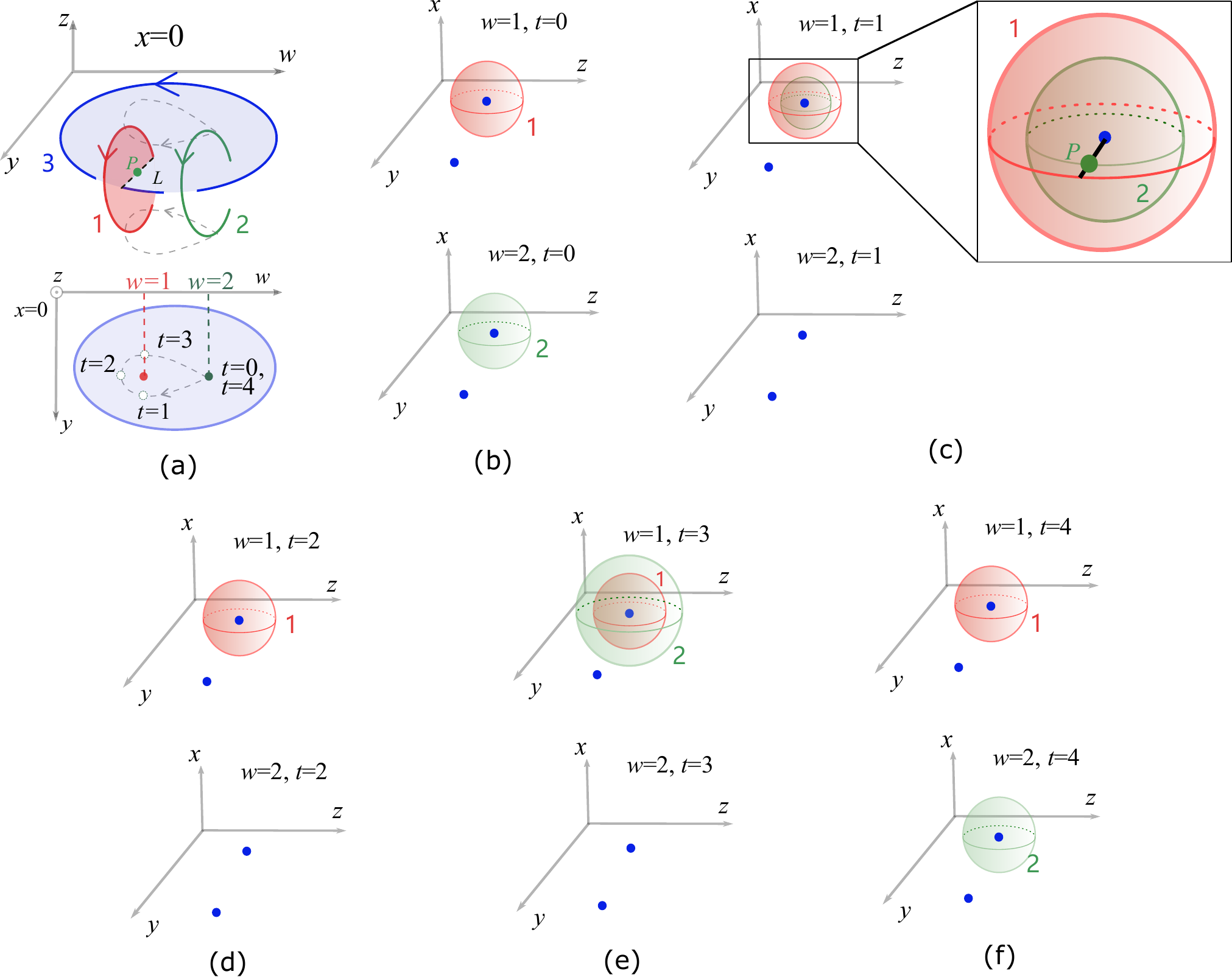}
  \caption{Loop-membrane-membrane braiding described by $BAdA$ term viewed in $3$D space. (a) Viewed in $ywz$-space with $x=0$, the two membranes ($1$ and $2$) appears as two loops that are linked with the loop excitation (loop $3$, blue) respectively. Loop $3$ and membrane $1$ are assumed to be static in space. This loop-membrane-membrane braiding looks like a $3$-loop braiding in $ywz$-space with $x=0$ while this $3$-loop braiding can be further viewed as a braiding of two anyons on the Seifert surface of loop $3$. The $\Omega_{3}$ in eq.~(\ref{eq_vev_B3A1dA2}) is projected to the $2$D Seifert surface of loop $3$. $\Xi_1$ is projected to the $3$D Seifert surface of static membrane $1$. Since in this $ywz$-space with $x=0$, static membrane $1$ appears as a static loop, $\Xi_1$'s projection shows as the this loop's $2$D Seifert surface. $\omega_2$ is projected to the closed spatial trajectory of membrane $2$ (loop $2$ in this $3$D space). We can see that these manifolds intersect at point $P$ (green solid circle, also refer to figure.~\ref{fig_AAdA}). This means that $\left|\#\left(\Omega_{3}\cap\Xi_{1}\cap\omega_{2}\right)\right|=1$ with a sign determined by orientation. (b)-(f) Two membranes viewed in $yzx$-space with $w=1$ and $w=2$ respectively at different moments. The position of membrane $2$ at different $t$'s are also labeled in (a). In these $3$D spaces, loop $3$ appears as two points (blue solid circles). In $yzx$-space with $w=1$, $\Omega_3$ appears as the $1$D Seifert surface of two points (loop $3$); $\Xi_1$ appears the $3$D Seifert surface of static membrane $3$; $\omega_2$ is the spatial trajectory of membrane $2$ that is the union of membrane $2$'s locations at $t=1$ and $t=3$ thus being discrete. The zoomed-in picture in (c) shows that these manifolds resulted from projection intersect at the point $P$ (green solid circle), corresponding to that illustrated in (a). This also means that $\left|\#\left(\Omega_{3}\cap\Xi_{1}\cap\omega_{2}\right)\right|=1$ with a sign determined by orientation.}
  \label{fig_BAdA}
\end{figure}

Now we move back to the $BAdA$ twisted term and the braiding of one
loop and two membranes it describes. For a $B^{3}A^{1}dA^{2}$ twisted
term, the phase shift from the braiding process is related to $\#\left(\Omega_{3}\cap\Xi_{1}\cap\omega_{3}\right)$,
see eq.~(\ref{eq_vev_B3A1dA2}), where $\Omega_{3}$ is a $3$D Seifert hypersurface,
$\Xi_{3}$ is a $4$D Seifert hypersurface, and $\omega_{3}$ is a
closed $3$D world-volume. Similarly, we can project $\Omega_{3}\cap\Xi_{1}\cap\omega_{3}$
from $5$D spacetime to $4$-dimensional space: (1) $\Omega_{3}$
is projected to be a $2$D Seifert surface that can be viewed as the
Seifert surface of a static loop; (2) $\Xi_{1}$ appears as a $3$D
Seifert hypersurface that bounds a static membrane; (3) world-volume
$\omega_{3}$ is projected to be a closed spatial trajectory of a
moving membrane. In $4$D space, we see this braiding process as follows.
The initial configuration is that a loop is linked with two membranes. If viewed in $3$D space, see figure~\ref{fig_BAdA}, a loop is linked with another
two loops (two membranes are saw as two loops in $3$D space).
Then one membrane moves, passing through another membrane, such that in
$3$D space a $3$-loop braiding is observed as shown in figure~\ref{fig_BAdA}.

The $3$-loop braiding described by $AAdA$ term can be understood
as a braiding of two anyons on the Seifert surface of the base loop,
see figure~\ref{fig_AAdA}. The latter is captured by the $AdA$ term, i.e., Chern-Simons term in $3$D. Since $BAdA$ and $AAdA$ terms
share a same $AdA$ part, we argue that the braiding process described
by $BAdA$ can be reduced to a braiding of two anyons on a Seifert
surface of the loop excitation, as shown in figure~\ref{fig_BAdA}.

\emph{$BBA$ twisted topological term}. An example of mixed $BF$ theory with $BBA$ term is
\begin{equation}
S=S_{BF}+S_{BBA}=\int\frac{N_{1}}{2\pi}\widetilde{B}^{1}dB^{1}+\frac{N_{2}}{2\pi}\widetilde{B}^{2}dB^{2}+\frac{N_{3}}{2\pi}C^{3}dA^{3}+qB^{1}B^{2}A^{3}
\end{equation}
where $q$ is a quantized and periodic coefficient: $q=\frac{pN_{1}N_{2}N_{3}}{\left(2\pi\right)^{2}N_{123}}, p\in\mathbb{Z}_{N_{123}}$.
The gauge transformations are
\begin{align}
B^{1}\rightarrow B^{1}+dV^{1},\  & \widetilde{B}^{1}\rightarrow\widetilde{B}^{1}+d\widetilde{V}^{1}+\frac{2\pi q}{N_{1}}\left(V^{2}A^{3}+B^{2}\chi^{3}+V^{2}d\chi^{3}\right),\nonumber \\
B^{2}\rightarrow B^{2}+dV^{2},\  & \widetilde{B}^{2}\rightarrow\widetilde{B}^{2}+d\widetilde{V}^{2}+\frac{2\pi q}{N_{2}}\left(V^{1}A^{3}+B^{1}\chi^{3}+V^{1}d\chi^{3}\right),\nonumber \\
A^{3}\rightarrow A^{3}+d\chi^{3},\  & C^{3}\rightarrow C^{3}+dT^{3}+\frac{2\pi q}{N_{3}}\left(-V^{1}B^{2}-B^{1}V^{2}-\frac{1}{2}V^{1}dV^{2}-\frac{1}{2}dV^{1}V^{2}\right),
\end{align}
where $V^{i}$ and $\widetilde{V}^{i}$ are different 1-form gauge parameters.
The gauge-invariant observable is
\begin{align}
\mathcal{W}= & \exp\left\{ {\rm i}\int_ {\sigma_{1}}e_{1}\left[\widetilde{B}^{1}-\frac{1}{2}\frac{2\pi q}{N_{1}}\left(d^{-1}B^{2}A^{3}+d^{-1}A^{3}B^{2}\right)\right]\right.\nonumber\\
 & +{\rm i}\int_{\sigma_{2}}e_{2}\left[\widetilde{B}^{2}-\frac{1}{2}\frac{2\pi q}{N_{2}}\left(d^{-1}B^{1}A^{3}+d^{-1}A^{3}B^{1}\right)\right]\nonumber\\
 & +\left.{\rm i}\int_{\omega_{3}}e_{3}\left[C^{3}-\frac{1}{2}\frac{2\pi q}{N_{3}}\left(d^{-1}B^{1}B^{2}+d^{-1}B^{2}B^{1}\right)\right]\right\}
\end{align}
and its expectation value is
\begin{align}
\left\langle \mathcal{W}\right\rangle = & \exp\left\{ -\frac{{\rm i}2\pi pe_{1}e_{2}e_{3}}{N_{123}}\#\left(\Omega_{1}\cap\Omega_{2}\cap\Xi_{3}\right)\right.\nonumber\\
 & +\frac{{\rm i}\pi pe_{1}e_{2}e_{3}}{N_{123}}\left[\#\left(\sigma_{1}\cap\Xi_{3}\cap\nu_{2} |_{\sigma_{1}}\cap\Omega_{2}\right)+\#\left(\sigma_{1}\cap\Omega_{2}\cap\mu_{3}|_{\sigma_{1}}\cap\Xi_{3}\right)\right]\nonumber\\
 & +\frac{{\rm i}\pi pe_{1}e_{2}e_{3}}{N_{123}}\left[\#\left(\sigma_{2}\cap\Xi_{3}\cap\nu_{1} |_{\sigma_{2}}\cap\Omega_{1}\right)+\#\left(\sigma_{2}\cap\Omega_{1}\cap\mu_{3}|_{\sigma_{2}}\cap\Xi_{3}\right)\right]\nonumber\\
 & +\left.\frac{{\rm i}\pi pe_{1}e_{2}e_{3}}{N_{123}}\left[\#\left(\omega_{3}\cap\Omega_{1}\cap\nu_{1} |_{\omega_{3}}\cap\Omega_{1}\right)+\#\left(\omega_{3}\cap\Omega_{2}\cap\nu_{2} |_{\omega_{3}}\cap\Omega_{2}\right)\right]\right\}.
 \label{eq_vev_BBA}
\end{align}
For a braiding process of two loops and one membrane, the phase shift should be obtained via counting the intersections of loops' world-sheets and membrane's world-volume. Eq.~(\ref{eq_vev_BBA}) just produces this phase shift once we see $\sigma_i$ ($\omega_i$) as closed world-sheet (world-volume) of loop (membrane) and notice that $\partial \Omega_i=\sigma_i$ and $\partial \Xi_i=\omega_i$. In this sense, we can say that the $BBA$ term corresponds to the braiding of two loops and one membrane.

\emph{$AAAB$ twisted topological term}. Finally we consider the $AAAB$ twisted term. An example of mixed $BF$ theory with $AAAB$ is
\begin{equation}
S=S_{BF}+S_{AAAB}=\int\sum_{i=1}^{3}\frac{N_{i}}{2\pi}C^{i}dA^{i}+\frac{N_{4}}{2\pi}\widetilde{B}^{4}dB^{4}+qA^{1}A^{2}A^{3}B^{4},
\label{eq_action_AAAB}
\end{equation}
with $q=\frac{pN_{1}N_{2}N_{3}N_{4}}{\left(2\pi\right)^{3}N_{1234}}$, $p\in\mathbb{Z}_{N_{1234}}$.
The gauge transformations are
\begin{align}
C^{i}\rightarrow & C^{i}+dT^{i}\nonumber \\
 & +\frac{2\pi q}{N_{i}}\sum_{j,k}\epsilon^{ijk}\left(-\chi^{j}A^{k}B^{4}-\frac{1}{2}A^{j}A^{k}V^{4}-\frac{1}{2}\chi^{j}d\chi^{k}B^{4}-\chi^{j}A^{k}dV^{4}+\frac{1}{2}d\chi^{j}\chi^{k}dV^{4}\right),\nonumber \\
\widetilde{B}^{4}\rightarrow & \widetilde{B}^{4}+d\widetilde{V}^{4}\nonumber \\
 & +\frac{2\pi q}{N_{4}}\sum_{i,j,k}\epsilon^{4ijk}\left(-\frac{1}{2}\chi^{i}A^{j}A^{k}+\frac{1}{2}A^{i}\chi^{j}d\chi^{k}-\frac{1}{6}\chi^{i}d\chi^{j}d\chi^{k}\right).
\end{align}
The gauge-invariant observable is
\begin{align}
\mathcal{W}= & \exp \left\{ {\rm i}\sum_{i=1}^{3}\int_{\omega_{i}}e_{i}\left[C^{i}-\frac{1}{3}\frac{2\pi q}{N_{i}}\sum_{j,k}\epsilon^{ijk}\left(B^{4}A^{j}d^{-1}A^{k}-\frac{1}{2}A^{j}A^{k}d^{-1}B^{4}\right)\right]\right.\nonumber\\
 & +\left.{\rm i}\int_{\sigma_{4}}e_{4}\left(\widetilde{B}^{4}-\frac{1}{6}\frac{2\pi q}{N_{4}}\sum_{i,j,k}\epsilon^{ijk}A^{i}A^{j}d^{-1}A^{k}\right)\right\}.
\end{align}
Its expectation value is
\begin{align}
\left\langle \mathcal{W}\right\rangle = & \exp\left\{ \frac{{\rm i}2\pi p\prod_{i=1}^{4}e_{i}}{N_{1234}}\#\left(\Xi_{1}\cap\Xi_{2}\cap\Xi_{3}\cap\Omega_{4}\right)\right.\nonumber\\
 & -\frac{{\rm i}2\pi p\prod_{i=1}^{4}e_{i}}{3N_{1234}}\sum_{i,j,k}\epsilon^{ijk}\left[\#\left(\omega_{i}\cap\Omega_{4}\cap\Xi_{j}\cap\mu_{k} |_{\omega_{i}}\cap\Xi_{k}\right)-\frac{1}{2}\#\left(\omega_{i}\cap\Xi_{j}\cap\Xi_{k}\cap\nu_{4} |_{\omega_{i}}\cap\Omega_{4}\right)\right]\nonumber\\
 & -\left.\frac{{\rm i}2\pi p\prod_{i=1}^{4}e_{i}}{6N_{1234}}\sum_{i,j,k}\epsilon^{ijk}\#\left(\sigma_{4}\cap\Xi_{i}\cap\Xi_{j}\cap\mu_{k} |_{\omega_{i}}\cap\Xi_{k}\right)\right\}.
 \label{eq_vev_AAAB}
\end{align}
In eq.~(\ref{eq_vev_AAAB}), $\sigma_i$ stands for $2$D closed world-sheet of loop, $\omega_i$ stands for $3$D closed world-volume of membrane; $\partial \Xi_i=\omega_i$, $\partial\Omega_i=\sigma_i$. If we consider a braiding of three membranes and one loop, eq.~(\ref{eq_vev_AAAB}) just counts the intersections of these excitations' world-sheets and world-volume in a gauge-invariant way and gives the phase shift. Thus we believe that this $AAAB$ term along with its gauge-invariant observable describes the braiding of three membranes and one loop.

Before closing this section, we point out that the $BC$ term is also a possible $5$-form twisted
term but it is not taken into consideration in the present paper. The $BC$
term in $5$D $BF$ theory looks like the $BB$ term in $4$D. Both of which are quadratic terms just like $BF$ terms.
So, inclusion of $BC$ term may complicate the analysis. In $4$D, $BB$ term can drastically change the gauge group, which means that the coefficients of $BF$ terms cannot uniquely determine the gauge group $G$ \cite{PhysRevB.99.235137,Kapustin2014}. We regard that $BC$ term may have similar effect. In addition, $BB$ term in $4$D can change the self-statistics (bosonic or fermionic) of particles~\cite{bti2,string7,Kapustin2014,2012FrPhy...7..150W,kb2015,PhysRevB.99.235137}. $BC$ term may also play important role in transmuting self-statistics of particles. Since we are only interested in $5$D topological orders with all bosons in this paper, we do not take $BC$ term into account. But it will be definitely exciting to incorporate $BC$ term in the future and study canonical quantization and equation of mention in the presence of source terms \cite{PhysRevB.99.235137}.

\section{TQFTs in $5$ dimensions:
within and beyond Dijkgraaf-Witten cohomological classification}\label{sec_tqft_classification}
In section~\ref{sec_topo_excitation_braid}, we have studied type-I and mixed $BF$
theories with \emph{single} twisted term {as well as} the corresponding braiding
processes in $5$D topological orders. It is natural to ask what if
more twisted terms are {considered in these} $BF$ theories. Similar to the case
in $4$D~\cite{zhang2021compatible}, the compatibility of twisted terms in $5$D should also be
considered. Using the technique developed in ref.~\cite{zhang2021compatible}, we can exhaust $5$D
TQFT actions with all allowed twisted terms once the gauge group is given.
Then, we study the classification of TQFT actions, which is the main
purpose of this section. The complexity of classification of $5$D
TQFT actions mostly comes from the fact that there exist two types of $BF$
terms in $5$D. For a given gauge group $G=\prod_{i=1}^{n}\mathbb{Z}_{N_{i}}$,
since each $\mathbb{Z}_{N_{i}}$ cyclic gauge subgroup can be encoded
in one of {two} types of $BF$ terms, different distributions of $\mathbb{Z}_{N_{i}}$
to type-I and type-II $BF$ terms lead to different twisted terms. Once
the $BF$ terms are determined, all possible twisted terms can
be figured out. An action with all allowed twisted terms which are detailed in sec.~\ref{subsec_type_1_BF_a_twist} and sec.~\ref{subsec_mixed_BF_a_twist}, can be written and its classification can be obtained by counting the coefficients of twisted terms. In the remaining part of this
section, we will discuss classification of some $BF$ theories
with multiple twisted terms for $G=\mathbb{Z}_{N_{1}}$ (sec.~\ref{subsec_class_1Zn}), $G=\prod_{i=1}^{2}\mathbb{Z}_{N_{i}}$ (sec.~\ref{subsec_class_Zn1Zn2}), $G=\prod_{i=1}^{3}\mathbb{Z}_{N_{i}}$ (sec.~\ref{subsec_class_3Zn}), and $G=\prod_{i=1}^{n}\mathbb{Z}_{N_{i}}$ with $n\geq 4$ (sec.~\ref{subsec_class_4Zn}). We find that
only a part of type-I $BF$ theories with twisted terms are consistent with Dijkgraaf-Witten cohomological classification. The other part of type-I $BF$ theories and all mixed $BF$ theories are totally beyond the group cohomology classification.

Some notations used in this section need to be explained here before we move forward.
We use a set $\left\{ N_{i}\right\} _{i=1}^{n}=\left\{ N_{1},\cdots,N_{n}\right\}$
to label the gauge group $G=\prod_{i=1}^{n}\mathbb{Z}_{N_{i}}$. Then,
a subset $\alpha \subset\left\{ N_{i}\right\} _{i=1}^{n}$ is introduced
to denote which $\mathbb{Z}_{N_{i}}$ gauge subgroup is encoded in
type-I $BF$ term. For example, consider $\left\{ N_{i}\right\} _{i=1}^{5}$ and
let $\alpha=\left\{ N_1,N_2,N_3\right\} $, in this case the $BF$ terms are $\sum_{i=1}^{3}\frac{N_{i}}{2\pi}C^{i}dA^{i}+\sum_{i=4}^{5}\frac{N_{i}}{2\pi}\widetilde{B}^{i}dB^{i}$;
if we let $\alpha'=\left\{ N_2,N_4,N_5\right\} $, the $BF$ terms are $\sum_{i=2,4,5}\frac{N_{i}}{2\pi}C^{i}dA^{i}+\sum_{i=1,3}\frac{N_{i}}{2\pi}\widetilde{B}^{i}dB^{i}$
instead. Naturally, $\alpha$ and $\alpha'$ {lead to different twisted terms, hence different
TQFT actions and classifications}. However, $\alpha$ and $\alpha'$ {have the same cardinality}, the two different TQFT actions
and classifications can be connected by {rearranging the indices, as can be seen in the examples below}.
Therefore, for simplicity, we only present results for one configuration
of $\alpha$ when the cardinality of the set $\alpha$ is fixed.

\subsection{$G=\mathbb{Z}_{N_{1}}$}\label{subsec_class_1Zn}

If $\alpha=\emptyset$, the action is
\begin{equation}
S=\int\frac{N_{1}}{2\pi}\widetilde{B}^{1}dB^{1}.
\end{equation}
 {Obviously, the coefficient $N_i$ is fixed, resulting in only one TQFT. The classification is denoted as ``$\mathbb{Z}_1$''.}

When $G=\mathbb{Z}_{N_{1}}$, the only nontrivial choice of $\alpha$ is $\alpha=\left\{ N_1\right\}$.
The action is
\begin{equation}
S=\int\frac{N_{1}}{2\pi}C^{1}dA^{1}+\bigg\langle A^{1}dA^{1}dA^{1}\bigg\rangle,
\label{eq_Zn1_DW}
\end{equation}
whose classification is $\mathbb{Z}_{N_{1}}$.\footnote{We use $\bigg\langle \text{twisted term $1$},\text{twisted term $2$}, \cdots\bigg\rangle$ to simply denote a summation of twisted terms that appear in actions. In this notation,   all coefficients of twisted terms, which are properly quantized as mentioned in section~\ref{subsec_type_1_BF_a_twist} and~\ref{subsec_mixed_BF_a_twist}, are omitted for the notational convenience.} We point out that the action (\ref{eq_Zn1_DW}) is classified by the \nth{5} cohomology group $H^{5}\left(\mathbb{Z}_{N_1},\mathbb{R}/\mathbb{Z}\right)$.
Classification of actions when $G=\mathbb{Z}_{N_{1}}$
is summarized in table~\ref{tab_classi_1-3Zn}.
We conclude that there are in total $1+N_{1}$ different $BF$ theories
when $G=\mathbb{Z}_{N_{1}}$.

\begin{table}[tbp]
\centering{}%
\resizebox{\textwidth}{!}{
\begin{tabular}{|c|c|c|c|}
\hline
$G$  & $\alpha$ & Classification & \tabularnewline
\hline
\multirow{2}{*}{$\mathbb{Z}_{N_{1}}$} & $\emptyset$ & $\mathbb{Z}_{1}$ & \tabularnewline
\cline{2-4} \cline{3-4} \cline{4-4}
 & $\left\{ N_{1}\right\} $ & $\mathbb{Z}_{N_{1}}$ & $\star$\tabularnewline
\hline
\multirow{4}{*}{${\displaystyle \prod_{i=1}^{2}\mathbb{Z}_{N_{i}}}$} & $\emptyset$ & $\mathbb{Z}_{1}$ & \tabularnewline
\cline{2-4} \cline{3-4} \cline{4-4}
 & $\left\{ N_{1}\right\} $ & $\mathbb{Z}_{N_{1}}\times\left(\mathbb{Z}_{N_{12}}\right)^{2}$ & \tabularnewline
\cline{2-4} \cline{3-4} \cline{4-4}
 & $\left\{ N_{2}\right\} $ & $\mathbb{Z}_{N_{2}}\times\left(\mathbb{Z}_{N_{12}}\right)^{2}$ & \tabularnewline
\cline{2-4} \cline{3-4} \cline{4-4}
 & $\left\{ N_{1},N_{2}\right\} $ & $\mathbb{Z}_{N_{1}}\times\mathbb{Z}_{N_{2}}\times\left(\mathbb{Z}_{N_{12}}\right)^{2}$ & $\star$\tabularnewline
\hline
\multirow{11}{*}{${\displaystyle \prod_{i=1}^{3}\mathbb{Z}_{N_{i}}}$} & $\emptyset$ & $\mathbb{Z}_{1}$ & \tabularnewline
\cline{2-4} \cline{3-4} \cline{4-4}
 & $\left\{ N_{1}\right\} $ & $\mathbb{Z}_{N_{1}}\times\left(\mathbb{Z}_{N_{12}}\right)^{2}\times\left(\mathbb{Z}_{N_{13}}\right)^{2}\times\mathbb{Z}_{N_{123}}$ & \tabularnewline
\cline{2-4} \cline{3-4} \cline{4-4}
 & $\left\{ N_{2}\right\} $ & $\mathbb{Z}_{N_{2}}\times\left(\mathbb{Z}_{N_{12}}\right)^{2}\times\left(\mathbb{Z}_{N_{23}}\right)^{2}\times\mathbb{Z}_{N_{123}}$ & \tabularnewline
\cline{2-4} \cline{3-4} \cline{4-4}
 & $\left\{ N_{3}\right\} $ & $\mathbb{Z}_{N_{3}}\times\left(\mathbb{Z}_{N_{13}}\right)^{2}\times\left(\mathbb{Z}_{N_{23}}\right)^{2}\times\mathbb{Z}_{N_{123}}$ & \tabularnewline
\cline{2-4} \cline{3-4} \cline{4-4}
 & $\left\{ N_{1},N_{2}\right\} $ & $\mathbb{Z}_{N_{1}}\times\mathbb{Z}_{N_{2}}\times{\displaystyle \left(\mathbb{Z}_{N_{12}}\right)^{2}\times\left(\mathbb{Z}_{N_{13}}\right)^{2}\times\left(\mathbb{Z}_{N_{23}}\right)^{2}}\times\left(\mathbb{Z}_{N_{123}}\right)^{2}$ & \tabularnewline
\cline{2-4} \cline{3-4} \cline{4-4}
 & $\left\{ N_{1},N_{3}\right\} $ & $\mathbb{Z}_{N_{1}}\times\mathbb{Z}_{N_{3}}\times\left(\mathbb{Z}_{N_{13}}\right)^{2}\times\left(\mathbb{Z}_{N_{12}}\right)^{2}\times\left(\mathbb{Z}_{N_{23}}\right)^{2}\times\left(\mathbb{Z}_{N_{123}}\right)^{2}$ & \tabularnewline
\cline{2-4} \cline{3-4} \cline{4-4}
 & $\left\{ N_{2},N_{3}\right\} $ & $\mathbb{Z}_{N_{2}}\times\mathbb{Z}_{N_{3}}\times\left(\mathbb{Z}_{N_{23}}\right)^{2}\times\left(\mathbb{Z}_{N_{12}}\right)^{2}\times\left(\mathbb{Z}_{N_{13}}\right)^{2}\times\left(\mathbb{Z}_{N_{123}}\right)^{2}$ & \tabularnewline
\cline{2-4} \cline{3-4} \cline{4-4}
 & \multirow{4}{*}{$\left\{ N_{1},N_{2},N_{3}\right\} $} & no $AAC$: $\prod_{i=1}^{3}\mathbb{Z}_{N_{i}}\times\prod_{1\leq i<j\leq3}\left(\mathbb{Z}_{N_{ij}}\right)^{2}\times\left(\mathbb{Z}_{N_{123}}\right)^{4}$ & $\star$\tabularnewline
\cline{3-4} \cline{4-4}
 &  & with $A^{1}A^{2}C^{3}$: $\left(\mathbb{Z}_{N_{123}}\setminus\left\{ 0\right\} \right)\times\mathbb{Z}_{N_{1}}\times\mathbb{Z}_{N_{2}}\times\left(\mathbb{Z}_{N_{12}}\right)^{2}$ & \tabularnewline
\cline{3-4} \cline{4-4}
 &  & with $A^{2}A^{3}C^{1}$: $\left(\mathbb{Z}_{N_{123}}\setminus\left\{ 0\right\} \right)\times\mathbb{Z}_{N_{2}}\times\mathbb{Z}_{N_{3}}\times\left(\mathbb{Z}_{N_{23}}\right)^{2}$ & \tabularnewline
\cline{3-4} \cline{4-4}
 &  & with $A^{3}A^{1}C^{2}$: $\left(\mathbb{Z}_{N_{123}}\setminus\left\{ 0\right\} \right)\times\mathbb{Z}_{N_{1}}\times\mathbb{Z}_{N_{3}}\times\left(\mathbb{Z}_{N_{13}}\right)^{2}$ & \tabularnewline
\hline
\end{tabular}
}
\caption{Classification of $BF$ theories with twisted terms for different
gauge groups $G$, which are detailed in section \ref{subsec_class_1Zn},
\ref{subsec_class_Zn1Zn2}, and \ref{subsec_class_3Zn}. $\alpha$ is a set, which denotes $\mathbb{Z}_{N_{i}}$ gauge subgroups that are encoded in
the type-I $BF$ term, e.g., $\alpha=\left\{ N_{1},N_{2}\right\} $ means
$\frac{N_{1}}{2\pi}C^{1}dA^{1}+\frac{N_{2}}{2\pi}C^{2}dA^{2}$.
Only a part of $5$D $BF$ theories, denoted by a $\star$ symbol,
are consistent with Dijkgraaf-Witten cohomological classification, i.e., $H^{5}\left(G,\mathbb{R}/\mathbb{Z}\right)$.
The remaining are beyond the group cohomology classification.
When $G=\prod_{i=1}^{3}\mathbb{Z}_{N_{i}}$ and $\alpha=\left\{ N_{1},N_{2},N_{3}\right\} $,
depending on different choices of $AAC$ term, the actions and their classification
are different, as discussed in section \ref{subsec_class_3Zn}. The
reason is that $AAC$ term may be incompatible with other twisted
terms, hence cannot be included in the action.
\label{tab_classi_1-3Zn}}
\end{table}

\subsection{$G=\mathbb{Z}_{N_{1}}\times\mathbb{Z}_{N_{2}}$}\label{subsec_class_Zn1Zn2}
If $\alpha=\emptyset$, the action is
\begin{equation}
S=\int\sum_{i=1}^{2}\frac{N_{i}}{2\pi}\widetilde{B}^{i}dB^{i}
\end{equation}
whose classification is $\mathbb{Z}_1$. In fact, when $\alpha=\emptyset$, the action can only be the type-II $BF$ theory without twisted terms, whose classification is automatically equal to $\mathbb{Z}_1$. For $\alpha \neq \emptyset$, the TQFT actions are mixed $BF$ theories or type-I $BF$ theories with twisted terms.

If $\alpha =\left\{ N_1\right\}$,
the action is
\begin{equation}
S=\int\frac{N_{1}}{2\pi}C^{1}dA^{1}+\frac{N_{2}}{2\pi}\widetilde{B}^{2}dB^{2}+\bigg\langle  A^{1}dA^{1}dA^{1},B^{2}B^{2}A^{1},B^{2}A^{1}dA^{1}\bigg\rangle,
\end{equation}
whose classification is $\mathbb{Z}_{N_{1}}\times\left(\mathbb{Z}_{N_{12}}\right)^{2}$.
For the case of $\alpha=\left\{ N_2\right\}$,
one just needs to switch the indices $1$ and $2$ to obtain the corresponding classification: $\mathbb{Z}_{N_{2}}\times\left(\mathbb{Z}_{N_{12}}\right)^{2}$,
as shown in table~\ref{tab_classi_1-3Zn}.

If $\alpha =\left\{N_1, N_2\right\}$,
the action is
\begin{equation}
S=\int\sum_{i=1}^{2}\frac{N_{i}}{2\pi}C^{i}dA^{i}+\bigg\langle A^{1}dA^{1}dA^{1},A^{2}dA^{2}dA^{2},A^{1}dA^{2}dA^{1},A^{2}dA^{1}dA^{2}\bigg\rangle ,
\label{eq_Zn1Zn2_DW}
\end{equation}
which is classified by the \nth{5} cohomology group of $\mathbb{Z}_{N_1}\times\mathbb{Z}_{N_2}$: $H^{5}\left(\mathbb{Z}_{N_{1}}\times\mathbb{Z}_{N_{2}},\mathbb{R}/\mathbb{Z}\right)=\mathbb{Z}_{N_{1}}\times\mathbb{Z}_{N_{2}}\times\left(\mathbb{Z}_{N_{12}}\right)^{2}$.

The number of different $BF$ theories for $G=\mathbb{Z}_{N_{1}}\times\mathbb{Z}_{N_{2}}$
can be obtained as follows. For $\alpha=\emptyset$, there is one
$BF$ theory. For $\alpha=\left\{ N_{1}\right\} $ , there are $\left|\mathbb{Z}_{N_{1}}\times\left(\mathbb{Z}_{N_{12}}\right)^{2}\right|=N_{1}\left(N_{12}\right)^{2}$
different $BF$ theories. For $\alpha=\left\{ N_{2}\right\} $ and
$\alpha=\left\{ N_{1},N_{2}\right\} $, this number is $N_{2}\left(N_{12}\right)^{2}$
and $N_{1}N_{2}\left(N_{12}\right)^{2}$, respectively. Therefore,
the total number of different $BF$ theories for $G=\mathbb{Z}_{N_{1}}\times\mathbb{Z}_{N_{2}}$
is $1+N_{1}\left(N_{12}\right)^{2}+N_{2}\left(N_{12}\right)^{2}+N_{1}N_{2}\left(N_{12}\right)^{2}$.

\subsection{$G=\mathbb{Z}_{N_{1}}\times\mathbb{Z}_{N_{2}}\times\mathbb{Z}_{N_{3}}$}\label{subsec_class_3Zn}
Follow the same line of thinking, we can obtain
the classification  of $BF$ theories for $G=\prod_{i=1}^{3}\mathbb{Z}_{N_{i}}$, which is collected in table~\ref{tab_classi_1-3Zn}.

If $\alpha=\emptyset$, the action is type-II $BF$ theory without
twisted terms, whose classification is simply $\mathbb{Z}_1$.

If $\alpha =\left\{ N_1\right\} $,
the action is
\begin{equation}
S=\int\frac{N_{1}}{2\pi}C^{1}dA^{1}+\sum_{i=2,3}\frac{N_{i}}{2\pi}\widetilde{B}^{i}dB^{i}+\left\langle A^{1}dA^{1}dA^{1},\sum_{i=2,3}B^{i}B^{i}A^{1}+B^{i}A^{1}dA^{1},B^{2}B^{3}A^{1}\right\rangle ,
\end{equation}
whose classification is $\mathbb{Z}_{N_{1}}\times\left(\mathbb{Z}_{N_{12}}\right)^{2}\times\left(\mathbb{Z}_{N_{13}}\right)^{2}\times\mathbb{Z}_{N_{123}}$. For the case of $\alpha=\left\{ N_2\right\}$ or $\alpha= \left\{ N_3\right\} $, the results are similar, as shown in table~\ref{tab_classi_1-3Zn}.

If $\alpha=\left\{ N_1,N_2\right\} $, the action is
\begin{align}
S= & \int\sum_{i=1}^{2}\frac{N_{i}}{2\pi}C^{i}dA^{i}+\frac{N_{3}}{2\pi}\widetilde{B}^{3}dB^{3}+\left\langle \sum_{i=1,2}A^{i}dA^{i}dA^{i},A^{1}dA^{2}dA^{1},A^{2}dA^{1}dA^{2},\right.\nonumber\\
 & \left.\sum_{i=1,2}\left(B^{3}B^{3}A^{i}+B^{3}A^{i}dA^{i}\right),A^{1}A^{2}dB^{3},B^{3}A^{2}dA^{1}\right\rangle,
\end{align}
whose classification is $\mathbb{Z}_{N_{1}}\times\mathbb{Z}_{N_{2}}\times\prod_{1\leq i<j\leq3}\left(\mathbb{Z}_{N_{ij}}\right)^{2}\times\left(\mathbb{Z}_{N_{123}}\right)^{2}$. For other choices of $\alpha$, the corresponding classifications are listed in table~\ref{tab_classi_1-3Zn}.

If $\alpha=\left\{ N_1,N_2,N_3\right\} $, we get a type-I $BF$ theory with twists.
Possible twisted terms include $AAC$, $AdAdA$, and $AAAdA$. Due to the
possible incompatibility between $AAC$ and $AdAdA$ or $AAAdA$ terms~\cite{zhang2021compatible},
we need to treat these twisted terms carefully to obtain correct classifications.
First we do not consider $AAC$ terms in this type-I $BF$ theory.
The action is
\begin{align}
S= & \int\sum_{i=1}^{3}\frac{N_{i}}{2\pi}C^{i}dA^{i}+\left\langle \sum_{i=1}^{3}A^{i}dA^{i}dA^{i},\sum_{1\leq i<j\leq3}\left(A^{i}dA^{j}dA^{i}+A^{j}dA^{i}dA^{j}\right),\right.\nonumber\\
 & \left.A^{1}dA^{2}dA^{3},\sum_{i=1}^{3}A^{1}A^{2}A^{3}dA^{i}\right\rangle ,
\label{eq_action_3Zn_DW}
\end{align}
being classified by the \nth{5} cohomology group of $\prod_{i=1}^{3}\mathbb{Z}_{N_i}$:
\begin{equation}
H^{5}\left(\prod_{i=1}^{3}\mathbb{Z}_{N_{i}},\mathbb{R}/\mathbb{Z}\right)=\prod_{i=1}^{3}\mathbb{Z}_{N_{i}}\times\prod_{1\leq i<j\leq3}\left(\mathbb{Z}_{N_{ij}}\right)^{2}\times\left(\mathbb{Z}_{N_{123}}\right)^{4}.\label{eq_classi_3Zn_DW}
\end{equation}
Then we consider an $AAC$ term in the action. Without loss of generality,
we add the $A^{1}A^{2}C^{3}$ term in twisted terms. We point out
that $A^{2}A^{3}C^{1}$ and $A^{3}A^{1}C^{2}$ are also possible $AAC$
terms, yet neither of them is compatible with $A^{1}A^{2}C^{3}$ \cite{zhang2021compatible}.
If $A^{2}A^{3}C^{1}$ or $A^{3}A^{1}C^{2}$, instead of $A^{1}A^{2}C^{3}$,
is included in twisted terms, the discussion and result are similar, as shown in table~\ref{tab_classi_1-3Zn}.
Since $AdAdA$ and $AAAdA$ with $A^{3}$ or $dA^{3}$ are incompatible
with $A^{1}A^{2}C^{3}$ \cite{zhang2021compatible}, the action is
\begin{equation}
S=\int\sum_{i=1}^{3}\frac{N_{i}}{2\pi}C^{i}dA^{i}+\frac{pN_{1}N_{2}N_{3}}{\left(2\pi\right)^{2}N_{123}}A^{1}A^{2}C^{3}+\left\langle \sum_{i=1,2}A^{i}dA^{i}dA^{i},A^{1}dA^{2}dA^{1},A^{2}dA^{1}dA^{2}\right\rangle \label{eq_action_3Zn_A1A2C3}
\end{equation}
with $p\in\mathbb{Z}_{N_{123}}\setminus \left\{0\right\}$. The reason for this incompatibility is that either $A^3$ or $C^3$ has to be the Lagrange multiplier, meaning that they cannot appear in twisted terms in the same time. More concretely, if the action consists of $A^1 A^2 C^3$ and other twisted terms with $A^3$ or $dA^3$, it would unavoidably break the gauge-invariance, just like the illegitimate $A^1 A^2 C^2$ term discussed in ``\emph{$AAC$ twisted topological term}'' in section~\ref{subsec_type_1_BF_a_twist}. The classification of action (\ref{eq_action_3Zn_A1A2C3})
is
\begin{equation}
\left(\mathbb{Z}_{N_{123}}\setminus\left\{ 0\right\} \right)\times\mathbb{Z}_{N_{1}}\times\mathbb{Z}_{N_{2}}\times\left(\mathbb{Z}_{N_{12}}\right)^{2},\label{eq_classi_3Zn_A1A2C3}
\end{equation}
where $\mathbb{Z}_{N_{123}}\setminus\left\{ 0\right\} $ is the set
obtained by removing the identity element $\left\{ 0\right\} $ from
the cyclic group $\mathbb{Z}_{N_{123}}$, corresponding to the existence
of $\frac{pN_{1}N_{2}N_{3}}{\left(2\pi\right)^{2}N_{123}}A^{1}A^{2}C^{3}$, i.e., $p\neq0$. Therefore (\ref{eq_classi_3Zn_A1A2C3}) is not a group
any more but it is still a set:
\begin{align}
 & \left(\mathbb{Z}_{N_{123}}\setminus\left\{ 0\right\} \right)\times\mathbb{Z}_{N_{1}}\times\mathbb{Z}_{N_{2}}\times\left(\mathbb{Z}_{N_{12}}\right)^{2}\nonumber \\
= & \left\{ \left(x_{1},x_{2},x_{3},x_{4},x_{5}\right)|\ x_{1}\in\mathbb{Z}_{N_{123}}\setminus\left\{ 0\right\} ,x_{2}\in\mathbb{Z}_{N_{1}},x_{3}\in\mathbb{Z}_{N_{2}},x_{4}\in\mathbb{Z}_{N_{12}},x_{5}\in\mathbb{Z}_{N_{12}}\right\} .
\end{align}
Each element is given by a group of 5 numbers, i.e., $\left(x_{1},x_{2},x_{3},x_{4},x_{5}\right)$, denoting
a choice of periodic coefficients of twisted terms in action~(\ref{eq_action_3Zn_A1A2C3}):
$x_{1}$ for $A^{1}A^{2}C^{3}$, $x_{2}$ for $A^{1}dA^{1}dA^{1}$,
$\dots$, and $x_{5}$ for $A^{2}dA^{1}dA^{2}$.
There are in total $\left(N_{123}-1\right)N_{1}N_{2}\left(N_{12}\right)^{2}$
elements in (\ref{eq_classi_3Zn_A1A2C3}), which means
$\left(N_{123}-1\right)N_{1}N_{2}\left(N_{12}\right)^{2}$ different
actions when $G=\prod_{i=1}^{3}\mathbb{Z}_{N_{i}}$, $\alpha=\left\{ N_{1},N_{2},N_{3}\right\} $,
and the $A^{1}A^{2}C^{3}$ term is considered. Compare to the classification
of action~(\ref{eq_action_3Zn_DW}), we see that the existence of $A^{1}A^{2}C^{3}$
term indeed excludes some twisted terms.

From the example above we see that even if $\alpha$ is fixed, the classification and number of different
$BF$ theories may still depend on the choice of $AAC$ terms. Generally
speaking, though the $BF$ terms, i.e., $\alpha$, are determined,
there are still different combinations of compatible twisted terms
\citep{zhang2021compatible} which lead to different $BF$ theories
and classification. This feature is more obvious and important when
the $G$ contains more cyclic subgroups, e.g., cases discussed in section \ref{subsec_class_4Zn}. The complete classification of $BF$ theories for $G=\prod_{i=1}^{3}\mathbb{Z}_{N_{i}}$
is given in table \ref{tab_classi_1-3Zn}. In order to find out the total
number of $BF$ theories,
one just need to sum up the corresponding cardinality of each classification
in table \ref{tab_classi_1-3Zn}.

\subsection{$G=\prod_{i=1}^{4}\mathbb{Z}_{N_{i}}$ and generalization to $G=\prod_{i=1}^{n}\mathbb{Z}_{N_{i}}$ with $n\geq 5$}\label{subsec_class_4Zn}
In this section, we investigate $BF$ theories and their classifications when the gauge group is $G=\prod_{i=1}^{4}\mathbb{Z}_{N_{i}}$. In this case, there are much more twisted terms and complicated compatibility issues, which makes a long list of $BF$ theories as well as their classification. For sake of simplicity, we aim to provide some typical examples of $BF$ theories and classifications for $G=\prod_{i=1}^{4}\mathbb{Z}_{N_{i}}$ which indicate regularities for generalization to all $BF$ theories and classifications.

If $\alpha=\emptyset$, the $BF$ theories can only be
\begin{equation}
S=\int\sum_{i=1}^{4}\frac{N_{i}}{2\pi}\widetilde{B}^{i}dB^{i}
\end{equation}
with the classification of $\mathbb{Z}_1$.

If $\alpha=\left\{ N_1\right\}$, the action is
\begin{align}
S= & \int\frac{N_{1}}{2\pi}C^{1}dA^{1}+\sum_{i=2}^{4}\frac{N_{i}}{2\pi}\widetilde{B}^{i}dB^{i}\nonumber\\
 & +\left\langle A^{1}dA^{1}dA^{1},\sum_{i=2}^{4}\left(B^{i}B^{i}A^{1}+B^{i}A^{1}dA^{1}\right),\sum_{2\leq i<j\leq 4}^{4}B^{i}B^{j}A^{1}\right\rangle ,
\end{align}
whose classification is $\mathbb{Z}_{N_{1}}\times\prod_{i=2}^{4}\left(\mathbb{Z}_{N_{1i}}\right)^{2}\times\mathbb{Z}_{N_{123}}\times\mathbb{Z}_{N_{124}}\times\mathbb{Z}_{N_{134}}$.

If $\alpha=\left\{ N_1,N_2\right\} $, the action is
\begin{align}
S= & \int\sum_{i=1}^{2}\frac{N_{i}}{2\pi}C^{i}dA^{i}+\sum_{i=3}^{4}\frac{N_{i}}{2\pi}\widetilde{B}^{i}dB^{i}+\left\langle \sum_{i=1}^{2}A^{i}dA^{i}dA^{i},A^{1}dA^{2}dA^{1},A^{2}dA^{1}dA^{2},B^{3}B^{4}A^{1},\right.\nonumber\\
 & \left.B^{3}B^{4}A^{2},\sum_{i=3,4}\left(B^{i}B^{i}A^{1}+B^{i}A^{1}dA^{1}+B^{i}B^{i}A^{2}+B^{i}A^{2}dA^{2}+A^{1}A^{2}dB^{i}+B^{i}A^{2}dA^{1}\right)\right\rangle ,
\end{align}
whose classification is $\mathbb{Z}_{N_{1}}\times\mathbb{Z}_{N_{2}}\times\left(\mathbb{Z}_{N_{12}}\right)^{2}\times\mathbb{Z}_{N_{134}}\times\mathbb{Z}_{N_{234}}\times{\prod_{i=3}^{4}\left[\left(\mathbb{Z}_{N_{1i}}\right)^{2}\times\left(\mathbb{Z}_{N_{2i}}\right)^{2}\times\left(\mathbb{Z}_{N_{12i}}\right)^{2}\right]}$.

If $\alpha=\left\{ N_{1},N_{2},N_{3}\right\} $, $BF$ terms are $\sum_{i=1}^{3}\frac{N_{i}}{2\pi}C^{i}dA^{i}+\frac{N_{4}}{2\pi}\widetilde{B}^{4}dB^{4}$.
Possible twisted terms include $AAC$, $AdAdA$, $AAAdA$, $BBA$,
$BAdA$, and $AAdB$. Once again, we need to take care of the incompatibility
of $AAC$ and other twisted terms, as discussed in section \ref{subsec_class_3Zn}.
There are two situations: the action includes $AAC$ term or not.
\begin{enumerate}
\item If there is no $AAC$ term in twisted terms, the action is
\begin{align}
S= & \int\sum_{i=1}^{3}\frac{N_{i}}{2\pi}C^{i}dA^{i}+\frac{N_{4}}{2\pi}\widetilde{B}^{4}dB^{4}+\left\langle \sum_{i=1}^{3}A^{i}dA^{i}dA^{i},\sum_{1\leq i<j\leq3}\left(A^{i}dA^{j}dA^{i}+A^{j}dA^{i}dA^{j}\right),\right.\nonumber\\
 & A^{1}dA^{2}dA^{3},\sum_{i=1}^{3}A^{1}A^{2}A^{3}dA^{i},\sum_{i=1}^{3}\left(B^{4}B^{4}A^{i}+B^{4}A^{i}dA^{i}\right),\nonumber\\
 & \left.\sum_{1\leq i<j\leq3}\left(A^{i}A^{j}dB^{4}+B^{4}A^{j}dA^{i}\right),A^{1}A^{2}A^{3}B^{4}\right\rangle ,
\end{align}
whose classification is ${ \prod_{i=1}^{3}\mathbb{Z}_{N_{i}}}\times{\prod_{1\leq i<j\leq3}\left(\mathbb{Z}_{N_{ij}}\right)^{2}}\times\left(\mathbb{Z}_{N_{123}}\right)^{4}\times{ \prod_{i=1}^{3}\left(\mathbb{Z}_{N_{i4}}\right)^{2}}\times{ \prod_{1\leq i<j\leq3}^{3}\left(\mathbb{Z}_{N_{ij4}}\right)^{2}}\times\mathbb{Z}_{N_{1234}}$.
\item If one $AAC$ term is added to the action, without loss of generality, let it be $A^{1}A^{2}C^{3}$, the action is
\begin{align}
S= & \int\sum_{i=1}^{3}\frac{N_{i}}{2\pi}C^{i}dA^{i}+\frac{N_{4}}{2\pi}\widetilde{B}^{4}dB^{4}+\frac{pN_{1}N_{2}N_{3}}{\left(2\pi\right)^{2}N_{123}}A^{1}A^{2}C^{3}+\left\langle \sum_{i=1,2}A^{i}dA^{i}dA^{i},\right.\nonumber\\
 & \left.A^{1}dA^{2}dA^{1},A^{2}dA^{1}dA^{2},\sum_{i=1,2}\left(B^{4}B^{4}A^{i}+B^{4}A^{i}dA^{i}\right),A^{1}A^{2}dB^{4},B^{4}A^{2}dA^{1}\right\rangle
\label{eq_action_4Zn_A1A2C3}
\end{align}
with $p\in\mathbb{Z}_{N_{123}}\setminus \left\{ 0\right\} $. As mentioned in section \ref{subsec_class_3Zn}, when there are only
$3$ elements in $\alpha$, different $AAC$ terms are incompatible with
each other. Therefore, only \emph{one} $AAC$ term can be added to the action when $\left|\alpha\right|=3$, e.g., $\alpha=\left\{N_1 ,N_2 ,N_3\right\}$. The classification of action (\ref{eq_action_4Zn_A1A2C3})
is
\begin{equation}
\left(\mathbb{Z}_{N_{123}}\setminus\left\{ 0\right\} \right)\times\prod_{i=1}^{2}\mathbb{Z}_{N_{i}}\times\left(\mathbb{Z}_{N_{12}}\right)^{2}\times\left(\mathbb{Z}_{N_{14}}\right)^{2}\times\left(\mathbb{Z}_{N_{24}}\right)^{2}\times\left(\mathbb{Z}_{N_{124}}\right)^{2}.
\end{equation}
Compared to action (\ref{eq_action_3Zn_A1A2C3}), the difference between
their classifications is due to extra twisted terms resulted from $\widetilde{B}^{4}dB^{4}$
and type-I $BF$ terms. Appendix~\ref{appendix_4Zn_class_all_AAC} presents the corresponding action and classification if other $AAC$ term is considered.
\end{enumerate}

If $\alpha=\left\{ N_{1},N_{2},N_{3},N_{4}\right\} $, the action
is a type-I $BF$ theory with twisted terms. Similarly, we need to study two
situations in which the action consists of $AAC$ terms or not.
First, we consider the action without any $AAC$ terms,
\begin{align}
S= & \int\sum_{i=1}^{4}\frac{N_{i}}{2\pi}C^{i}dA^{1}+\left\langle \sum_{i=1}^{4}A^{i}dA^{i}dA^{i},\sum_{1\leq i<j\leq4}\left(A^{i}dA^{j}dA^{i}+A^{j}dA^{i}dA^{j}\right),\sum_{1\leq i<j<k\leq4}A^{i}dA^{j}dA^{k},\right.\nonumber\\
 & \left.\sum_{1\leq i<j<k\leq4}\left(\sum_{l=i,j,k}A^{i}A^{j}A^{k}dA^{l}\right),A^{1}A^{2}A^{3}dA^{4},A^{3}A^{2}A^{4}dA^{1},A^{1}A^{3}A^{4}dA^{2}\right\rangle ,
\label{eq_action_4Zn_DW}
\end{align}
whose classification is
\begin{equation}
\prod_{i=1}^{4}\mathbb{Z}_{N_{i}}\times\prod_{1\leq i<j\leq4}\left(\mathbb{Z}_{N_{ij}}\right)^{2}\times\prod_{1\leq i<j<k\leq4}\left(\mathbb{Z}_{N_{ijk}}\right)^{4}\times\left(\mathbb{Z}_{N_{1234}}\right)^{3}=H^{5}\left(\prod_{i=1}^{4}\mathbb{Z}_{N_{i}},\mathbb{R}/\mathbb{Z}\right),
\label{eq:H^5_4Zn}
\end{equation}
{same as} the result obtained from Dijkgraaf-Witten model.
Then we study the case in which $AAC$ terms are added to the action.
Since there are $4$ type-I $BF$ terms, some $AAC$ terms may be
compatible, unlike the above case of $\alpha=\left\{ N_{1},N_{2},N_{3}\right\}$. We first discuss the action and its classification when only one $AAC$
term is considered. Then, following the compatibility principle \citep{zhang2021compatible},
we add more allowed $AAC$ terms to the action and figure out the
corresponding classification.
\begin{enumerate}
\item Consider the action with only one $AAC$ term, without loss of generality, $A^{1}A^{2}C^{4}$,
\begin{align}
S= & \int\sum_{i=1}^{4}\frac{N_{i}}{2\pi}C^{i}dA^{1}+\frac{pN_{1}N_{2}N_{4}}{\left(2\pi\right)^{2}N_{124}}A^{1}A^{2}C^{4}+\left\langle \sum_{i=1}^{3}A^{i}dA^{i}dA^{i},\right.\nonumber\\
 & \left.\sum_{\substack{1\leq i<j\leq3}
}\left(A^{i}dA^{j}dA^{i}+A^{j}dA^{i}dA^{j}\right),A^{1}dA^{2}dA^{3},\sum_{i=1}^{3}A^{1}A^{2}A^{3}dA^{i}\right\rangle ,
\label{eq_4Zn_4a_A1A2C4}
\end{align}
where $p\in\mathbb{Z}_{N_{124}}\setminus\left\{ 0\right\} $. Its
classification is $\left(\mathbb{Z}_{N_{124}}\setminus\left\{ 0\right\} \right)\times{\prod_{i=1}^{3}\mathbb{Z}_{N_{i}}}\times{\prod_{1\leq i<j\leq3}\left(\mathbb{Z}_{N_{ij}}\right)^{2}}\times\left(\mathbb{Z}_{N_{123}}\right)^{4}$.
There are other $AAC$ terms compatible with $A^{1}A^{2}C^{4}$, e.g., $A^{1}A^{3}C^{4}$
and $A^{2}A^{3}C^{4}$ can also be added to action (\ref{eq_4Zn_4a_A1A2C4}).
\item If $A^{1}A^{2}C^{4}$ and $A^{1}A^{3}C^{4}$ are included in twisted
terms, the action is
\begin{align}
S= & \int\sum_{i=1}^{4}\frac{N_{i}}{2\pi}C^{i}dA^{1}+\frac{p N_{1}N_{2}N_{4}}{\left(2\pi\right)^{2}N_{124}}A^{1}A^{2}C^{4}+\frac{p'  N_{1}N_{3}N_{4}}{\left(2\pi\right)^{2}N_{134}}A^{1}A^{3}C^{4}+\left\langle \sum_{i=1}^{3}A^{i}dA^{i}dA^{i},\right.\nonumber\\
 & \left.\sum_{\substack{1\leq i<j\leq3}
}\left(A^{i}dA^{j}dA^{i}+A^{j}dA^{i}dA^{j}\right),A^{1}dA^{2}dA^{3},\sum_{i=1}^{3}A^{1}A^{2}A^{3}dA^{i}\right\rangle
\label{eq_4Zn_4a_A1A2C4+A1A3C4}
\end{align}
with $p\in\mathbb{Z}_{N_{124}}\setminus\left\{ 0\right\} $ and $p' \in\mathbb{Z}_{N_{134}}\setminus\left\{ 0\right\} $.
Its classification is $\left(\mathbb{Z}_{N_{124}}\setminus\left\{ 0\right\} \right)\times\left(\mathbb{Z}_{N_{134}}\setminus\left\{ 0\right\} \right)\times{\prod_{i=1}^{3}\mathbb{Z}_{N_{i}}}\times{\prod_{1\leq i<j\leq3}\left(\mathbb{Z}_{N_{ij}}\right)^{2}}\times\left(\mathbb{Z}_{N_{123}}\right)^{4}$.
\item If twisted terms include $A^{1}A^{2}C^{4}$, $A^{1}A^{3}C^{4}$, and
$A^{2}A^{3}C^{4}$, the classification of corresponding action is
${\prod_{1\leq i<j\leq3}\left(\mathbb{Z}_{N_{ij4}}\setminus\left\{ 0\right\} \right)}\times{\prod_{i=1}^{3}\mathbb{Z}_{N_{i}}}\times{\prod_{1\leq i<j\leq3}\left(\mathbb{Z}_{N_{ij}}\right)^{2}}\times\left(\mathbb{Z}_{N_{123}}\right)^{4}$.
\item On the other hand, $A^{1}A^{2}C^{3}$ is also compatible with $A^{1}A^{2}C^{4}$,
but not compatible with $A^{1}A^{3}C^{4}$ or $A^{2}A^{3}C^{4}$. The action with $A^{1}A^{2}C^{4}$
and $A^{1}A^{2}C^{3}$ is
\begin{align}
S =& \int\sum_{i=1}^{4}\frac{N_{i}}{2\pi}C^{i}dA^{1}+\frac{pN_{1}N_{2}N_{4}}{\left(2\pi\right)^{2}N_{124}}A^{1}A^{2}C^{4}+\frac{p' N_{1}N_{2}N_{3}}{\left(2\pi\right)^{2}N_{123}}A^{1}A^{2}C^{3}\nonumber \\
 & +\left\langle \sum_{i=1}^{2}A^{i}dA^{i}dA^{i},A^{1}dA^{2}dA^{1},A^{2}dA^{1}dA^{2}\right\rangle
\end{align}
 with $p\in\mathbb{Z}_{N_{124}}\setminus\left\{ 0\right\} $ and $p' \in\mathbb{Z}_{N_{123}}\setminus\left\{ 0\right\} $.
The corresponding classification is $\left(\mathbb{Z}_{N_{124}}\setminus\left\{ 0\right\} \right)\times\left(\mathbb{Z}_{N_{123}}\setminus\left\{ 0\right\} \right)\times{\prod_{i=1}^{2}\mathbb{Z}_{N_{i}}}\times\left(\mathbb{Z}_{N_{12}}\right)^{2}$.
\end{enumerate}
At last, we point out that if other $AAC$ term, e.g., $A^{2}A^{3}C^{4}$,
and its compatible $AAC$ terms are considered in the action, the action
and classification can be obtained by a similar manner. In appendix~\ref{appendix_4Zn_class_all_AAC}, we list actions and their classifications for all possible combinations of $AAC$ terms.

In section \ref{subsec_class_1Zn}, \ref{subsec_class_Zn1Zn2},
\ref{subsec_class_3Zn}, and \ref{subsec_class_4Zn}, we have studied
the classification of $BF$ theories for gauge group consisting of
up $4$ cyclic groups. Such discussion can be generalized to the cases of $G=\prod_{i=1}^{n}\mathbb{Z}_{N_{i}}$
with $n\geq5$. Some typical
$BF$ theories and their classifications are given in appendix \ref{appendix_example_5Zn},
which is helpful for considering general cases of $G=\prod_{i=1}^{5}\mathbb{Z}_{N_{i}}$.
In short, one needs to first determine $BF$ terms. Then, all possible twisted
terms can be found out according to $BF$ terms. Next, one should
check the compatibility between twisted terms, which can be done with
the guidance provided in ref.~\citep{zhang2021compatible}. Finally, an
action with all allowed twisted terms can be written and its classification
can be obtained by counting the coefficients of twisted terms.

We point out that when $G=\prod_{i=1}^{5}\mathbb{Z}_{N_{i}}$, $\alpha=\left\{ N_{1},N_{2},N_{3},N_{4},N_{5}\right\} $,
and no $AAC$ terms are taken in account, the action is
\begin{align}
S= & \int\sum_{i=1}^{5}\frac{N_{i}}{2\pi}C^{i}dA^{i}+\left\langle \sum_{i=1}^{5}A^{i}dA^{i}dA^{i},\sum_{1\leq i<j\leq5}\left(A^{i}dA^{j}dA^{i}+A^{j}dA^{i}dA^{j}\right),\right.\nonumber\\
 & \sum_{1\leq i<j<k\leq5}\left(A^{i}dA^{j}dA^{k}+\sum_{l=i,j,k}A^{i}A^{j}A^{k}dA^{l}\right),\nonumber\\
 & \left.\sum_{1\leq i<j<k<l\leq5}\left(A^{i}A^{j}A^{k}dA^{l}+A^{k}A^{j}A^{l}dA^{i}+A^{i}A^{k}A^{l}dA^{j}\right),A^{1}A^{2}A^{3}A^{4}A^{5}\right\rangle ,
\end{align}
which is classified by
\begin{align}
H^{5}\left(\prod_{i=1}^{5}\mathbb{Z}_{N_{i}},\mathbb{R}/\mathbb{Z}\right)= & \prod_{i=1}^{5}\mathbb{Z}_{N_{i}}\times\prod_{1\leq i<j\leq5}\left(\mathbb{Z}_{N_{ij}}\right)^{2}\times\prod_{1\leq i<j<k\leq5}\left(\mathbb{Z}_{N_{ijk}}\right)^{4}\nonumber \\
 & \times\prod_{1\leq i<j<k<l\leq5}\left(\mathbb{Z}_{N_{ijkl}}\right)^{3}\times\mathbb{Z}_{N_{12345}}.
\end{align}
Once $AAC$ terms are considered or for other choices of $\alpha$, the action's
classification is totally beyond that obtained from group cohomology, which is an important feature of TQFT in $5$D.

\section{Conclusions and outlook}\label{section_concl_main}
In summary, we study $5$D topological orders from the field-theoretical aspect. In $5$D topological orders, topological excitations include
particles, loops, and membranes, whose braiding processes are complicated
and not been fully understood yet. With the help of TQFT, we write down topological terms, including
$BF$ terms and twisted terms, in $5$D. More concretely, there are
\emph{two} types of $BF$ terms in $5$D, unlike the case in $3$D
or $4$D. Such two types of $BF$
terms are studied in details. By combining $BF$ terms and twisted
terms, we write down TQFT actions that are invariant   under gauge transformations. For each TQFT action,
we construct the gauge-invariant Wilson operator whose expectation
value can be expressed as intersection patterns of geometric objects
in $5$D, as detailed in section \ref{subsec_2_types_BF}, \ref{subsec_type_1_BF_a_twist},
and \ref{subsec_mixed_BF_a_twist}. These results are
obtained from field theory in a gauge-invariant fashion, and correspond to link invariants of links formed by closed spacetime trajectories of topological excitations {in $5$D}. In addition, the
observable phase of braiding process is given by the Wilson operator.
In section \ref{sec_tqft_classification}, we study classifications
of TQFT action consisting of $BF$ terms and twisted terms. Depending
on the type of $BF$ term, these TQFT actions are dubbed   type-I
or mixed $BF$ theories. We find that, only some of type-I $BF$ theories
is classified by group cohomology, i.e., consistent with Dijkgraaf-Witten model. For
other type-I $BF$ theories and all mixed $BF$ theories, their classifications
are beyond Dijkgraaf-Witten  {cohomological} classification. Table \ref{tab_classi_1-3Zn} summarizes
classification of $BF$ theories for $G=\mathbb{Z}_{N_{1}}$, $G=\mathbb{Z}_{N_{1}}\times\mathbb{Z}_{N_{2}}$,
and $G=\prod_{i=1}^{3}\mathbb{Z}_{N_{i}}$. Some interesting questions still remain open, which are left for future study:
\begin{enumerate}
 \item Gauge transformations for TQFT actions with twisted terms usually
contain shift terms. How to understand these shift terms from mathematical
perspective (e.g., fibre bundle)? We hope future work could give more insight to this problem.

\item

It should be noticed that the classification of topological orders is not identical to that of TQFTs. For example, consider a $3+1$D all-boson topological order (all particle excitations are bosons) with gauge group $G=\mathbb{Z}_2 \times \mathbb{Z}_2$, we can write down a TQFT action for it: $S\sim \int \sum_{i=1}^{2}\frac{N_i}{2\pi}B^i dA^i+q_{1}A^{1}A^{2}dA^{1}+q_{2}A^{1}A^{2}dA^{2}$. This gauge theory is classified by $H^4\left(G,\mathbb{R}/\mathbb{Z}\right)=\mathbb{Z}_{2}\times\mathbb{Z}_{2}$. However, this topological order is actually classified by $H^4_{\rm Aut}\left(G\right)=H^4\left(G,\mathbb{R}/\mathbb{Z}\right)/{\rm Aut}=\mathbb{Z}_{2}$, where ``${\rm Aut}$'' stands for the automorphism of $G$. This can be understood by the redundancy of relabelling the two $\mathbb{Z}_2$ fluxes. For all TQFTs within cohomology classification~\cite{lantian3dto1}, we expect that group automorphism can be used to find the classification of topological orders. For other TQFTs, more careful considerations are needed for the classification of corresponding topological orders. Physically, in order to find out classification of $5$D topological orders, one should consider the  {inequivalent} data sets formed by physical observables.  In the future, it would be important to thoroughly study observables in $5$D topological orders and find out the complete classification.

\item  The world we live in is a $4$D spacetime. Nevertheless, the $4$D
anomalous theory cannot exist alone unless it appears as the boundary
theory of a $5$D topologically ordered state, which is the phenomenon of gravitational anomaly \cite{Wen_grav_prd2013,string8,2016arXiv161008645Y}.  Since $5$D $\prod_{i}^{n}\mathbb{Z}_{N_{i}}$
gauge theories are investigated in this paper, it would be interesting
to study the relation between $5$D theories and $4$D anomalous theories.
For the purpose, we need to consider TQFTs on a manifold with boundary.

\item In this paper, linking number in $5$D is obtained via a field-theoretical
approach, whose topological invariance is guaranteed by gauge invariance
of Wilson operator. Our work may shed light on the study of links
and knots in higher dimensions, which still call for  {joint efforts} from physicists
and mathematicians.

\item By adding global symmetry in $5$D topological orders, we can study  symmetry fractionalization on membranes, based on the field-theoretical framework in ref.~\cite{2018arXiv180101638N} where symmetry fractionalization on loops is characterized and classified. Putting the theory on a manifold with boundary may lead to anomalous symmetry fractionalization patterns, which may answer the question asked at the end of ref.~\cite{2016arXiv161008645Y}.
\item By noting that $K$-matrix Chern-Simons theory (i.e., $\frac{K^{IJ}}{4\pi}\int A^IdA^J$ with an symmetric integer matrix $K$) can be used to describe all Abelian topological orders in 3D spacetime, It is interesting to consider $K$-matrix $BdB$ theory with the action $S=\frac{K^{IJ}}{4\pi}B^IdB^J$ where $K$ is an antisymmetric integer matrix. The above BF term $Bd\tilde{B}$ considered in this paper is just the off-diagonal term in this theory. Then, the boundary theory and canonical quantization of this action can be studied systematically.

\end{enumerate}

\acknowledgments

 We sincerely thank Ling-Yan Hung and Wenliang Li for very helpful suggestions on the second   version of this manuscript. This work was supported by   Sun Yat-sen University  Talent Plan,     Guangdong Basic and Applied Basic Research Foundation under Grant No.~2020B1515120100, National Natural Science Foundation of China (NSFC) Grant (No.~11847608 \& No.~12074438). During preparation of this version,  we notice a recent   work \cite{chen2021loops} in which  loop statistics and boundary in $\Z_N$ $2$-form topological orders in 5D spacetime are  investigated.

\appendix

\section{Quantization and periodicity of twisted terms\label{sec_quantization_of_coefficient}}

In this appendix, we will derive the quantized and periodic coefficients
for specific twisted terms as examples. The basic guiding principle is that the TQFT action should satisfy \emph{large gauge invariance} and \emph{flux identification} \cite{PhysRevLett.114.031601}.
These derivations can be easily generalized to other twisted terms.

\subsection{Twisted terms from type-I $BF$ terms}
\emph{Example: $AAC$ term studied in section~\ref{subsec_type_1_BF_a_twist}}.

Consider $G=\mathbb{Z}_{N_{1}}\times\mathbb{Z}_{N_{2}}\times\mathbb{Z}_{N_{3}}$.
The TQFT action is
\begin{equation}
S=S_{BF}+S_{AAC}=\int\sum_{i=1}^{3}\frac{N_{i}}{2\pi}C^{i}dA^{i}+qA^{1}A^{2}C^{3}.\label{eq_S_CdA+AAC-1}
\end{equation}
After integrating out the Lagrange multipliers $C^{1},$$C^{2}$ and
$A^{3}$, the action $S$ reduces to $S_{AAC}=\int q A^1 A^2 C^3$, where the fields
$A^{1}$, $A^{2}$ and $C^{3}$ are set to be closed with $\oint A^{1}\in\frac{2\pi}{N_{1}}\mathbb{Z}_{N_{1}}$,
$\oint A^{2}\in\frac{2\pi}{N_{2}}\mathbb{Z}_{N_{2}}$ and $\oint C^{3}\in\frac{2\pi}{N_{3}}\mathbb{Z}_{N_{3}}$.
Consider putting this action on the spacetime manifold $\mathcal{M}=\mathbb{S}^{1}\times\mathbb{S}^{1}\times\mathbb{S}^{3}$.
Under the gauge transformations $A^{1,2}\rightarrow A^{1,2}+d\chi^{1,2}$
and $C^{3}\rightarrow C^{3}+dT^{3}$, $S_{AAC}$ changes as
\begin{align}
S_{AAC}\rightarrow S_{AAC}'= & \int_{\mathcal{M}}q\left(A^{1}+d\chi^{1}\right)\left(A^{2}+d\chi^{2}\right)\left(C^{3}+dT^{3}\right)\nonumber \\
= & S_{AAC}+\int_{\mathcal{M}}q\left(A^{1}d\chi^{2}C^{3}+d\chi^{1}A^{2}C^{3}+A^{1}A^{2}dT^{3}\right)\nonumber \\
 & +\int_{\mathcal{M}}q\left(A^{1}d\chi^{1}dT^{3}+d\chi^{1}A^{2}dT^{3}\right)+\int_{\mathcal{M}}qd\chi^{1}d\chi^{2}dT^{3}\nonumber \\
\equiv & S_{AAC}+\Delta S_{AAC}^{\left(1\right)}+\Delta S_{AAC}^{\left(2\right)}+\Delta S_{AAC}^{\left(3\right)}.
\end{align}
The \emph{large gauge invariance} requires that $\Delta S_{AAC}=\Delta S_{AAC}^{\left(1\right)}+\Delta S_{AAC}^{\left(2\right)}+\Delta S_{AAC}^{\left(3\right)}$ should be $0\mod 2\pi$.
Suppose that
\begin{align}
\oint_{\mathbb{S}^{1}}A^{1}=\frac{2\pi n_{1}}{N_{1}}, & \oint_{\mathbb{S}^{1}}d\chi^{1}=2\pi m_{1},\\
\oint_{\mathbb{S}^{1}}A^{2}=\frac{2\pi n_{2}}{N_{2}}, & \oint_{\mathbb{S}^{1}}d\chi^{2}=2\pi m_{2},\\
\oint_{\mathbb{S}^{3}}C^{3}=\frac{2\pi n_{3}}{N_{3}}, & \oint_{\mathbb{S}^{3}}dT^{3}=2\pi m_{3},
\end{align}
where $n_{i}$ and $m_{i}$ $\left(i=1,2,3\right)$ are integers.
Then
\begin{align}
\int_{\mathcal{M}}A^{1}d\chi^{2}C^{3}= & \oint_{S^{1}}A^{1}\oint_{S^{1}}d\chi^{2}\oint_{S^{3}}C^{3}=\frac{\left(2\pi\right)^{3}n_{1}m_{2}n_{3}}{N_{1}N_{3}},\\
\int_{\mathcal{M}}d\chi^{1}A^{2}C^{3}= & \oint_{S^{1}}d\chi^{1}\oint_{S^{1}}A^{2}\oint_{S^{3}}C^{3}=\frac{\left(2\pi\right)^{3}m_{1}n_{2}n_{3}}{N_{2}N_{3}},\\
\int_{\mathcal{M}}A^{1}A^{2}dT^{3}= & \oint_{S^{1}}A^{1}\oint_{S^{1}}A^{2}\oint_{S^{3}}dT^{3}=\frac{\left(2\pi\right)^{3}n_{1}n_{2}m_{3}}{N_{1}N_{2}},\\
\int_{\mathcal{M}}A^{1}d\chi^{1}dT^{3}= & \oint_{S^{1}}A^{1}\oint_{S^{1}}d\chi^{1}\oint_{S^{3}}dT^{3}=\frac{\left(2\pi\right)^{3}n_{1}m_{2}m_{3}}{N_{1}},\\
\int_{\mathcal{M}}d\chi^{1}A^{2}dT^{3}= & \oint_{S^{1}}d\chi^{1}\oint_{S^{2}}A^{2}\oint_{S^{3}}dT^{3}=\frac{\left(2\pi\right)^{3}m_{1}n_{2}m_{3}}{N_{2}},\\
\int_{\mathcal{M}}d\chi^{1}d\chi^{2}dT^{3}= & \oint_{S^{1}}d\chi^{1}\oint_{S^{1}}d\chi^{2}\oint_{S^{3}}dT^{3}=\left(2\pi\right)^{3}m_{1}m_{2}m_{3}.
\end{align}
The large gauge invariance of $S_{AAC}$ is guaranteed by the following constraints
\begin{align}
\Delta S_{AAC}^{\left(1\right)}= & \left(2\pi\right)^{3}\cdot\left(\frac{qn_{1}m_{2}n_{3}}{N_{1}N_{3}}+\frac{qm_{1}n_{2}n_{3}}{N_{2}N_{3}}+\frac{qn_{1}n_{2}m_{3}}{N_{1}N_{2}}\right)=0\mod2\pi,\\
\Delta S_{AAC}^{\left(2\right)}= & \left(2\pi\right)^{3}\cdot\left(\frac{qn_{1}m_{2}m_{3}}{N_{1}}+\frac{qm_{1}n_{2}m_{3}}{N_{2}}\right)=0\mod2\pi,\\
\Delta S_{AAC}^{\left(3\right)}= & \left(2\pi\right)^{3}\cdot qm_{1}m_{2}m_{3}=0\mod2\pi.
\end{align}
For arbitrary value of integers $n_i$ and $m_i$, the above constraints are satisfied by quantizing the coefficient $q$:
\begin{equation}
q=\frac{pN_{1}N_{2}N_{3}}{\left(2\pi\right)^{2}N_{123}},p\in\mathbb{Z},
\end{equation}
where $N_{123}$ is the greatest common divisor of $N_{1}$, $N_{2}$
and $N_{3}$.

So far, we have determined the quantization of $q$. Next, we find out the periodicity of $q$ by \emph{flux identification}. When the $\mathbb{Z}_{N_{i}}$ flux
is inserted as $n_{i}$ multiple units of $\frac{2\pi}{N_{i}}$, we
have
\begin{equation}
\int qA^{1}A^{2}C^{3}=q\oint_{S^{1}}A^{1}\oint_{S^{1}}A^{2}\oint_{S^{3}}C^{3}=\frac{p N_1 N_2 N_3}{\left(2\pi\right)^2 N_{123}}\frac{2\pi n_1}{N_1}\frac{2\pi n_2}{N_2}\frac{2\pi n_3}{N_3}=\frac{2\pi pn_{1}n_{2}n_{3}}{N_{123}}.
\end{equation}
For arbitrary value of $n_{1}n_{2}n_{3}$, $\exp\left({\rm i}\int qA^{1}A^{2}C^{3}\right)$
is invariant when $\frac{2\pi p}{N_{123}}$ shifts by $2\pi$, i.e., $p\rightarrow p+N_{123}$, which implies that $p$ should be identified with $p+N_{123}$.

Combined with $p\in\mathbb{Z}$, we conclude
that the quantization and periodic condition on the coefficient $q$
is:
\begin{equation}
q=\frac{pN_{1}N_{2}N_{3}}{\left(2\pi\right)^{2}N_{123}},p\in\mathbb{Z}_{N_{123}}.
\end{equation}
In the above example, we have shown how to get quantization of the coefficient of the $AAC$ twisted term based on large gauge invariance and flux identification. The similar derivation can be applied on other twisted terms.

\emph{Example: $AdAdA$ term $\left(A^{i}dA^{i}dA^{i},A^{i}dA^{i}dA^{j},A^{i}dA^{j}dA^{k}\right)$ studied in section~\ref{subsec_type_1_BF_a_twist}}.

\emph{Consider} $G=\mathbb{Z}_{N_1}$.  The TQFT action is
\begin{equation}
S=S_{BF}+S_{AdAdA}=\int\frac{N_{1}}{2\pi}C^{1}dA^{1}+qA^{1}dA^{1}dA^{1}
\end{equation}
and the gauge transformation is
\begin{align}
A^{1}\rightarrow & A^{1}+d\chi^{1},\nonumber\\
C^{1}\rightarrow & C^{1}+dT^{1}.
\end{align}
After integrating out the Lagrange multipliers $C^{1}$, the action $S$ reduces to $S_{AdAdA}=\int q A^1 dA^1 dA^1$, where the field $A^{1}$ is set to be closed with $\oint A^{1}\in\frac{2\pi}{N_{1}}\mathbb{Z}_{N_{1}}$.
After the gauge transformations,
\begin{equation}
S_{AdAdA}\rightarrow S_{AdAdA}+\int qd\chi^{1}dA^{1}dA^{1}.
\end{equation}
The large gauge invariance requires that
\begin{equation}
\Delta S_{AdAdA}=\int qd\chi^{1}dA^{1}dA^{1}=0\mod2\pi.
\end{equation}
Consider putting this TQFT on the spacetime manifold $\mathbb{S}^{1}\times\mathbb{S}^{2}\times\mathbb{S}^{2}$ and assuming
\begin{equation}
\oint_{S^{1}}d\chi^{1}=2\pi m_{1},  \oint_{S^{2}}dA^{1}=2\pi n_{a1},
\end{equation}
we have
\begin{equation}
\Delta S_{AdAdA}=\int qd\chi^{1}dA^{1}dA^{1}=q\oint_{S^{1}}d\chi^{1}\oint_{S^{2}}dA^{1}\oint_{S^{2}}dA^{1}=\left(2\pi\right)^{3} qm_{1}n_{a1}n_{a1}=0\mod2\pi,
\end{equation}
which indicates
\begin{equation}
q=\frac{p}{\left(2\pi\right)^{2}},p\in\mathbb{Z}.
\end{equation}
Now we consider the periodicity of $q$. Consider the $\mathbb{Z}_{N_{1}}$
flux is inserted as $n_{1}$ multiple units of $\frac{2\pi}{N_{1}}$,
we have
\begin{equation}
\int qA^{1}dA^{1}dA^{1}=\frac{p}{\left(2\pi\right)^{2}}\oint_{S^{1}}A^{1}\oint_{S^{2}}dA^{1}\oint_{S^{2}}dA^{1}=\frac{2\pi pn_{1}n_{a}n_{a}}{N_{1}}.
\end{equation}
For arbitrary values of $n_{1}$ and $n_{a}$, the partition function
should be invariant under a shift by $2\pi$, which means $\exp\left({\rm i}\int qA^{1}dA^{1}dA^{1}\right)\simeq\exp\left({\rm i}\int qA^{1}dA^{1}dA^{1}+2\pi {\rm i}\right)$,
thus we have
\begin{equation}
\frac{2\pi pn_{1}n_{a}n_{a}}{N_{1}}\simeq\frac{2\pi pn_{1}n_{a}n_{a}}{N_{1}}+2\pi\Rightarrow p\simeq p+N_{1}.
\end{equation}
In conclusion, the coefficient $q$ of $A^1 dA^1 dA^1$ twisted term is
\begin{equation}
q=\frac{p}{\left(2\pi\right)^{2}},p\in\mathbb{Z}_{N_{1}}.
\end{equation}

\emph{Consider} $G=\mathbb{Z}_{N_{1}}\times\mathbb{Z}_{N_{2}}$.
Possible twisted terms are $A^{1}dA^{2}dA^{1}$, $A^{1}dA^{1}dA^{2}$
$A^{1}dA^{2}dA^{2}$, $A^{2}dA^{1}dA^{2}$, $A^{2}dA^{2}dA^{1}$ and
$A^{2}dA^{1}dA^{1}$. Notice that
\begin{align}
A^{1}dA^{2}dA^{1}= & A^{1}dA^{1}dA^{2},\\
A^{2}dA^{1}dA^{2}= & A^{2}dA^{2}dA^{1},
\end{align}
 and
\begin{align}
d\left(A^{1}A^{2}dA^{1}\right)= & dA^{1}A^{2}dA^{1}-A^{1}dA^{2}dA^{1},\\
d\left(A^{1}A^{2}dA^{2}\right)= & dA^{1}A^{2}dA^{2}-A^{1}dA^{2}dA^{2}.
\end{align}
The linearly independent twisted terms are $A^{1}dA^{2}dA^{1}$
and $A^{2}dA^{1}dA^{2}$.
Take $A^{1}dA^{2}dA^{1}$ as an example:
\begin{equation}
S_{A^{1}dA^{2}dA^{1}}=\int\sum_{i=1}^{2}\frac{N_{i}}{2\pi}C^{i}dA^{i}+qA^{1}dA^{2}dA^{1}.
\end{equation}
Similarly, integrating out Lagrange multipliers $C^1$ and $C^2$ implies constraints that $A^1$ and $A^2$ are closed with $\oint A^1 \in\frac{2\pi}{N_1}\mathbb{Z}_{N_1}$ and $\oint A^2 \in\frac{2\pi}{N_2}\mathbb{Z}_{N_2}$.

The large gauge invariance,
\begin{equation}
\Delta S_{A^{1}dA^{2}dA^{1}}=\int qd\chi^{1}dA^{2}dA^{1}=q\left(2\pi\right)^{3}m_{1}n_{a2}n_{a1}=0\mod2\pi,
\end{equation}
requires that
\begin{equation}
q=\frac{p}{\left(2\pi\right)^{2}},p\in\mathbb{Z}.
\end{equation}
According to flux identification, $\int qA^{1}dA^{2}dA^{1}=\frac{2\pi pn_{1}n_{a2}n_{a1}}{N_{1}}$ indicates that
\begin{equation}
\frac{2\pi p}{N_1}\simeq\frac{2\pi p}{N_1}+2\pi\Rightarrow p\simeq p+N_1\Rightarrow p\in\mathbb{Z}_{N_{1}}.
\end{equation}
Notice that
\begin{equation}
d\left(A^{1}A^{2}dA^{1}\right)=dA^{1}A^{2}dA^{1}-A^{1}dA^{2}dA^{1},
\end{equation}
we can view $A^{1}dA^{2}dA^{1}$ and $A^{2}dA^{1}dA^{1}$ as the same
topological term up to a boundary term. Thus
\begin{align}
\exp\left({\rm i}\int qA^{1}dA^{2}dA^{1}\right)= & \exp\left[{\rm i}\int qdA^{1}A^{2}dA^{1}-{\rm i}\int qd\left(A^{1}A^{2}dA^{1}\right)\right]\nonumber\\
= & \exp\left(\frac{{\rm i}2\pi pn_{a1}n_{2}n_{a1}}{N_{2}}\right),
\end{align}
which tells us that
\begin{equation}
p\simeq p+N_{2}\Rightarrow p\in\mathbb{Z}_{N_{2}}.
\end{equation}
Together with $p\in\mathbb{Z}_{N_{1}}$, the coefficient $q$ of $A^{1}dA^{2}dA^{1}$ twisted term
is
\begin{equation}
q=\frac{p}{\left(2\pi\right)^{2}},p\in\mathbb{Z}_{N_{12}}.
\end{equation}

\emph{Consider} $G=\mathbb{Z}_{N_{1}}\times\mathbb{Z}_{N_{2}}\times\mathbb{Z}_{N_{3}}$.
The $AdAdA$ term can be $A^{1}dA^{2}dA^{3}$, $A^{2}dA^{3}dA^{1}$, or $A^{3}dA^{1}dA^{2}$.
But we will see that these 3 twisted terms are not linearly independent.
Consider the TQFT action
\begin{equation}
S=\int\sum_{i=1}^{3}\frac{N_i}{2\pi}C^{i}dA^{i}+q A^{1}dA^{2}dA^{3}
\end{equation}
with the gauge transformations
\begin{align}
A^{i}\rightarrow & A^{i}+d\chi^{i},\nonumber\\
C^{i}\rightarrow & C^{i}+dT^{i}.
\end{align}
Integrating out $C^i$ reduces the action $S$ to $\int q A^1 dA^2 dA^3$ where the fields $A^i$ are enforced to be closed with $\oint A^i \in \frac{2\pi}{N_i}\mathbb{Z}_{N_i}$.

The quantization of coefficient $q $ can be derived in the same manner:
\begin{equation}
q =\frac{p }{\left(2\pi\right)^{2}},p \in\mathbb{Z}_{N_{1}}.
\end{equation}
On the other hand, notice that
\begin{align}
d\left(A^{1}A^{2}dA^{3}\right)= & A^{2}dA^{3}dA^{1}-A^{1}dA^{2}dA^{3},\\
d\left(A^{1}dA^{2}A^{3}\right)= & A^{3}dA^{1}dA^{2}-A^{1}dA^{2}dA^{3};
\end{align}
up to boundary terms we have
\begin{align}
\exp\left({\rm i}\int qA^{1}dA^{2}dA^{3}\right)= & \exp\left[{\rm i}\int qA^{2}dA^{3}dA^{1}-{\rm i}\int qd\left(A^{1}A^{2}dA^{3}\right)\right]\nonumber \\
= & \exp\left(\frac{2\pi{\rm i}pn_{2}n_{a3}n_{a1}}{N_{2}}\right)\label{eq_flux_id_A1dA2dA3-1}
\end{align}
and
\begin{align}
\exp\left({\rm i}\int qA^{1}dA^{2}dA^{3}\right)= & \exp\left[{\rm i}\int qA^{3}dA^{1}dA^{2}-{\rm i}\int qd\left(A^{1}dA^{2}A^{3}\right)\right]\nonumber \\
= & \exp\left(\frac{2\pi ipn_{3}n_{a1}n_{a2}}{N_{3}}\right).\label{eq_flux_id_A1dA2dA3-2}
\end{align}
The flux identification of eq. (\ref{eq_flux_id_A1dA2dA3-1}) and
eq. (\ref{eq_flux_id_A1dA2dA3-2}) respectively leads to
\begin{align}
p\simeq & p+N_{2},\\
p\simeq & p+N_{3}.
\end{align}
Together with $p \simeq p +N_{1}$, we can see that the period of
$p $ is $N_{123}$, i.e.,
\begin{equation}
p \simeq p +N_{123}.
\end{equation}
Therefore, the coefficient $q $ of topological term $A^{1}dA^{2}dA^{3}$
is
\begin{equation}
q =\frac{p }{\left(2\pi\right)^{2}},p \in\mathbb{Z}_{N_{123}}.
\end{equation}

\emph{Example: $AAAdA$ term $\left(A^{i}A^{j}A^{k}dA^{i},A^{i}A^{j}A^{k}dA^{l}\right)$ studied in section~\ref{subsec_type_1_BF_a_twist}}.

\emph{Consider} $G=\prod_{i=1}^{4}\mathbb{Z}_{N_i}$. Possible twisted terms: $A^{2}A^{3}A^{4}dA^{1}$, $A^{3}A^{4}A^{1}dA^{2}$,
$A^{4}A^{1}A^{2}dA^{3}$ and $A^{1}A^{2}A^{3}dA^{4}$. We should notice that
these 4 twisted terms are not linearly independent.\footnote{
$d\left(A^{1}A^{2}A^{3}A^{4}\right)=dA^{1}A^{2}A^{3}A^{4}-A^{1}dA^{2}A^{3}A^{4}+A^{1}A^{2}dA^{3}A^{4}-A^{1}A^{2}A^{3}dA^{4}$.}
Take $A^{1}A^{2}A^{3}dA^{4}$ as an example, the TQFT action is
\begin{equation}
S=S_{BF}+S_{AAAdA}=\int\sum_{i=1}^{4}\frac{N_{i}}{2\pi}C^{i}dA^{i}+qA^{1}A^{2}A^{3}dA^{4}
\end{equation}
and the gauge transformations are
\begin{align}
A^{i}\rightarrow & A^{i}+d\chi^{i},\nonumber \\
C^{1}\rightarrow & C^{1}+dT^{1}+\frac{2\pi q}{N_{1}}\left(A^{2}\chi^{3}dA^{4}-A^{3}\chi^{2}dA^{4}-\frac{1}{2}\chi^{2}d\chi^{3}dA^{4}+\frac{1}{2}\chi^{3}d\chi^{2}dA^{4}\right),\nonumber \\
C^{2}\rightarrow & C^{2}+dT^{2}+\frac{2\pi q}{N_{2}}\left(-A^{1}\chi^{3}dA^{4}+A^{3}\chi^{1}dA^{4}+\frac{1}{2}\chi^{1}d\chi^{3}dA^{4}-\frac{1}{2}\chi^{3}d\chi^{1}dA^{4}\right),\nonumber \\
C^{3}\rightarrow & C^{3}+dT^{3}+\frac{2\pi q}{N_{3}}\left(A^{1}\chi^{2}dA^{4}-A^{2}\chi^{1}dA^{4}-\frac{1}{2}\chi^{1}d\chi^{2}dA^{4}+\frac{1}{2}\chi^{2}d\chi^{1}dA^{4}\right),\nonumber \\
C^{4}\rightarrow & C^{4}+dT^{4}.
\end{align}
Once we integrate out the Lagrange multipliers $C^i$, we obtain constraints on $A^i$: $dA^i=0$ and $\oint A^i\in \frac{2\pi}{N_i}\mathbb{Z}_{N_i}$.
Now we determine the coefficient $q$. First we find out the quantization
of $q$. Under the gauge transformations $A^{i}\rightarrow A^{i}+d\chi^{i}$
$\left(i=1,2,3,4\right)$,
\begin{equation}
S_{AAAdA}\rightarrow S_{AAAdA}+\Delta S_{AAAdA}^{\left(1\right)}+\Delta S_{AAAdA}^{\left(2\right)}+\Delta S_{AAAdA}^{\left(3\right)},
\end{equation}
where
\begin{align}
\Delta S_{AAAdA}^{\left(1\right)}= & \int q\left(d\chi^{1}A^{2}A^{3}dA^{4}+A^{1}d\chi^{2}A^{3}dA^{4}+A^{1}A^{2}d\chi^{3}dA^{4}\right),\\
\Delta S_{AAAdA}^{\left(2\right)}= & \int q\left(d\chi^{1}d\chi^{2}A^{3}dA^{4}+d\chi^{1}A^{2}d\chi^{3}dA^{4}+A^{1}A^{2}d\chi^{3}dA^{4}\right),\\
\Delta S_{AAAdA}^{\left(3\right)}= & \int qd\chi^{1}d\chi^{2}d\chi^{3}dA^{4}.
\end{align}
The large gauge invariance requires that $\Delta S_{AAAdA}=0\mod2\pi$,
i.e.,
\begin{align}
\Delta S{}_{AAAdA}^{\left(1\right)}= & \frac{\left(2\pi\right)^{4}q m_{1}n_{2}n_{3}n_{a4}}{N_{2}N_{3}}+\frac{\left(2\pi\right)^{4}qn_{1}m_{2}n_{3}n_{a4}}{N_{1}N_{3}}+\frac{\left(2\pi\right)^{4}q n_{1}n_{2}m_{3}n_{a4}}{N_{1}N_{2}}=0\mod2\pi,\\
\Delta S{}_{AAAdA}^{\left(2\right)}= & \frac{\left(2\pi\right)^{4}q m_{1}m_{2}n_{3}n_{a4}}{N_{3}}+\frac{\left(2\pi\right)^{4}q m_{1}n_{2}m_{3}n_{a4}}{N_{2}}+\frac{\left(2\pi\right)^{4}q n_{1}m_{2}m_{3}n_{a4}}{N_{1}}=0\mod2\pi,\\
\Delta S{}_{AAAdA}^{\left(3\right)}= & \left(2\pi\right)^{4}q m_{1}m_{2}m_{3}n_{a4}.
\end{align}
These constraints lead to
\begin{equation}
q =\frac{p N_{1}N_{2}N_{3}}{\left(2\pi\right)^{3}N_{123}},p \in\mathbb{Z}.
\end{equation}
Next we find out the period of $q $. We have
\begin{equation}
\int q A^{1}A^{2}A^{3}dA^{4}=\frac{p N_{1}N_{2}N_{3}}{\left(2\pi\right)^{3}N_{123}}\cdot\frac{2\pi n_{1}}{N_{1}}\cdot\frac{2\pi n_{2}}{N_{2}}\cdot\frac{2\pi n_{3}}{N_{3}}\cdot2\pi n_{a4}=\frac{2\pi p n_{1}n_{2}n_{3}n_{a4}}{N_{123}}.
\end{equation}
The flux identification tells us that no matter what values $n_{i}$
and $n_{a4}$ are, the partition function is invariant under a shift
by $2\pi$, which means
\begin{align}
\exp\left({\rm i}\int qA^{1}A^{2}A^{3}dA^{4}\right)= & \exp\left[\frac{{\rm i}2\pi pn_{1}n_{2}n_{3}n_{a4}}{N_{123}}\right]\simeq\exp\left[\frac{{\rm i}2\pi\left(p+1\right)n_{1}n_{2}n_{3}n_{a4}}{N_{123}}\right].
\end{align}
leading to
\begin{equation}
p \simeq p +N_{123}.
\end{equation}
However, there is another constraint on the period of $p $. Notice
that
\begin{equation}
d\left(A^{1}A^{2}A^{3}A^{4}\right)=dA^{1}A^{2}A^{3}A^{4}-A^{1}dA^{2}A^{3}A^{4}+A^{1}A^{2}dA^{3}A^{4}-A^{1}A^{2}A^{3}dA^{4},
\end{equation}
up boundary terms we have
\begin{align}
 & \exp\left[{\rm i}\int q\left(A^{2}A^{3}A^{4}dA^{1}-A^{3}A^{4}A^{1}dA^{2}+A^{4}A^{1}A^{2}dA^{3}\right)-{\rm i}\int qd\left(A^{1}A^{2}A^{3}A^{4}\right)\right]\nonumber \\
= & \exp\left[\frac{{\rm i}2\pi pN_{1}n_{2}n_{3}n_{4}n_{a1}}{N_{4}N_{123}}-\frac{{\rm i}2\pi pN_{2}n_{3}n_{4}n_{1}n_{a2}}{N_{4}N_{123}}+\frac{{\rm i}2\pi pN_{3}n_{4}n_{1}n_{2}n_{a3}}{N_{4}N_{123}}\right]\nonumber \\
= & \exp\left[\frac{{\rm i}2\pi pn_{4}}{N_{4}}\cdot\left(\frac{N_{1}n_{2}n_{3}n_{a1}-N_{2}n_{3}n_{1}n_{a2}+N_{3}n_{1}n_{2}n_{a3}}{N_{123}}\right)\right]\nonumber \\
\simeq & \exp\left[\frac{{\rm i}2\pi\left(p+1\right)n_{4}}{N_{4}}\cdot\left(\frac{N_{1}n_{2}n_{3}n_{a1}-N_{2}n_{3}n_{1}n_{a2}+N_{3}n_{1}n_{2}n_{a3}}{N_{123}}\right)\right],
\end{align}
which means that
\begin{equation}
p \simeq p +N_{4}.
\end{equation}
Together with $p \simeq p +N_{123}$, the period of $p $ is given
by $\gcd\left(N_{123},N_{4}\right)=N_{1234}$. In conclusion, the
coefficient $q $ of $A^{1}A^{2}A^{3}dA^{4}$ is
\begin{equation}
q= \frac{p N_{1}N_{2}N_{3}}{\left(2\pi\right)^{3}N_{123}},p \in\mathbb{Z}_{N_{1234}}.
\end{equation}

\emph{Example: $AAAAA$ term studied in section~\ref{subsec_type_1_BF_a_twist}}.

This twisted term is possible when there are at
least $5$ $\mathbb{Z}_{N_{i}}$ gauge subgroups, e.g., $\prod_{i=1}^{5}\mathbb{Z}_{N_{i}}$, $A^{1}A^{2}A^{3}A^{4}A^{5}$.
The action is
\begin{equation}
S=S_{BF}+S_{AAAAA}=\int\sum_{i=1}^{5}\frac{N_{i}}{2\pi}C^{i}dA^{i}+qA^{1}A^{2}A^{3}A^{4}A^{5}.
\end{equation}
The large gauge invariance and flux identification conditions indicate
the quantization and periodicity of $q$ in a similar manner:
\begin{equation}
q=\frac{pN_{1}N_{2}N_{3}N_{4}N_{5}}{\left(2\pi\right)^{4}N_{12345}},p\in\mathbb{Z}_{N_{12345}}.
\end{equation}

\subsection{Twisted terms from mixture of type-I and type-II $BF$ terms}
\emph{Examples: $BAdA$ term and $AAdB$ term studied in section~\ref{subsec_mixed_BF_a_twist}}.

Consider $G=\mathbb{Z}_{N_{1}}\times\mathbb{Z}_{N_{2}}\times\mathbb{Z}_{N_{3}}$. Notice that $d\left(B^{3}A^{1}A^{2}\right)=dB^{3}A^{1}A^{2}+B^{3}dA^{1}A^{2}-B^{3}A^{1}dA^{2}$,
thus only two of them are linearly independent. Take the $B^{3}A^{1}dA^{2}$
term as an example, the TQFT action is
\begin{equation}
S=\int\sum_{i=1}^{2}\frac{N_{i}}{2\pi}C^{i}dA^{i}+\frac{N_{3}}{2\pi}\widetilde{B}^{3}dB^{3}+q B^{3}A^{1}dA^{2}
\end{equation}
which is gauge-invariant under
\begin{align}
A^{1}\rightarrow A^{1}+d\chi^{1},\  & C^{1}\rightarrow C^{1}+dT^{1}-\frac{2\pi q }{N_{1}}dV^{3}A^{2}\nonumber \\
A^{2}\rightarrow A^{2}+d\chi^{2},\  & C^{2}\rightarrow C^{2}+dT^{2}\nonumber \\
B^{3}\rightarrow B^{3}+dV^{3},\  & \widetilde{B}^{3}\rightarrow\widetilde{B}^{3}+d\widetilde{V}^{3}-\frac{2\pi q }{N_{3}}d\chi^{1}A^{2}.
\end{align}
After integrating out $C^1$, $C^2$ and $\widetilde{B}^3$, the action reduces to $\int q B^3 A^1 dA^2$ where the fields are set to be closed with $\oint A^1\in \frac{2\pi}{N_1}\mathbb{Z}_{N_1}$, $\oint A^2\in \frac{2\pi}{N_2}\mathbb{Z}_{N_2}$ and $\oint B^3\in \frac{2\pi}{N_3}\mathbb{Z}_{N_3}$.
In a similar manner, it can be derived that the coefficient $q$
of $B^{3}A^{1}dA^{2}$ is
\begin{equation}
q =\frac{p N_{3}N_{1}}{\left(2\pi\right)^{2}N_{13}},p \in\mathbb{Z}_{N_{13}}.
\end{equation}
However, there is another constraint on the coefficient $q$. Noticed that
\begin{equation}
d\left(B^{3}A^{1}A^{2}\right)=dB^{3}A^{1}A^{2}+B^{3}dA^{1}A^{2}-B^{3}A^{1}dA^{2},
\end{equation}
we have
\begin{align}
\exp\left({\rm i}\int qB^{3}A^{1}dA^{2}\right)= & \exp\left[{\rm i}\int qA^{1}A^{2}dB^{2}+{\rm i}\int qB^{3}A^{2}dA^{1}-{\rm i}\int qd\left(B^{3}A^{1}A^{2}\right)\right]\nonumber \\
= & \exp\left[\frac{2\pi{\rm i}pN_{3}N_{1}n_{1}n_{2}}{N_{13}\cdot N_{1}N_{2}}+\frac{2\pi{\rm i}pN_{3}N_{1}n_{3}n_{2}}{N_{13}\cdot N_{3}N_{2}}\right]\nonumber \\
= & \exp\left[\frac{2\pi{\rm i}pn_{2}}{N_{2}}\frac{n_{1}N_{3}+n_{3}N_{1}}{N_{13}}\right].
\end{align}
Since $\frac{n_{1}N_{3}+n_{3}N_{1}}{N_{13}}$ is an integer, $n_{2}\cdot\frac{n_{1}N_{3}+n_{3}N_{1}}{N_{13}}$
is also an integer. In order to keep $\mathcal{Z}$ invariant for arbitrary
$n_{1}$, $n_{2}$ and $n_{3}$, $\frac{2\pi p }{N_{2}}$ is required
to be identical to $\frac{2\pi p }{N_{2}}+2\pi$, i.e.,
\begin{equation}
\frac{2\pi p }{N_{2}}\simeq\frac{2\pi p }{N_{2}}+2\pi\Rightarrow p \simeq p +N_{2}.
\end{equation}
Together with $p \in\mathbb{Z}_{N_{13}}$, we can conclude that the
actual period of $p $ is
\begin{equation}
\gcd\left(N_{13},N_{2}\right)=N_{123}.
\end{equation}
Thus the quantization and period of the coefficient $q $ of $B^{3}A^{1}dA^{2}$
are actually given by
\begin{equation}
q =\frac{p N_{3}N_{1}}{\left(2\pi\right)^{2}N_{13}},p \in\mathbb{Z}_{N_{123}}
\end{equation}
If we consider the $B^{3}A^{2}dA^{1}$ term and a TQFT action
\begin{equation}
S_{1}=\int\sum_{i=1}^{2}\frac{N_{i}}{2\pi}C^{i}dA^{i}+\frac{N_{3}}{2\pi}\widetilde{B}^{3}dB^{3}+q_{1}B^{3}A^{2}dA^{1},
\end{equation}
the gauge transformations are
\begin{align}
A^{1}\rightarrow & A^{1}+d\chi^{1},\nonumber\\
A^{2}\rightarrow & A^{2}+d\chi^{2},\nonumber\\
B^{3}\rightarrow & B^{3}+dV^{3},\nonumber\\
\widetilde{B}^{3}\rightarrow & \widetilde{B}^{3}+d\widetilde{V}^{3}-\frac{2\pi q_{1} }{N_{3}}  d\chi^{2}A^{1},\nonumber\\
C^{1}\rightarrow & C^{1}+dT^{1},\nonumber\\
C^{2}\rightarrow & C^{2}+dT^{2}- \frac{2\pi q_{1}}{N_{2}}  dV^{3}A^{1};
\end{align}
the coefficient $q_{1}$ is determined in a similar manner:
\begin{equation}
q_{1}=\frac{p_{1}N_{3}N_{2}}{\left(2\pi\right)^{2}N_{23}},p_{1}\in\mathbb{Z}_{N_{123}}.
\end{equation}
If we consider the $A^{1}A^{2}dB^{3}$ term and a TQFT action
\begin{equation}
S_{2}=\int\sum_{i=1}^{2}\frac{N_{i}}{2\pi}C^{i}dA^{i}+\frac{N_{3}}{2\pi}\widetilde{B}^{3}dB^{3}+q_{2}A^{1}A^{2}dB^{3},
\end{equation}
the gauge transformations are
\begin{align}
A^{1}\rightarrow & A^{1}+d\chi^{1},\nonumber\\
A^{2}\rightarrow & A^{2}+d\chi^{2},\nonumber\\
B^{3}\rightarrow & B^{3}+dV^{3},\nonumber\\
\widetilde{B}^{3}\rightarrow & \widetilde{B}^{3}+d\widetilde{V}^{3},\nonumber\\
C^{1}\rightarrow & C^{1}+dT^{1}+ \frac{2\pi q_{2}}{N_{1}}  d\chi^{2}B^{3},\nonumber\\
C^{2}\rightarrow & C^{2}+dT^{2}- \frac{2\pi q_{2}}{N_{2}}  d\chi^{1}B^{3};
\end{align}
the coefficient $q_{2}$ is
\begin{equation}
q_{2}=\frac{p_{2}N_{1}N_{2}}{\left(2\pi\right)^{2}N_{12}},p_{2}\in\mathbb{Z}_{N_{123}}.
\end{equation}

\section{Gauge-invariance of $S=S_{BF}+S_{AAC}$ and its Wilson operator\label{appendix_gauge_invariance_AAC}}

In this appendix, we derive the gauge transformations for the action~(\ref{eq_S_CdA+AAC}),
studied in section \ref{subsec_type_1_BF_a_twist},
\begin{equation}
S=S_{BF}+S_{AAC}=\int\sum_{i=1}^{3}\frac{N_{i}}{2\pi}C^{i}dA^{i}+qA^{1}A^{2}C^{3}\label{eq_action_appendix_AAC}
\end{equation}
with $q=\frac{pN_{1}N_{2}N_{3}}{\left(2\pi\right)^{2}N_{123}}$ and
$p\in\mathbb{Z}_{N_{123}}$. The basic idea can be applied on other
actions studied in this paper. Then we verify that the corresponding
Wilson operator
\begin{align}
\mathcal{W}= & \exp\left\{ {\rm i}\int_{\omega_{1}}e_{1}\left[C^{1}+\frac{1}{2}\frac{2\pi q}{N_{1}}\left(d^{-1}A^{2}C^{3}-d^{-1}C^{3}A^{2}\right)\right]\right.\nonumber \\
 & +{\rm i}\int_{\omega_{2}}e_{2}\left[C^{2}+\frac{1}{2}\frac{2\pi q}{N_{2}}\left(d^{-1}C^{3}A^{1}-d^{-1}A^{1}C^{3}\right)\right]\nonumber \\
 & +\left.{\rm i}\int_{\gamma_{3}}e_{3}\left[A^{3}+\frac{1}{2}\frac{2\pi q}{N_{3}}\left(d^{-1}A^{1}A^{2}-d^{-1}A^{2}A^{1}\right)\right]\right\}
\end{align}
is indeed gauge-invariant under the gauge transformations.

In this action $S=S_{BF}+S_{AAC}+\int\sum_{i=1}^{3}\frac{N_{i}}{2\pi}C^{i}dA^{i}+qA^{1}A^{2}C^{3}$,
$C^{1}$, $C^{2}$ and $A^{3}$ serve as Lagrange multipliers which
respectively enforce the $\mathbb{Z}_{N_{1}}$, $\mathbb{Z}_{N_{2}}$
and $\mathbb{Z}_{N_{3}}$ cyclic group structures. These cyclic group
structures imply that the gauge transformations of $A^{1}$, $A^{2}$
and $C^{3}$ are
\begin{align}
A^{1}\rightarrow & A^{1}+d\chi^{1},\\
A^{2}\rightarrow & A^{2}+d\chi^{2},\\
C^{3}\rightarrow & C^{3}+dT^{3},
\end{align}
where $\chi^{i}$ and $T^{i}$ are $0$-form and $2$-form gauge parameters.
In order to compensate the variation of $S_{AAC}$ after gauge transformation,
the Lagrange multipliers $C^{1}$, $C^{2}$ and $A^{3}$ transform
with extra shift terms:
\begin{align}
C^{1}\rightarrow & C^{1}+dT^{1}+X^{1},\\
C^{2}\rightarrow & C^{2}+dT^{2}+X^{2},\\
A^{3}\rightarrow & A^{3}+d\chi^{3}+Y^{3}.
\end{align}
Our goal is to determine these shift terms $X^{1}$, $X^{2}$ and
$Y^{3}$. After gauge transformation, the action changes as
\begin{align}
 & \int\sum_{i=1}^{3}\frac{N_{i}}{2\pi}C^{i}dA^{i}+qA^{1}A^{2}C^{3}\nonumber \\
\rightarrow & \int\sum_{i=1}^{2}\frac{N_{i}}{2\pi}\left(C^{i}+dT^{i}+X^{i}\right)dA^{i}+\frac{N_{3}}{2\pi}\left(A^{3}+d\chi^{3}+Y^{3}\right)dC^{3}\nonumber \\
 & +q\left(A^{1}+d\chi^{1}\right)\left(A^{2}+d\chi^{2}\right)\left(C^{3}+dT^{3}\right)\nonumber \\
= & \int\sum_{i=1}^{3}\frac{N_{i}}{2\pi}C^{i}dA^{i}+qA^{1}A^{2}C^{3}+\frac{N_{1}}{2\pi}dT^{1}dA^{1}+\frac{N_{2}}{2\pi}dT^{2}dA^{2}+\frac{N_{3}}{2\pi}d\chi^{3}dC^{3}\nonumber \\
 & +\frac{N_{1}}{2\pi}X^{1}dA^{1}+\frac{N_{2}}{2\pi}X^{2}dA^{2}+\frac{N_{3}}{2\pi}Y^{3}dC^{3}\nonumber \\
 & +q\left(d\chi^{1}A^{2}C^{3}+A^{1}d\chi^{2}C^{3}+d\chi^{1}d\chi^{2}C^{3}A^{1}A^{2}dT^{3}+d\chi^{1}A^{2}dT^{3}+A^{1}d\chi^{2}dT^{3}\right)\nonumber \\
 & +qd\chi^{1}d\chi^{2}dT^{3}.
\end{align}
Because we only concern the definition of gauge transformations here,
we can drop the boundary terms and obtain
\begin{align}
\Delta S= & \int\frac{N_{1}}{2\pi}X^{1}dA^{1}+\frac{N_{2}}{2\pi}X^{2}dA^{2}+\frac{N_{3}}{2\pi}Y^{3}dC^{3}\nonumber \\
 & +q\left(d\chi^{1}A^{2}C^{3}+A^{1}d\chi^{2}C^{3}+d\chi^{1}d\chi^{2}C^{3}\right)\nonumber \\
 & +q\left(A^{1}A^{2}dT^{3}+d\chi^{1}A^{2}dT^{3}+A^{1}d\chi^{2}dT^{3}\right).
\end{align}
Gauge-invariance requires $\Delta S=\int\left(\text{total derivatives}\right)$
thus $\Delta S$ can be neglected. Noticed that
\begin{align}
d\chi^{1}A^{2}C^{3}= & d\left(\chi^{1}A^{2}C^{3}\right)-\chi^{1}C^{3}dA^{2}+\chi^{1}A^{2}dC^{3},\\
A^{1}d\chi^{2}C^{3}= & -d\left(A^{1}\chi^{2}C^{3}\right)+\chi^{2}C^{3}dA^{1}-A^{1}\chi^{2}dC^{3},\\
d\chi^{1}d\chi^{2}C^{3}= & \frac{1}{2}d\left(\chi^{1}d\chi^{2}C^{3}\right)-\frac{1}{2}d\left(d\chi^{1}\chi^{2}C^{3}\right)+\frac{1}{2}\chi^{1}d\chi^{2}dC^{3}-\frac{1}{2}\chi^{2}d\chi^{1}dC^{3},\\
A^{1}A^{2}dT^{3}= & d\left(A^{1}A^{2}T^{3}\right)-A^{2}T^{3}dA^{1}+A^{1}T^{3}dA^{2},\\
d\chi^{1}A^{2}dT^{3}= & d\left(\chi^{1}A^{2}dT^{3}\right)-\chi^{1}dT^{3}dA^{2},\\
A^{1}d\chi^{2}dT^{3}= & -d\left(A^{1}\chi^{2}dT^{3}\right)+\chi^{2}dT^{3}dA^{1},
\end{align}
we see that for
\begin{align}
X^{1}= & -\frac{2\pi q}{N_{1}}\left(\chi^{2}C^{3}-A^{2}T^{3}+\chi^{2}dT^{3}\right),\\
X^{2}= & \frac{2\pi q}{N_{2}}\left(\chi^{1}C^{3}-A^{1}T^{3}+\chi^{1}dT^{3}\right),\\
Y^{3}= & -\frac{2\pi q}{N_{3}}\left[\left(\chi^{1}A^{2}+\frac{1}{2}\chi^{1}d\chi^{2}\right)-\left(\chi^{2}A^{1}+\frac{1}{2}\chi^{2}d\chi^{1}\right)\right],
\end{align}
$\Delta S$ can be written as
\begin{align}
\Delta S= & \int q\left[d\left(\chi^{1}A^{2}C^{3}\right)-d\left(A^{1}\chi^{2}C^{3}\right)+\frac{1}{2}d\left(\chi^{1}d\chi^{2}C^{3}\right)-\frac{1}{2}d\left(d\chi^{1}\chi^{2}C^{3}\right)\right]\nonumber \\
 & +q\left[d\left(A^{1}A^{2}T^{3}\right)+d\left(\chi^{1}A^{2}dT^{3}\right)-d\left(A^{1}\chi^{2}dT^{3}\right)\right]\nonumber \\
= & \int\left(\text{total derivatives}\right),
\end{align}
which ensure the action (\ref{eq_action_appendix_AAC}) is invariant
after gauge transformation.Therefore, we conclude that the gauge transformations
for $S=\int\sum_{i=1}^{3}\frac{N_{i}}{2\pi}C^{i}dA^{i}+qA^{1}A^{2}C^{3}$
are
\begin{align}
A^{1}\rightarrow A^{1}+d\chi^{1},\  & C^{1}\rightarrow C^{1}+dT^{1}-\frac{2\pi q}{N_{1}}\left(\chi^{2}C^{3}-A^{2}T^{3}+\chi^{2}dT^{3}\right),\nonumber \\
A^{2}\rightarrow A^{2}+d\chi^{2},\  & C^{2}\rightarrow C^{2}+dT^{2}+\frac{2\pi q}{N_{2}}\left(\chi^{1}C^{3}-A^{1}T^{3}+\chi^{1}dT^{3}\right),\nonumber \\
C^{3}\rightarrow C^{3}+dT^{3},\  & A^{3}\rightarrow A^{3}+d\chi^{3}-\frac{2\pi q}{N_{3}}\left[\left(\chi^{1}A^{2}+\frac{1}{2}\chi^{1}d\chi^{2}\right)-\left(\chi^{2}A^{1}+\frac{1}{2}\chi^{2}d\chi^{1}\right)\right],\label{eq_GT_appendix_AAC}
\end{align}
same as eq.~(\ref{eq_GT_AAC}) in section \ref{subsec_type_1_BF_a_twist}.

Next, we verify that the Wilson operator for this action,
\begin{align}
\mathcal{W}= & \exp\left\{ {\rm i}\int_{\omega_{1}}e_{1}\left[C^{1}+\frac{1}{2}\frac{2\pi q}{N_{1}}\left(d^{-1}A^{2}C^{3}-d^{-1}C^{3}A^{2}\right)\right]\right.\nonumber \\
 & +{\rm i}\int_{\omega_{2}}e_{2}\left[C^{2}+\frac{1}{2}\frac{2\pi q}{N_{2}}\left(d^{-1}C^{3}A^{1}-d^{-1}A^{1}C^{3}\right)\right]\nonumber \\
 & +\left.{\rm i}\int_{\gamma_{3}}e_{3}\left[A^{3}+\frac{1}{2}\frac{2\pi q}{N_{3}}\left(d^{-1}A^{1}A^{2}-d^{-1}A^{2}A^{1}\right)\right]\right\} ,\label{eq_wilson_AAC}
\end{align}
is invariant under the gauge transformations. This Wilson operator
is composed of three similar terms. Below we show that $C^{1}+\frac{1}{2}\frac{2\pi q}{N_{1}}\left(d^{-1}A^{2}C^{3}-d^{-1}C^{3}A^{2}\right)$
is invariant under transformation (\ref{eq_GT_appendix_AAC}); the
other components of $\mathcal{W}$ can be proven gauge-invariant in
a similar way. After gauge transformation (\ref{eq_GT_appendix_AAC}),
\begin{align}
 & \int_{\omega_{1}}C^{1}+\frac{1}{2}\frac{2\pi q}{N_{1}}\left(d^{-1}A^{2}C^{3}-d^{-1}C^{3}A^{2}\right)\nonumber \\
\rightarrow & \int_{\omega_{1}}C^{1}+dT^{1}-\frac{2\pi q}{N_{1}}\left(\chi^{2}C^{3}-A^{2}T^{3}+\chi^{2}dT^{3}\right)\nonumber \\
 & +\frac{1}{2}\frac{2\pi q}{N_{1}}\left(d^{-1}A^{2}+\chi^{2}\right)\left(C^{3}+dT^{3}\right)-\frac{1}{2}\frac{2\pi q}{N_{1}}\left(d^{-1}C^{3}+T^{3}\right)\left(A^{2}+d\chi^{2}\right)\nonumber \\
= & \int_{\omega_{1}}C^{1}+dT^{1}-\frac{2\pi q}{N_{1}}\left(\chi^{2}C^{3}-A^{2}T^{3}+\chi^{2}dT^{3}\right)\nonumber \\
 & +\frac{1}{2}\frac{2\pi q}{N_{1}}\left(d^{-1}A^{2}C^{3}+\chi^{2}C^{3}+d^{-1}A^{2}dT^{3}+\chi^{2}dT^{3}\right)-\frac{1}{2}\frac{2\pi q}{N_{1}}\left(d^{-1}C^{3}A^{2}+T^{3}A^{2}+d^{-1}C^{3}d\chi^{2}+T^{3}d\chi^{2}\right)\nonumber \\
= & \int_{\omega_{1}}C^{1}+\frac{1}{2}\frac{2\pi q}{N_{1}}\left(d^{-1}A^{2}C^{3}-d^{-1}C^{3}A^{2}\right)+dT^{1}\nonumber \\
 & -\frac{1}{2}\frac{2\pi q}{N_{1}}\chi^{2}C^{3}+\frac{1}{2}\frac{2\pi q}{N_{1}}A^{2}T^{3}-\frac{1}{2}\frac{2\pi q}{N_{1}}\chi^{2}dT^{3}-\frac{1}{2}\frac{2\pi q}{N_{1}}T^{3}d\chi^{2}+\frac{1}{2}\frac{2\pi q}{N_{1}}d^{-1}A^{2}dT^{3}-\frac{1}{2}\frac{2\pi q}{N_{1}}d^{-1}C^{3}d\chi^{2}.
\end{align}
Because $d\left(\chi^{2}T^{3}\right)=d\chi^{2}T^{3}+\chi^{2}dT^{3}=T^{3}d\chi^{2}+\chi^{2}dT^{3}$,
\begin{equation}
\int_{\omega_{1}}\left(-\frac{1}{2}\frac{2\pi q}{N_{1}}\chi^{2}dT^{3}-\frac{1}{2}\frac{2\pi q}{N_{1}}T^{3}d\chi^{2}\right)=-\frac{1}{2}\frac{2\pi q}{N_{1}}\int_{\omega_{1}}\left(\chi^{2}dT^{3}+T^{3}d\chi^{2}\right)=-\frac{1}{2}\frac{2\pi q}{N_{1}}\int_{\omega_{1}}d\left(\chi^{2}T^{3}\right).
\end{equation}
Meanwhile, using
\begin{align}
\int_{\omega_{1}}d\left(d^{-1}A^{2}T^{3}\right)=0= & \int_{\omega_{1}}A^{2}T^{3}+\int_{\omega_{1}}d^{-1}A^{2}dT^{3},\\
\int_{\omega_{1}}d\left(d^{-1}C^{3}\chi^{2}\right)=0= & \int_{\omega_{1}}C^{3}\chi^{2}+\int_{\omega_{1}}d^{-1}C^{3}d\chi^{2},
\end{align}
we have
\begin{align}
\int_{\omega_{1}}\frac{1}{2}\frac{2\pi q}{N_{1}}A^{2}T^{3}+\int_{\omega_{1}}\frac{1}{2}\frac{2\pi q}{N_{1}}d^{-1}A^{2}dT^{3}= & \frac{1}{2}\frac{2\pi q}{N_{1}}\left(\int_{\omega_{1}}A^{2}T^{3}+\int_{\omega_{1}}d^{-1}A^{2}dT^{3}\right)=0\\
\int_{\omega_{1}}\left[-\frac{1}{2}\frac{2\pi q}{N_{1}}\left(\chi^{2}C^{3}\right)-\frac{1}{2}\frac{2\pi q}{N_{1}}d^{-1}C^{3}d\chi^{2}\right]= & -\frac{1}{2}\frac{2\pi q}{N_{1}}\left(\int_{\omega_{1}}\chi^{2}C^{3}+\int_{\omega_{1}}d^{-1}C^{3}d\chi^{2}\right)=0.
\end{align}
Therefore, we can see that after gauge transformation,
\begin{align}
\int_{\omega_{1}}C^{1}+\frac{1}{2}\frac{2\pi q}{N_{1}}\left(d^{-1}A^{2}C^{3}-d^{-1}C^{3}A^{2}\right)\rightarrow & \int_{\omega_{1}}C^{1}+\frac{1}{2}\frac{2\pi q}{N_{1}}\left(d^{-1}A^{2}C^{3}-d^{-1}C^{3}A^{2}\right)\nonumber \\
 & +\int_{\omega_{1}}\text{boundary terms},
\end{align}
i.e., $\exp\left\{ {\rm i}\int_{\omega_{1}}e_{1}\left[C^{1}+\frac{1}{2}\frac{2\pi q}{N_{1}}\left(d^{-1}A^{2}C^{3}-d^{-1}C^{3}A^{2}\right)\right]\right\} $
is invariant. Similarly, $\mathcal{W}$ can be verified as gauge-invariant.

Last but not least, it should be emphasized that the idea of derivation
of gauge transformation can be generalized to all actions discussed
in this paper, as long as we carefully deal with the shift terms for
the gauge fields serving as Lagrange multipliers.

\section{Classification for all combinations of $AAC$ terms when
$G=\prod_{i=1}^{4}\mathbb{Z}_{N_{i}}$\label{appendix_4Zn_class_all_AAC}}

When the gauge group is $G=\prod_{i=1}^{4}\mathbb{Z}_{N_{i}}$, the
action and corresponding classification depend on the which $AAC$
terms are considered if $\alpha=\left\{ N_{1},N_{2},N_{3}\right\} $
or $\alpha=\left\{ N_{1},N_{2},N_{3},N_{4}\right\} $. In this appendix,
we list all classifications for each compatible combination of $AAC$
terms and other twisted terms when $G=\prod_{i=1}^{4}\mathbb{Z}_{N_{i}}$.

In main text, for $\alpha=\left\{ N_{1},N_{2},N_{3}\right\} $, we
have demonstrate the situation in which the $A^{1}A^{2}C^{3}$ term
is included in the action:
\begin{align}
S= & \int\sum_{i=1}^{3}\frac{N_{i}}{2\pi}C^{i}dA^{i}+\frac{N_{4}}{2\pi}\widetilde{B}^{4}dB^{4}+\frac{pN_{1}N_{2}N_{3}}{\left(2\pi\right)^{2}N_{123}}A^{1}A^{2}C^{3}+\left\langle \sum_{i=1,2}A^{i}dA^{i}dA^{i},\right.\nonumber\\
 & \left.A^{1}dA^{2}dA^{1},A^{2}dA^{1}dA^{2},\sum_{i=1,2}\left(B^{4}B^{4}A^{i}+B^{4}A^{i}dA^{i}\right),A^{1}A^{2}dB^{4},B^{4}A^{2}dA^{1}\right\rangle
\label{eq_action_4Zn_A1A2C3-1}
\end{align}
with $p\in\mathbb{Z}_{N_{123}}\setminus\left\{ 0\right\} $, whose
classification is
\begin{equation}
\left(\mathbb{Z}_{N_{123}}\setminus\left\{ 0\right\} \right)\times\prod_{i=1}^{2}\mathbb{Z}_{N_{i}}\times\left(\mathbb{Z}_{N_{12}}\right)^{2}\times\left(\mathbb{Z}_{N_{14}}\right)^{2}\times\left(\mathbb{Z}_{N_{24}}\right)^{2}\times\left(\mathbb{Z}_{N_{124}}\right)^{2}.
\end{equation}
In fact, besides $A^{1}A^{2}C^{3}$, possible linearly independent
$AAC$ terms are $A^{2}A^{3}C^{1}$ and $A^{3}A^{1}C^{2}$. If $A^{2}A^{3}C^{1}$,
instead of $A^{1}A^{2}C^{3}$, is added to the action, the action
is
\begin{align}
S= & \int\sum_{i=1}^{3}\frac{N_{i}}{2\pi}C^{i}dA^{i}+\frac{N_{4}}{2\pi}\widetilde{B}^{4}dB^{4}+\frac{pN_{1}N_{2}N_{3}}{\left(2\pi\right)^{2}N_{123}}A^{2}A^{3}C^{1}+\left\langle  \sum_{i=2,3}A^{i}dA^{i}dA^{i},\right.\nonumber \\
 & \left.A^{2}dA^{3}dA^{2},A^{3}dA^{2}dA^{3},\sum_{i=2,3}\left(B^{4}B^{4}A^{i}+B^{4}A^{i}dA^{i}\right),A^{2}A^{3}dB^{4},B^{4}A^{3}dA^{2}\right\rangle
\end{align}
with $p\in\mathbb{Z}_{N_{123}}\setminus\left\{ 0\right\} $ and its
classification is
\begin{equation}
\left(\mathbb{Z}_{N_{123}}\setminus\left\{ 0\right\} \right)\times\prod_{i=2,3}\mathbb{Z}_{N_{i}}\times\left(\mathbb{Z}_{N_{23}}\right)^{2}\times\left(\mathbb{Z}_{N_{24}}\right)^{2}\times\left(\mathbb{Z}_{N_{34}}\right)^{2}\times\left(\mathbb{Z}_{N_{234}}\right)^{2}.
\end{equation}
If $A^{3}A^{1}C^{2}$ appears in the action, the action is
\begin{align}
S= & \int\sum_{i=1}^{3}\frac{N_{i}}{2\pi}C^{i}dA^{i}+\frac{N_{4}}{2\pi}\widetilde{B}^{4}dB^{4}+\frac{pN_{1}N_{2}N_{3}}{\left(2\pi\right)^{2}N_{123}}A^{3}A^{1}C^{2}+\left\langle  \sum_{i=1,3}A^{i}dA^{i}dA^{i},\right.\nonumber \\
 & \left.A^{3}dA^{1}dA^{3},A^{1}dA^{3}dA^{1},\sum_{i=1,3}\left(B^{4}B^{4}A^{i}+B^{4}A^{i}dA^{i}\right),A^{3}A^{1}dB^{4},B^{4}A^{1}dA^{3}\right\rangle
\end{align}
with $p\in\mathbb{Z}_{N_{123}}\setminus\left\{ 0\right\} $ and its
classification is
\begin{equation}
\left(\mathbb{Z}_{N_{123}}\setminus\left\{ 0\right\} \right)\times\prod_{i=1,3}\mathbb{Z}_{N_{i}}\times\left(\mathbb{Z}_{N_{13}}\right)^{2}\times\left(\mathbb{Z}_{N_{14}}\right)^{2}\times\left(\mathbb{Z}_{N_{34}}\right)^{2}\times\left(\mathbb{Z}_{N_{134}}\right)^{2}.
\end{equation}

For $\alpha=\left\{ N_{1},N_{2},N_{3},N_{4}\right\} $, there are
more possible $AAC$ terms and more of them are compatible
thus can be added to the action simultaneously. In the main text, we have
shown the classifications for cases in which $A^{1}A^{2}C^{4}$, $A^{1}A^{2}C^{4}+A^{1}A^{3}C^{4}$,
$A^{1}A^{2}C^{4}+A^{1}A^{3}C^{4}+A^{2}A^{3}C^{4}$, and $A^{1}A^{2}C^{4}+A^{1}A^{2}C^{3}$
are included in actions respectively. Below we exhaust all possible
combinations of $AAC$ terms and give corresponding classifications.
\begin{description}
\item [{$A^{2}A^{3}C^{1}$}] The action is
\begin{align}
S= & \int\sum_{i=1}^{4}\frac{N_{i}}{2\pi}C^{i}dA^{1}+\frac{pN_{2}N_{3}N_{1}}{\left(2\pi\right)^{2}N_{123}}A^{2}A^{3}C^{1}+\left\langle  \sum_{i=2,3,4}A^{i}dA^{i}dA^{i},\right.\nonumber\\
 & \left.\sum_{\substack{2\leq i<j\leq4}
}\left(A^{i}dA^{j}dA^{i}+A^{j}dA^{i}dA^{j}\right),A^{2}dA^{3}dA^{4},\sum_{i=2,3,4}A^{2}A^{3}A^{4}dA^{i}\right\rangle
\end{align}
with $p\in\mathbb{Z}_{N_{123}}\setminus\left\{ 0\right\} $. Its classification
is
\begin{equation}
\left(\mathbb{Z}_{N_{123}}\setminus\left\{ 0\right\} \right)\times\prod_{i=2}^{4}\mathbb{Z}_{N_{i}}\times\prod_{2\leq i<j\leq4}\left(\mathbb{Z}_{N_{ij}}\right)^{2}\times\left(\mathbb{Z}_{N_{234}}\right)^{4}.
\end{equation}
\item [{$A^{2}A^{4}C^{1}$}] The classification is $\left(\mathbb{Z}_{N_{124}}\setminus\left\{ 0\right\} \right)\times{\displaystyle \prod_{i=2}^{4}\mathbb{Z}_{N_{i}}}\times{\displaystyle \prod_{2\leq i<j\leq4}}\left(\mathbb{Z}_{N_{ij}}\right)^{2}\times\left(\mathbb{Z}_{N_{234}}\right)^{4}$.
\item [{$A^{3}A^{4}C^{1}$}] The classification is $\left(\mathbb{Z}_{N_{134}}\setminus\left\{ 0\right\} \right)\times{\displaystyle \prod_{i=2}^{4}\mathbb{Z}_{N_{i}}}\times{\displaystyle \prod_{2\leq i<j\leq4}}\left(\mathbb{Z}_{N_{ij}}\right)^{2}\times\left(\mathbb{Z}_{N_{234}}\right)^{4}$.
\item [{$A^{2}A^{3}C^{1}+A^{2}A^{4}C^{1}$}] The action is
\begin{align}
S= & \int\sum_{i=1}^{4}\frac{N_{i}}{2\pi}C^{i}dA^{1}+\frac{pN_{2}N_{3}N_{1}}{\left(2\pi\right)^{2}N_{123}}A^{2}A^{3}C^{1}+\frac{p' N_{2}N_{4}N_{1}}{\left(2\pi\right)^{2}N_{124}}A^{2}A^{4}C^{1}+\left\langle \sum_{i=2,3,4}A^{i}dA^{i}dA^{i},\right.\nonumber\\
 & \left.\sum_{\substack{2\leq i<j\leq4}
}\left(A^{i}dA^{j}dA^{i}+A^{j}dA^{i}dA^{j}\right),A^{2}dA^{3}dA^{4},\sum_{i=2,3,4}A^{2}A^{3}A^{4}dA^{i}\right\rangle
\end{align}
with $p\in\mathbb{Z}_{N_{123}}\setminus\left\{ 0\right\} $ and $p' \in\mathbb{Z}_{N_{124}}\setminus\left\{ 0\right\} $.
Its classification is
\begin{equation}
\left(\mathbb{Z}_{N_{123}}\setminus\left\{ 0\right\} \right)\times\left(\mathbb{Z}_{N_{124}}\setminus\left\{ 0\right\} \right)\times\prod_{i=2}^{4}\mathbb{Z}_{N_{i}}\times\prod_{2\leq i<j\leq4}\left(\mathbb{Z}_{N_{ij}}\right)^{2}\times\left(\mathbb{Z}_{N_{234}}\right)^{4}.
\end{equation}
\item [{$A^{2}A^{3}C^{1}+A^{3}A^{4}C^{1}$}] The classification is $\left(\mathbb{Z}_{N_{123}}\setminus\left\{ 0\right\} \right)\times\left(\mathbb{Z}_{N_{134}}\setminus\left\{ 0\right\} \right)\times{\displaystyle \prod_{i=2}^{4}\mathbb{Z}_{N_{i}}}\times{\displaystyle \prod_{2\leq i<j\leq4}}\left(\mathbb{Z}_{N_{ij}}\right)^{2}\times\left(\mathbb{Z}_{N_{234}}\right)^{4}$.
\item [{$A^{2}A^{4}C^{1}+A^{3}A^{4}C^{1}$}] The classification is $\left(\mathbb{Z}_{N_{124}}\setminus\left\{ 0\right\} \right)\times\left(\mathbb{Z}_{N_{134}}\setminus\left\{ 0\right\} \right)\times{\displaystyle \prod_{i=2}^{4}}\mathbb{Z}_{N_{i}}\times{\displaystyle \prod_{2\leq i<j\leq4}}\left(\mathbb{Z}_{N_{ij}}\right)^{2}\times\left(\mathbb{Z}_{N_{234}}\right)^{4}$.
\item [{$A^{2}A^{3}C^{1}+A^{2}A^{4}C^{1}+A^{3}A^{4}C^{1}$}] The classification
is $\left(\mathbb{Z}_{N_{123}}\setminus\left\{ 0\right\} \right)\times\left(\mathbb{Z}_{N_{124}}\setminus\left\{ 0\right\} \right)\times\left(\mathbb{Z}_{N_{134}}\setminus\left\{ 0\right\} \right)\times{\displaystyle \prod_{i=2}^{4}\mathbb{Z}_{N_{i}}}\times{\displaystyle \prod_{2\leq i<j\leq4}}\left(\mathbb{Z}_{N_{ij}}\right)^{2}\times\left(\mathbb{Z}_{N_{234}}\right)^{4}.$
\item [{$A^{3}A^{1}C^{2}$}] The action is
\begin{align}
S= & \int\sum_{i=1}^{4}\frac{N_{i}}{2\pi}C^{i}dA^{1}+\frac{pN_{3}N_{1}N_{2}}{\left(2\pi\right)^{2}N_{123}}A^{3}A^{1}C^{2}+\left\langle  \sum_{i=1,3,4}A^{i}dA^{i}dA^{i},\right.\nonumber\\
 & \left.\sum_{\substack{1\leq i<j\leq4\\
i,j\neq2
}
}\left(A^{i}dA^{j}dA^{i}+A^{j}dA^{i}dA^{j}\right),A^{1}dA^{3}dA^{4},\sum_{i=1,3,4}A^{1}A^{3}A^{4}dA^{i}\right\rangle
\end{align}
with $p\in\mathbb{Z}_{N_{123}}\setminus\left\{ 0\right\} $. Its classification
is
\begin{equation}
\left(\mathbb{Z}_{N_{123}}\setminus\left\{ 0\right\} \right)\times\prod_{i=1,3,4}\mathbb{Z}_{N_{i}}\times\prod_{\substack{1\leq i<j\leq4\\
i,j\neq2
}
}\left(\mathbb{Z}_{N_{ij}}\right)^{2}\times\left(\mathbb{Z}_{N_{134}}\right)^{4}.
\end{equation}
\item [{$A^{4}A^{1}C^{2}$}] The classification is $\left(\mathbb{Z}_{N_{124}}\setminus\left\{ 0\right\} \right)\times{\displaystyle \prod_{i=1,3,4}}\mathbb{Z}_{N_{i}}\times{\displaystyle \prod_{\substack{1\leq i<j\leq4\\
i,j\neq2
}
}}\left(\mathbb{Z}_{N_{ij}}\right)^{2}\times\left(\mathbb{Z}_{N_{134}}\right)^{4}.$
\item [{$A^{3}A^{4}C^{2}$}] The classification is $\left(\mathbb{Z}_{N_{234}}\setminus\left\{ 0\right\} \right)\times{\displaystyle \prod_{i=1,3,4}}\mathbb{Z}_{N_{i}}\times{\displaystyle \prod_{\substack{1\leq i<j\leq4\\
i,j\neq2
}
}}\left(\mathbb{Z}_{N_{ij}}\right)^{2}\times\left(\mathbb{Z}_{N_{134}}\right)^{4}.$
\item [{$A^{3}A^{1}C^{2}+A^{4}A^{1}C^{2}$}] The classification is $\left(\mathbb{Z}_{N_{123}}\setminus\left\{ 0\right\} \right)\times\left(\mathbb{Z}_{N_{124}}\setminus\left\{ 0\right\} \right)\times{\displaystyle \prod_{i=1,3,4}}\mathbb{Z}_{N_{i}}\times{\displaystyle \prod_{\substack{1\leq i<j\leq4\\
i,j\neq2
}
}}\left(\mathbb{Z}_{N_{ij}}\right)^{2}\times\left(\mathbb{Z}_{N_{134}}\right)^{4}.$
\item [{$A^{3}A^{1}C^{2}+A^{3}A^{4}C^{2}$}] The classification is $\left(\mathbb{Z}_{N_{123}}\setminus\left\{ 0\right\} \right)\times\left(\mathbb{Z}_{N_{234}}\setminus\left\{ 0\right\} \right)\times{\displaystyle \prod_{i=1,3,4}}\mathbb{Z}_{N_{i}}\times{\displaystyle \prod_{\substack{1\leq i<j\leq4\\
i,j\neq2
}
}}\left(\mathbb{Z}_{N_{ij}}\right)^{2}\times\left(\mathbb{Z}_{N_{134}}\right)^{4}.$
\item [{$A^{3}A^{4}C^{2}+A^{4}A^{1}C^{2}$}] The classification is $\left(\mathbb{Z}_{N_{234}}\setminus\left\{ 0\right\} \right)\times\left(\mathbb{Z}_{N_{124}}\setminus\left\{ 0\right\} \right)\times{\displaystyle \prod_{i=1,3,4}}\mathbb{Z}_{N_{i}}\times{\displaystyle \prod_{\substack{1\leq i<j\leq4\\
i,j\neq2
}
}}\left(\mathbb{Z}_{N_{ij}}\right)^{2}\times\left(\mathbb{Z}_{N_{134}}\right)^{4}.$
\item [{$A^{3}A^{1}C^{2}+A^{4}A^{1}C^{2}+A^{3}A^{4}C^{2}$}] The classification
is $\left(\mathbb{Z}_{N_{123}}\setminus\left\{ 0\right\} \right)\times\left(\mathbb{Z}_{N_{124}}\setminus\left\{ 0\right\} \right)\times\left(\mathbb{Z}_{N_{234}}\setminus\left\{ 0\right\} \right)\times{\displaystyle \prod_{i=1,3,4}}\mathbb{Z}_{N_{i}}\times{\displaystyle \prod_{\substack{1\leq i<j\leq4\\
i,j\neq2
}
}}\left(\mathbb{Z}_{N_{ij}}\right)^{2}\times\left(\mathbb{Z}_{N_{134}}\right)^{4}.$
\item [{$A^{1}A^{2}C^{3}$}] The action is
\begin{align}
S= & \int\sum_{i=1}^{4}\frac{N_{i}}{2\pi}C^{i}dA^{1}+\frac{pN_{1}N_{2}N_{3}}{\left(2\pi\right)^{2}N_{123}}A^{1}A^{2}C^{3}+\left\langle  \sum_{i=1,2,4}A^{i}dA^{i}dA^{i},\right.\nonumber\\
 & \left.\sum_{\substack{1\leq i<j\leq4\\
i,j\neq3
}
}\left(A^{i}dA^{j}dA^{i}+A^{j}dA^{i}dA^{j}\right),A^{1}dA^{2}dA^{4},\sum_{i=1,2,4}A^{1}A^{2}A^{4}dA^{i}\right\rangle
\end{align}
with $p\in\mathbb{Z}_{N_{123}}\setminus\left\{ 0\right\} $. Its classification
is
\begin{equation}
\left(\mathbb{Z}_{N_{123}}\setminus\left\{ 0\right\} \right)\times\prod_{i=1,2,4}\mathbb{Z}_{N_{i}}\times\prod_{\substack{1\leq i<j\leq4\\
i,j\neq3
}
}\left(\mathbb{Z}_{N_{ij}}\right)^{2}\times\left(\mathbb{Z}_{N_{124}}\right)^{4}.
\end{equation}
\item [{$A^{1}A^{4}C^{3}$}] The classification is $\left(\mathbb{Z}_{N_{134}}\setminus\left\{ 0\right\} \right)\times{\displaystyle \prod_{i=1,2,4}}\mathbb{Z}_{N_{i}}\times{\displaystyle \prod_{\substack{1\leq i<j\leq4\\
i,j\neq3
}
}}\left(\mathbb{Z}_{N_{ij}}\right)^{2}\times\left(\mathbb{Z}_{N_{124}}\right)^{4}$.
\item [{$A^{2}A^{4}C^{3}$}] The classification is $\left(\mathbb{Z}_{N_{234}}\setminus\left\{ 0\right\} \right)\times{\displaystyle \prod_{i=1,2,4}}\mathbb{Z}_{N_{i}}\times{\displaystyle \prod_{\substack{1\leq i<j\leq4\\
i,j\neq3
}
}}\left(\mathbb{Z}_{N_{ij}}\right)^{2}\times\left(\mathbb{Z}_{N_{124}}\right)^{4}$.
\item [{$A^{1}A^{2}C^{3}+A^{1}A^{4}C^{3}$}] The classification is $\left(\mathbb{Z}_{N_{123}}\setminus\left\{ 0\right\} \right)\times\left(\mathbb{Z}_{N_{134}}\setminus\left\{ 0\right\} \right)\times{\displaystyle \prod_{i=1,2,4}}\mathbb{Z}_{N_{i}}\times{\displaystyle \prod_{\substack{1\leq i<j\leq4\\
i,j\neq3
}
}}\left(\mathbb{Z}_{N_{ij}}\right)^{2}\times\left(\mathbb{Z}_{N_{124}}\right)^{4}.$
\item [{$A^{1}A^{2}C^{3}+A^{2}A^{4}C^{3}$}] The classification is $\left(\mathbb{Z}_{N_{123}}\setminus\left\{ 0\right\} \right)\times\left(\mathbb{Z}_{N_{234}}\setminus\left\{ 0\right\} \right)\times{\displaystyle \prod_{i=1,2,4}}\mathbb{Z}_{N_{i}}\times{\displaystyle \prod_{\substack{1\leq i<j\leq4\\
i,j\neq3
}
}}\left(\mathbb{Z}_{N_{ij}}\right)^{2}\times\left(\mathbb{Z}_{N_{124}}\right)^{4}.$
\item [{$A^{2}A^{4}C^{3}+A^{1}A^{4}C^{3}$}] The classification is $\left(\mathbb{Z}_{N_{234}}\setminus\left\{ 0\right\} \right)\times\left(\mathbb{Z}_{N_{134}}\setminus\left\{ 0\right\} \right)\times{\displaystyle \prod_{i=1,2,4}}\mathbb{Z}_{N_{i}}\times{\displaystyle \prod_{\substack{1\leq i<j\leq4\\
i,j\neq3
}
}}\left(\mathbb{Z}_{N_{ij}}\right)^{2}\times\left(\mathbb{Z}_{N_{124}}\right)^{4}.$
\item [{$A^{1}A^{2}C^{3}+A^{1}A^{4}C^{3}+A^{2}A^{4}C^{3}$}] The classification
is $\left(\mathbb{Z}_{N_{123}}\setminus\left\{ 0\right\} \right)\times\left(\mathbb{Z}_{N_{134}}\setminus\left\{ 0\right\} \right)\times\left(\mathbb{Z}_{N_{234}}\setminus\left\{ 0\right\} \right)\times{\displaystyle \prod_{i=1,2,4}}\mathbb{Z}_{N_{i}}\times{\displaystyle \prod_{\substack{1\leq i<j\leq4\\
i,j\neq3
}
}}\left(\mathbb{Z}_{N_{ij}}\right)^{2}\times\left(\mathbb{Z}_{N_{124}}\right)^{4}.$
\item [{$A^{1}A^{2}C^{4}$}] The action is
\begin{align}
S= & \int\sum_{i=1}^{4}\frac{N_{i}}{2\pi}C^{i}dA^{1}+\frac{pN_{1}N_{2}N_{4}}{\left(2\pi\right)^{2}N_{124}}A^{1}A^{2}C^{4}+\left\langle  \sum_{i=1}^{3}A^{i}dA^{i}dA^{i},\right.\nonumber\\
 & \left.\sum_{\substack{1\leq i<j\leq3}
}\left(A^{i}dA^{j}dA^{i}+A^{j}dA^{i}dA^{j}\right),A^{1}dA^{2}dA^{3},\sum_{i=1}^{3}A^{1}A^{2}A^{3}dA^{i}\right\rangle
\end{align}
with $p\in\mathbb{Z}_{N_{124}}\setminus\left\{ 0\right\} $. Its classification
is
\begin{equation}
\left(\mathbb{Z}_{N_{124}}\setminus\left\{ 0\right\} \right)\times\prod_{i=1}^{3}\mathbb{Z}_{N_{i}}\times\prod_{1\leq i<j\leq3}\left(\mathbb{Z}_{N_{ij}}\right)^{2}\times\left(\mathbb{Z}_{N_{123}}\right)^{4}.
\end{equation}
\item [{$A^{1}A^{3}C^{4}$}] The classification is $\left(\mathbb{Z}_{N_{134}}\setminus\left\{ 0\right\} \right)\times{\displaystyle \prod_{i=1}^{3}}\mathbb{Z}_{N_{i}}\times{\displaystyle \prod_{1\leq i<j\leq3}}\left(\mathbb{Z}_{N_{ij}}\right)^{2}\times\left(\mathbb{Z}_{N_{123}}\right)^{4}.$
\item [{$A^{2}A^{3}C^{4}$}] The classification is $\left(\mathbb{Z}_{N_{234}}\setminus\left\{ 0\right\} \right)\times{\displaystyle \prod_{i=1}^{3}}\mathbb{Z}_{N_{i}}\times{\displaystyle \prod_{1\leq i<j\leq3}}\left(\mathbb{Z}_{N_{ij}}\right)^{2}\times\left(\mathbb{Z}_{N_{123}}\right)^{4}.$
\item [{$A^{1}A^{2}C^{4}+A^{1}A^{3}C^{4}$}] The classification is $\left(\mathbb{Z}_{N_{124}}\setminus\left\{ 0\right\} \right)\times\left(\mathbb{Z}_{N_{134}}\setminus\left\{ 0\right\} \right)\times{\displaystyle \prod_{i=1}^{3}}\mathbb{Z}_{N_{i}}\times{\displaystyle \prod_{1\leq i<j\leq3}}\left(\mathbb{Z}_{N_{ij}}\right)^{2}\times\left(\mathbb{Z}_{N_{123}}\right)^{4}.$
\item [{$A^{1}A^{2}C^{4}+A^{2}A^{3}C^{4}$}] The classification is $\left(\mathbb{Z}_{N_{124}}\setminus\left\{ 0\right\} \right)\times\left(\mathbb{Z}_{N_{234}}\setminus\left\{ 0\right\} \right)\times{\displaystyle \prod_{i=1}^{3}}\mathbb{Z}_{N_{i}}\times{\displaystyle \prod_{1\leq i<j\leq3}}\left(\mathbb{Z}_{N_{ij}}\right)^{2}\times\left(\mathbb{Z}_{N_{123}}\right)^{4}.$
\item [{$A^{2}A^{3}C^{4}+A^{1}A^{3}C^{4}$}] The classification is $\left(\mathbb{Z}_{N_{234}}\setminus\left\{ 0\right\} \right)\times\left(\mathbb{Z}_{N_{134}}\setminus\left\{ 0\right\} \right)\times{\displaystyle \prod_{i=1}^{3}}\mathbb{Z}_{N_{i}}\times{\displaystyle \prod_{1\leq i<j\leq3}}\left(\mathbb{Z}_{N_{ij}}\right)^{2}\times\left(\mathbb{Z}_{N_{123}}\right)^{4}.$
\item [{$A^{1}A^{2}C^{4}+A^{1}A^{3}C^{4}+A^{2}A^{3}C^{4}$}] The classification
is $\left(\mathbb{Z}_{N_{124}}\setminus\left\{ 0\right\} \right)\times\left(\mathbb{Z}_{N_{134}}\setminus\left\{ 0\right\} \right)\times\left(\mathbb{Z}_{N_{234}}\setminus\left\{ 0\right\} \right)\times{\displaystyle \prod_{i=1}^{3}}\mathbb{Z}_{N_{i}}\times{\displaystyle \prod_{1\leq i<j\leq3}}\left(\mathbb{Z}_{N_{ij}}\right)^{2}\times\left(\mathbb{Z}_{N_{123}}\right)^{4}.$
\item [{$A^{1}A^{2}C^{3}+A^{1}A^{2}C^{4}$}] The action is
\begin{align}
S & =\int\sum_{i=1}^{4}\frac{N_{i}}{2\pi}C^{i}dA^{1}+\frac{pN_{1}N_{2}N_{3}}{\left(2\pi\right)^{2}N_{123}}A^{1}A^{2}C^{3}+\frac{p N_{1}N_{2}N_{4}}{\left(2\pi\right)^{2}N_{124}}A^{1}A^{2}C^{4}+\nonumber \\
 & +\left\langle  \sum_{i=1}^{2}A^{i}dA^{i}dA^{i},A^{1}dA^{2}dA^{1},A^{2}dA^{1}dA^{2}\right\rangle
\end{align}
 with $p\in\mathbb{Z}_{N_{123}}\setminus\left\{ 0\right\} $ and $p \in\mathbb{Z}_{N_{124}}\setminus\left\{ 0\right\} $
. Its classification is
\begin{equation}
\left(\mathbb{Z}_{N_{123}}\setminus\left\{ 0\right\} \right)\times\left(\mathbb{Z}_{N_{124}}\setminus\left\{ 0\right\} \right)\times\prod_{i=1}^{2}\mathbb{Z}_{N_{i}}\times\left(\mathbb{Z}_{N_{12}}\right)^{2}.
\end{equation}
\item [{$A^{1}A^{3}C^{4}+A^{3}A^{1}C^{2}$}] The classification is $\left(\mathbb{Z}_{N_{134}}\setminus\left\{ 0\right\} \right)\times\left(\mathbb{Z}_{N_{123}}\setminus\left\{ 0\right\} \right)\times{\displaystyle \prod_{i=1,3}}\mathbb{Z}_{N_{i}}\times\left(\mathbb{Z}_{N_{13}}\right)^{2}.$
\item [{$A^{1}A^{4}C^{3}+A^{4}A^{1}C^{2}$}] The classification is $\left(\mathbb{Z}_{N_{134}}\setminus\left\{ 0\right\} \right)\times\left(\mathbb{Z}_{N_{124}}\setminus\left\{ 0\right\} \right)\times{\displaystyle \prod_{i=1,4}}\mathbb{Z}_{N_{i}}\times\left(\mathbb{Z}_{N_{14}}\right)^{2}.$
\item [{$A^{2}A^{3}C^{4}+A^{2}A^{3}C^{1}$}] The classification is $\left(\mathbb{Z}_{N_{234}}\setminus\left\{ 0\right\} \right)\times\left(\mathbb{Z}_{N_{123}}\setminus\left\{ 0\right\} \right)\times{\displaystyle \prod_{i=2,3}}\mathbb{Z}_{N_{i}}\times\left(\mathbb{Z}_{N_{23}}\right)^{2}.$
\item [{$A^{2}A^{4}C^{3}+A^{2}A^{4}C^{1}$}] The classification is $\left(\mathbb{Z}_{N_{234}}\setminus\left\{ 0\right\} \right)\times\left(\mathbb{Z}_{N_{124}}\setminus\left\{ 0\right\} \right)\times{\displaystyle \prod_{i=2,4}}\mathbb{Z}_{N_{i}}\times\left(\mathbb{Z}_{N_{24}}\right)^{2}.$
\item [{$A^{3}A^{4}C^{1}+A^{3}A^{4}C^{2}$}] The classification is $\left(\mathbb{Z}_{N_{134}}\setminus\left\{ 0\right\} \right)\times\left(\mathbb{Z}_{N_{234}}\setminus\left\{ 0\right\} \right)\times{\displaystyle \prod_{i=3,4}}\mathbb{Z}_{N_{i}}\times\left(\mathbb{Z}_{N_{34}}\right)^{2}.$
\end{description}

\section{Typical examples of $BF$ theories and classification when $G=\prod_{i=1}^{5}\mathbb{Z}_{N_i}$}\label{appendix_example_5Zn}
In this appendix, we give some representative examples of $BF$ theories and their classifications. It is tedious to list all $BF$ theories with twisted terms for $G=\prod_{i=1}^{5}\mathbb{Z}_{N_i}$ due to lots of combinations of compatible twisted terms. Nevertheless, the discussion in section~\ref{sec_tqft_classification} and appendix~\ref{appendix_4Zn_class_all_AAC} can be straightforward generalized to cases of gauge groups with arbitrary cyclic subgroups. Following the same line of thinking, we can figure out compatible twisted terms, write down the action, and find out the corresponding classification.

If $\alpha=\emptyset$, the action is
\begin{equation}
S=\int\sum_{i=1}^{5}\frac{N_{i}}{2\pi}\widetilde{B}^{i}dB^{i},
\end{equation}
whose classification is $\mathbb{Z}_{1}$.

If $\alpha=\left\{ N_1\right\} $,
the action is
\begin{align}
S= & \int\frac{N_{1}}{2\pi}C^{1}dA^{1}+\sum_{i=2}^{5}\frac{N_{i}}{2\pi}\widetilde{B}^{i}dB^{i}\nonumber\\
 & +\left\langle  A^{1}dA^{1}dA^{1},\sum_{i=2}^{5}B^{i}B^{i}A^{1},\sum_{i=2}^{5}B^{i}A^{1}dA^{1},\sum_{2\leq i<j\leq5}B^{i}B^{j}A^{1}\right\rangle  ,
\end{align}
whose classification of topological gauge theories is
$\mathbb{Z}_{N_{1}}\times{\displaystyle\prod_{i=2}^{5}\left(\mathbb{Z}_{N_{1i}}\right)^{2}}\times{\displaystyle\prod_{2\leq i<j\leq5}\mathbb{Z}_{N_{1ij}}}$.

If $\alpha =\left\{ N_1,N_2\right\}$, the action is
\begin{align}
S= & \int\sum_{i=1}^{2}\frac{N_{i}}{2\pi}C^{i}dA^{i}+\sum_{i=3}^{5}\frac{N_{i}}{2\pi}\widetilde{B}^{i}dB^{i}+\left\langle  \sum_{i=1}^{2}A^{i}dA^{i}dA^{i},A^{1}dA^{2}dA^{1},A^{2}dA^{1}dA^{2},\right.\nonumber\\
 & \left.\sum_{3\leq i<j\leq5} \sum_{k=1}^{2}B^{i}B^{j}A^{k},\sum_{i=3}^{5}\sum_{j=1}^{2}\left(B^{i}B^{i}A^{j}+B^{i}A^{j}dA^{j}\right),\sum_{i=3}^{5}\left(A^{1}A^{2}dB^{i}+B^{i}A^{2}dA^{1}\right)\right\rangle  ,
\end{align}
whose classification is
$\mathbb{Z}_{N_{1}}\times\mathbb{Z}_{N_{2}}\times\left(\mathbb{Z}_{N_{12}}\right)^{2}\times{\displaystyle\prod_{3\leq i<j\leq5}\left(\mathbb{Z}_{N_{1ij}}\times\mathbb{Z}_{N_{2ij}}\right)}\times{\displaystyle\prod_{i=3}^{5}\left(\mathbb{Z}_{N_{1i}}\times\mathbb{Z}_{N_{2i}}\right)^{2}}\times{\displaystyle\prod_{i=3}^{5}\left(\mathbb{Z}_{N_{12i}}\right)^{2}}$.

If $\alpha =\left\{ N_1,N_2,N_3\right\}$, we should be aware of keeping all twisted terms compatible with each other. Below we show two examples.
\begin{enumerate}
\item If we do not include $AAC$ terms in twisted terms, the action is
\begin{align}
S= & \int\sum_{i=1}^{3}\frac{N_{i}}{2\pi}C^{i}dA^{i}+\sum_{i=4}^{5}\frac{N_{i}}{2\pi}\widetilde{B}^{i}dB^{i}+\left\langle  \sum_{i=1}^{3}A^{i}dA^{i}dA^{i},\sum_{1\leq i<j\leq3}\left(A^{i}dA^{j}dA^{i}+A^{j}dA^{i}dA^{j}\right),\right.\nonumber\\
 & A^{1}dA^{2}dA^{3},\sum_{i=1}^{3}A^{1}A^{2}A^{3}dA^{i},\sum_{i=1}^{3}B^{4}B^{5}A^{i},\sum_{i=1}^{3}\sum_{j=4}^{5}\left(B^{j}B^{j}A^{i}+B^{j}A^{i}dA^{i}\right),\nonumber\\
 & \left.\sum_{1\leq i<j\leq3}\sum_{k=4}^{5}\left(A^{i}A^{j}dB^{k}+B^{k}A^{j}dA^{i}\right),\sum_{i=4}^{5}A^{1}A^{2}A^{3}B^{i}\right\rangle  ,
\end{align}
whose classification is $\displaystyle{\prod_{i=1}^{3}}\mathbb{Z}_{N_{i}}\times{\displaystyle \prod_{1\leq i<j\leq3}\left(\mathbb{Z}_{N_{ij}}\right)^{2}}\times\left(\mathbb{Z}_{N_{123}}\right)^{4}\times\displaystyle{\prod_{i=1}^{3}}\mathbb{Z}_{N_{i45}}\times\displaystyle{\prod_{i=1}^{3}}\left[\left(\mathbb{Z}_{N_{i4}}\right)^{2}\times\left(\mathbb{Z}_{N_{i5}}\right)^{2}\right]\times{\displaystyle \prod_{1\leq i<j\leq3}\left[\left(\mathbb{Z}_{N_{ij4}}\right)^{2}\times\left(\mathbb{Z}_{N_{ij5}}\right)^{2}\right]\times\prod_{i=4}^{5}\mathbb{Z}_{N_{123i}}}$.
\item If we add an $AAC$ term, e.g.,  $A^{1}A^{2}C^{3}$, to twisted terms, the action is
\begin{align}
S= & \int\sum_{i=1}^{3}\frac{N_{i}}{2\pi}C^{i}dA^{i}+\sum_{i=4}^{5}\frac{N_{i}}{2\pi}\widetilde{B}^{i}dB^{i}+\frac{pN_{1}N_{2}N_{3}}{\left(2\pi\right)^{2}N_{123}}A^{1}A^{2}C^{3}+\left\langle  \sum_{i=1}^{2}A^{i}dA^{i}dA^{i},A^{1}dA^{2}dA^{1},\right.\nonumber\\
 & \left.A^{2}dA^{1}dA^{2},\sum_{i=1}^{2}B^{4}B^{5}A^{i},\sum_{i=1}^{2}\sum_{j=4}^{5}\left(B^{j}B^{j}A^{i}+B^{j}A^{i}dA^{i}\right),\sum_{i=4}^{5}\left(A^{1}A^{2}dB^{i}+B^{i}A^{1}dA^{2}\right)\right\rangle
\end{align}
with $p\in\mathbb{Z}_{N_{123}}\setminus\left\{ 0\right\} $. Its classification
is $\left(\mathbb{Z}_{N_{123}}\setminus\left\{ 0\right\} \right)\times\prod_{i=1}^{2}\mathbb{Z}_{N_{i}}\times\left(\mathbb{Z}_{N_{12}}\right)^{2}\times\prod_{i=1}^{2}\mathbb{Z}_{N_{i45}}\times\prod_{i=1}^{2}\left( \mathbb{Z}_{N_{i4}} \times\mathbb{Z}_{N_{i5}}\right)^{2}\times\prod_{i=4}^{5}\left(\mathbb{Z}_{N_{12i}}\right)^{2}$.
\end{enumerate}
If $\alpha =\left\{ N_1,N_2,N_3,N_4\right\} $, we give an example for each situation.
\begin{enumerate}
\item If no $AAC$ terms appear in this mixed $BF$ theory, the action is
\begin{align}
S= & \int\sum_{i=1}^{4}\frac{N_{i}}{2\pi}C^{i}dA^{i}+\frac{N_{5}}{2\pi}\widetilde{B}^{5}dB^{5}+\bigg\langle\sum_{i=1}^{4}A^{i}dA^{i}dA^{i},\sum_{1\leq i<j\leq4}\left(A^{i}dA^{j}dA^{i}+A^{j}dA^{i}dA^{j}\right),\nonumber\\
 & \sum_{1\leq i<j<k\leq4}A^{i}dA^{j}dA^{k},\sum_{1\leq i<j<k\leq4}\left(\sum_{s=i,j,k}A^{i}A^{j}A^{k}dA^{s}\right),A^{1}A^{2}A^{3}dA^{4},A^{3}A^{2}A^{4}dA^{1},\nonumber\\
 & A^{1}A^{3}A^{4}dA^{2},\sum_{i=1}^{4}\left(B^{5}B^{5}A^{i}+B^{5}A^{i}dA^{i}\right),\sum_{1\leq i<j\leq4}\left(A^{i}A^{j}dB^{5}+B^{5}A^{j}dA^{i}\right),\nonumber\\
 & \sum_{1\leq i<j<k\leq4}A^{i}A^{j}A^{k}B^{5}\bigg\rangle,
\end{align}
whose classification is ${\displaystyle \prod_{i=1}^{4}\mathbb{Z}_{N_{i}}}\times{\displaystyle \prod_{1\leq i<j\leq4}\left(\mathbb{Z}_{N_{ij}}\right)^{2}}\times{\displaystyle \prod_{1\leq i<j<k\leq4}\left(\mathbb{Z}_{N_{ijk}}\right)^{4}}\times\left(\mathbb{Z}_{N_{1234}}\right)^{3}\times{\displaystyle \prod_{i=1}^{4}\left(\mathbb{Z}_{N_{i5}}\right)^{2}}\times{\displaystyle \prod_{1\leq i<j\leq4}\left(\mathbb{Z}_{N_{ij5}}\right)^{2}}\times{\displaystyle \prod_{1\leq i<j<k\leq4}\mathbb{Z}_{N_{ijk5}}}$.
\item If twisted terms include $A^{1}A^{2}C^{4}+A^{1}A^{3}C^{4}+A^{2}A^{3}C^{4}$ terms, the action is
\begin{align}
S= & \int\sum_{i=1}^{4}\frac{N_{i}}{2\pi}C^{i}dA^{i}+\frac{N_{5}}{2\pi}\widetilde{B}^{5}dB^{5}\nonumber\\
 & +\frac{pN_{1}N_{2}N_{4}}{\left(2\pi\right)^{2}N_{124}}A^{1}A^{2}C^{4}+\frac{p' N_{1}N_{3}N_{4}}{\left(2\pi\right)^{2}N_{134}}A^{1}A^{3}C^{4}+\frac{p''N_{2}N_{3}N_{4}}{\left(2\pi\right)^{2}N_{234}}A^{2}A^{3}C^{4}\nonumber\\
 & +\left\langle  \sum_{i=1}^{3}A^{i}dA^{i}dA^{i},\sum_{1\leq i<j\leq3}\left(A^{i}dA^{j}dA^{i}+A^{j}dA^{i}dA^{j}\right),A^{1}dA^{2}dA^{3},\sum_{i=1}^{3}A^{1}A^{2}A^{3}dA^{i},\right.\nonumber\\
 & \left.\sum_{i=1}^{3}\left(B^{5}B^{5}A^{i}+B^{5}A^{i}dA^{i}\right),\sum_{1\leq i<j\leq3}\left(A^{i}A^{j}dB^{5}+B^{5}A^{j}dA^{i}\right),A^{1}A^{2}A^{3}B^{5}\right\rangle
\end{align}
with $p\in\mathbb{Z}_{N_{124}}\setminus\left\{ 0\right\} $, $p' \in\mathbb{Z}_{N_{134}}\setminus\left\{ 0\right\} $, and
$p''\in\mathbb{Z}_{N_{234}}\setminus\left\{ 0\right\} $. Its classification
is ${\displaystyle \prod_{1\leq i<j\leq3}}\left(\mathbb{Z}_{N_{ij4}}\setminus\left\{ 0\right\} \right)\times{\displaystyle \prod_{i=1}^{3}\mathbb{Z}_{N_{i}}}\times{\displaystyle \prod_{1\leq i<j\leq3}\left(\mathbb{Z}_{N_{ij}}\right)^{2}}\times\left(\mathbb{Z}_{N_{123}}\right)^{4}\times{\displaystyle \prod_{i=1}^{3}\left(\mathbb{Z}_{N_{i5}}\right)^{2}}\times{\displaystyle \prod_{1\leq i<j\leq3}\left(\mathbb{Z}_{N_{ij5}}\right)^{2}}\times\mathbb{Z}_{N_{1235}}$.

\item If $A^{1}A^{2}C^{3}+A^{1}A^{2}C^{4}$ terms are added to twisted terms at the cost of excluding some other terms, the action is
\begin{align}
S= & \int\sum_{i=1}^{4}\frac{N_{i}}{2\pi}C^{i}dA^{i}+\frac{N_{5}}{2\pi}\widetilde{B}^{5}dB^{5}+\frac{pN_{1}N_{2}N_{3}}{\left(2\pi\right)^{2}N_{123}}A^{1}A^{2}C^{3}+\frac{p' N_{1}N_{2}N_{4}}{\left(2\pi\right)^{2}N_{124}}A^{1}A^{2}C^{4}\nonumber\\
 & +\left\langle \sum_{i=1}^{2}A^{i}dA^{i}dA^{i},A^{1}dA^{2}dA^{1},A^{2}dA^{1}dA^{2},\right.\nonumber\\
 & \left.\sum_{i=1}^{2}\left(B^{5}B^{5}A^{i}+B^{5}A^{i}dA^{i}\right),A^{1}A^{2}dB^{5},B^{5}A^{2}dA^{1}\right\rangle
\end{align}
with $p\in\mathbb{Z}_{N_{123}}\setminus\left\{ 0\right\} $ and $p' \in\mathbb{Z}_{N_{124}}\setminus\left\{ 0\right\} $.
Its classification is $\left(\mathbb{Z}_{N_{123}}\setminus\left\{ 0\right\} \right)\times\left(\mathbb{Z}_{N_{124}}\setminus\left\{ 0\right\} \right)\times\prod_{i=1}^{2}\mathbb{Z}_{N_{i}}\times\left(\mathbb{Z}_{N_{12}}\right)^{2}\times\prod_{i=1}^{2}\left(\mathbb{Z}_{N_{i5}}\right)^{2}\times\left(\mathbb{Z}_{N_{125}}\right)^{2}$.
\end{enumerate}

If $\alpha=\left\{ N_1,N_2,N_3,N_4,N_5\right\} $, i.e., type-I $BF$ theory, similar to previous discussion, different combinations of twisted terms result in different actions and classifications. Some examples are listed as follow.
\begin{enumerate}
\item If there is no  $AAC$ terms in the action, i.e.,
\begin{align}
S= & \int\sum_{i=1}^{5}\frac{N_{i}}{2\pi}C^{i}dA^{i}+\left\langle \sum_{i=1}^{5}A^{i}dA^{i}dA^{i},\sum_{1\leq i<j\leq5}\left(A^{i}dA^{j}dA^{i}+A^{j}dA^{i}dA^{j}\right),\right.\nonumber\\
 & \sum_{1\leq i<j<k\leq5}\left(A^{i}dA^{j}dA^{k}+\sum_{l=i,j,k}A^{i}A^{j}A^{k}dA^{l}\right),\nonumber\\
 & \left.\sum_{1\leq i<j<k<l\leq5}\left(A^{i}A^{j}A^{k}dA^{l}+A^{k}A^{j}A^{l}dA^{i}+A^{i}A^{k}A^{l}dA^{j}\right),A^{1}A^{2}A^{3}A^{4}A^{5}\right\rangle ,
\end{align}
whose classification is given by $H^{5}\left(\prod_{i=1}^{5}\mathbb{Z}_{N_i},\mathbb{R}/\mathbb{Z}\right)$, consistent with the Dijkgraaf-Witten cohomology classification:
\begin{equation}
\prod_{i=1}^{5}\mathbb{Z}_{N_{i}}\times\prod_{1\leq i<j\leq5}\left(\mathbb{Z}_{N_{ij}}\right)^{2}\times\prod_{1\leq i<j<k\leq5}\left(\mathbb{Z}_{N_{ijk}}\right)^{4}\times\prod_{1\leq i<j<k<l\leq5}\left(\mathbb{Z}_{N_{ijkl}}\right)^{3}\times\mathbb{Z}_{N_{12345}}.
\end{equation}
\item If we put $\sum_{1\leq i<j\leq4}A^{i}A^{j}C^{5}$
terms in twisted terms, the action is
\begin{align}
S= & \int\sum_{i=1}^{5}\frac{N_{i}}{2\pi}C^{i}dA^{i}+\sum_{1\leq i<j\leq4}\frac{p_{ij5}N_{i}N_{j}N_{5}}{\left(2\pi\right)^{2}N_{ij5}}A^{i}A^{j}C^{5}+\left\langle \sum_{i=1}^{4}A^{i}dA^{i}dA^{i},\right.\nonumber\\
 & \sum_{1\leq i<j\leq4}\left(A^{i}dA^{j}dA^{i}+A^{j}dA^{i}dA^{j}\right),\sum_{1\leq i<j<k\leq4}A^{i}dA^{j}dA^{k},\nonumber\\
 & \left.\sum_{1\leq i<j<k\leq4}\left(\sum_{l=i,j,k}A^{i}A^{j}A^{k}dA^{l}\right),A^{1}A^{2}A^{3}dA^{4},A^{3}A^{2}A^{4}dA^{1},A^{1}A^{3}A^{4}dA^{2}\right\rangle
\end{align}
where $p_{ij5}\in\mathbb{Z}_{N_{ij5}}\setminus\left\{ 0\right\} $, $1\leq i<j\leq4$.
Its classification is ${\displaystyle \prod_{1\leq i<j\leq4}}\left(\mathbb{Z}_{N_{ij5}}\setminus\left\{ 0\right\} \right)\times{\displaystyle \prod_{i=1}^{4}}\mathbb{Z}_{N_{i}}\times{\displaystyle \prod_{1\leq i<j\leq4}}\left(\mathbb{Z}_{N_{ij}}\right)^{2}\times{\displaystyle \prod_{1\leq i<j<k\leq4}}\left(\mathbb{Z}_{N_{ijk}}\right)^{4}\times\left(\mathbb{Z}_{N_{1234}}\right)^{3}$.
\item If $\sum_{1\leq i<j\leq3}\left(A^{i}A^{j}C^{4}+A^{i}A^{j}C^{5}\right)$ terms are considered, the action is
\begin{align}
S= & \int\sum_{i=1}^{5}\frac{N_{i}}{2\pi}C^{i}dA^{i}+\sum_{1\leq i<j\leq3}\left(\frac{p_{ij4}N_{i}N_{j}N_{4}}{\left(2\pi\right)^{2}N_{ij4}}A^{i}A^{j}C^{4}+\frac{p_{ij5}N_{i}N_{j}N_{5}}{\left(2\pi\right)^{2}N_{ij5}}A^{i}A^{j}C^{5}\right)\nonumber\\
 & +\left\langle \sum_{i=1}^{3}A^{i}dA^{i}dA^{i},\sum_{1\leq i<j\leq3}\left(A^{i}dA^{j}dA^{i}+A^{j}dA^{i}dA^{j}\right),A^{1}dA^{2}dA^{3},\sum_{i=1,2,3}A^{1}A^{2}A^{3}dA^{i}\right\rangle
\end{align}
where $p_{ij4}\in\mathbb{Z}_{N_{ij4}}\setminus\left\{ 0\right\} $
and $p_{ij5}\in\mathbb{Z}_{N_{ij5}}\setminus\left\{ 0\right\} $, $1\leq i<j\leq 3$.
Its classification is ${\displaystyle \prod_{1\leq i<j\leq3}}\left(\mathbb{Z}_{N_{ij4}}\setminus\left\{ 0\right\} \right)\times{\displaystyle \prod_{1\leq i<j\leq3}}\left(\mathbb{Z}_{N_{ij4}}\setminus\left\{ 0\right\} \right)\times{\displaystyle \prod_{i=1}^{3}}\mathbb{Z}_{N_{i}}\times{\displaystyle \prod_{1\leq i<j\leq3}}\left(\mathbb{Z}_{N_{ij}}\right)^{2}\times\left(\mathbb{Z}_{N_{123}}\right)^{4}$.
\item If $A^{1}A^{2}C^{3}+A^{1}A^{2}C^{4}+A^{1}A^{2}C^{5}$ terms are collected in twisted terms, the action is
\begin{equation}
S=\int\sum_{i=1}^{5}\frac{N_{i}}{2\pi}C^{i}dA^{i}+\sum_{i=3}^{5}\frac{p_{12i}N_{1}N_{2}N_{i}}{\left(2\pi\right)^{2}N_{12i}}A^{1}A^{2}C^{i}+\left\langle \sum_{i=1}^{2}A^{i}dA^{i}dA^{i},A^{1}dA^{2}dA^{1},A^{2}dA^{1}dA^{2}\right\rangle
\end{equation}
where $p_{12i}\in\mathbb{Z}_{N_{12i}}\setminus\left\{ 0\right\} $, $3\leq i\leq 5$.
Its classification is ${\displaystyle \prod_{i=1}^{3}}\left(\mathbb{Z}_{N_{12i}}\setminus\left\{ 0\right\} \right)\times{\displaystyle \prod_{i=1}^{2}}\mathbb{Z}_{N_{i}}\times\left(\mathbb{Z}_{N_{12}}\right)^{2}$.
\end{enumerate}

\bibliographystyle{JHEP}

\begin{thebibliography}{10}

\bibitem{Wen2019}
X.-G.~Wen, \emph{{Choreographed entanglement dances: Topological states of
  quantum matter}},
  \href{https://doi.org/10.1126/science.aal3099}{\emph{Science} (2019) }.

\bibitem{wen_stacking}
X.-G.~Wen, \emph{A theory of 2+1d bosonic topological orders},
  \href{https://doi.org/10.1093/nsr/nwv077}{\emph{Natl. Sci. Rev.} (2015) }
  [\href{https://arxiv.org/abs/arXiv:1506.05768}{{\ttfamily
  arXiv:1506.05768}}].

\bibitem{string1.5}
M.~Levin and X.-G.~Wen, \emph{\textit{Colloquium} : Photons and electrons as
  emergent phenomena},
  \href{https://doi.org/10.1103/RevModPhys.77.871}{\emph{Rev. Mod. Phys.}
  {\bfseries 77} (2005) 871}.

\bibitem{wenZootopoRMP}
X.-G.~Wen, \emph{Colloquium: Zoo of quantum-topological phases of matter},
  \href{https://doi.org/10.1103/RevModPhys.89.041004}{\emph{Rev. Mod. Phys.}
  {\bfseries 89} (2017) 041004}.

\bibitem{hartnoll2021quantum}
S.~Hartnoll, S.~Sachdev, T.~Takayanagi, X.~Chen, E.~Silverstein and J.~Sonner,
  \emph{Quantum connections}, {\emph{Nature Reviews Physics} {\bfseries 3}
  (2021) 391}.

\bibitem{KitaevTopoEE}
A.~Kitaev and J.~Preskill, \emph{Topological entanglement entropy},
  \href{https://doi.org/10.1103/PhysRevLett.96.110404}{\emph{Phys. Rev. Lett.}
  {\bfseries 96} (2006) 110404}.

\bibitem{Levin_Wen_TEE}
M.~Levin and X.-G.~Wen, \emph{Detecting topological order in a ground state
  wave function},
  \href{https://doi.org/10.1103/PhysRevLett.96.110405}{\emph{Phys. Rev. Lett.}
  {\bfseries 96} (2006) 110405}.

\bibitem{kong2020mathematical}
L.~Kong and H.~Zheng, \emph{A mathematical theory of gapless edges of 2d
  topological orders. part i}, {\emph{Journal of High Energy Physics}
  {\bfseries 2020} (2020) 1}.

\bibitem{kong2021mathematical}
L.~Kong and H.~Zheng, \emph{A mathematical theory of gapless edges of 2d
  topological orders. part ii}, {\emph{Nuclear Physics B} {\bfseries 966}
  (2021) 115384}.

\bibitem{sarma_08_TQC}
C.~Nayak, S.H.~Simon, A.~Stern, M.~Freedman and S.~Das~Sarma, \emph{Non-abelian
  anyons and topological quantum computation},
  \href{https://doi.org/10.1103/RevModPhys.80.1083}{\emph{Rev. Mod. Phys.}
  {\bfseries 80} (2008) 1083}.

\bibitem{Witten1989}
E.~Witten, \emph{Quantum field theory and the jones polynomial},
  \href{https://doi.org/10.1007/BF01217730}{\emph{Commun. Math. Phys.}
  {\bfseries 121} (1989) 351}.

\bibitem{wen_edge1995}
X.-G.~Wen, \emph{Topological orders and edge excitations in fractional quantum
  hall states}, \href{https://doi.org/10.1080/00018739500101566}{\emph{Advances
  in Physics} {\bfseries 44} (1995) 405}
  [\href{https://arxiv.org/abs/https://doi.org/10.1080/00018739500101566}{{\ttfamily
  https://doi.org/10.1080/00018739500101566}}].

\bibitem{Wen_grav_prd2013}
X.-G.~Wen, \emph{Classifying gauge anomalies through symmetry-protected trivial
  orders and classifying gravitational anomalies through topological orders},
  \href{https://doi.org/10.1103/PhysRevD.88.045013}{\emph{Phys. Rev. D}
  {\bfseries 88} (2013) 045013}.

\bibitem{string8}
L.~Kong and X.-G.~Wen, \emph{Braided fusion categories, gravitational
  anomalies, and the mathematical framework for topological orders in any
  dimensions},  2014.

\bibitem{string1}
M.A.~Levin and X.-G.~Wen, \emph{String-net condensation: A physical mechanism
  for topological phases},
  \href{https://doi.org/10.1103/PhysRevB.71.045110}{\emph{Phys. Rev. B}
  {\bfseries 71} (2005) 045110}.

\bibitem{wu84}
Y.-S.~Wu, \emph{General theory for quantum statistics in two dimensions},
  \href{https://doi.org/10.1103/PhysRevLett.52.2103}{\emph{Phys. Rev. Lett.}
  {\bfseries 52} (1984) 2103}.

\bibitem{CGW2010LRE}
X.~Chen, Z.-C.~Gu and X.-G.~Wen, \emph{Local unitary transformation, long-range
  quantum entanglement, wave function renormalization, and topological order},
  \href{https://doi.org/10.1103/PhysRevB.82.155138}{\emph{Phys. Rev. B}
  {\bfseries 82} (2010) 155138}.

\bibitem{Kitaev2009}
A.~Kitaev and C.~Laumann, \emph{{Topological phases and quantum computation}},
  \href{https://arxiv.org/abs/0904.2771}{{\ttfamily 0904.2771}}.

\bibitem{Kitaev2006}
A.~Kitaev, \emph{Anyons in an exactly solved model and beyond},
  \href{https://doi.org/10.1016/j.aop.2005.10.005}{\emph{Ann. Phys.} {\bfseries
  321} (2006) 2}.

\bibitem{2003AnPhy.303....2K}
A.Y.~{Kitaev}, \emph{{Fault-tolerant quantum computation by anyons}},
  \href{https://doi.org/10.1016/S0003-4916(02)00018-0}{\emph{Annals of Physics}
  {\bfseries 303} (2003) 2}
  [\href{https://arxiv.org/abs/quant-ph/9707021}{{\ttfamily
  quant-ph/9707021}}].

\bibitem{lantian3dto1}
T.~Lan, L.~Kong and X.-G.~Wen, \emph{Classification of
  $\mathbf{(}3+1\mathbf{)}\mathrm{D}$ bosonic topological orders: The case when
  pointlike excitations are all bosons},
  \href{https://doi.org/10.1103/PhysRevX.8.021074}{\emph{Phys. Rev. X}
  {\bfseries 8} (2018) 021074}.

\bibitem{lantian3dto2}
T.~Lan and X.-G.~Wen, \emph{Classification of $3+1\mathrm{D}$ bosonic
  topological orders (ii): The case when some pointlike excitations are
  fermions}, \href{https://doi.org/10.1103/PhysRevX.9.021005}{\emph{Phys. Rev.
  X} {\bfseries 9} (2019) 021005}.

\bibitem{yp18prl}
A.P.O.~Chan, P.~Ye and S.~Ryu, \emph{Braiding with borromean rings in
  ($3+1$)-dimensional spacetime},
  \href{https://doi.org/10.1103/PhysRevLett.121.061601}{\emph{Phys. Rev. Lett.}
  {\bfseries 121} (2018) 061601}.

\bibitem{ypdw}
X.~Wen, H.~He, A.~Tiwari, Y.~Zheng and P.~Ye, \emph{Entanglement entropy for
  (3+1)-dimensional topological order with excitations},
  \href{https://doi.org/10.1103/PhysRevB.97.085147}{\emph{Phys. Rev. B}
  {\bfseries 97} (2018) 085147}.

\bibitem{wang_levin1}
C.~Wang and M.~Levin, \emph{Braiding statistics of loop excitations in three
  dimensions},
  \href{https://doi.org/10.1103/PhysRevLett.113.080403}{\emph{Phys. Rev. Lett.}
  {\bfseries 113} (2014) 080403}.

\bibitem{bti2}
P.~Ye and Z.-C.~Gu, \emph{Vortex-line condensation in three dimensions: A
  physical mechanism for bosonic topological insulators},
  \href{https://doi.org/10.1103/PhysRevX.5.021029}{\emph{Phys. Rev. X}
  {\bfseries 5} (2015) 021029}.

\bibitem{jian_qi_14}
C.-M.~Jian and X.-L.~Qi, \emph{Layer construction of 3d topological states and
  string braiding statistics},
  \href{https://doi.org/10.1103/PhysRevX.4.041043}{\emph{Phys. Rev. X}
  {\bfseries 4} (2014) 041043}.

\bibitem{string5}
S.~Jiang, A.~Mesaros and Y.~Ran, \emph{Generalized modular transformations in
  $(3+1)\mathrm{D}$ topologically ordered phases and triple linking invariant
  of loop braiding},
  \href{https://doi.org/10.1103/PhysRevX.4.031048}{\emph{Phys. Rev. X}
  {\bfseries 4} (2014) 031048}.

\bibitem{PhysRevX.6.021015}
C.~Wang, C.-H.~Lin and M.~Levin, \emph{Bulk-boundary correspondence for
  three-dimensional symmetry-protected topological phases},
  \href{https://doi.org/10.1103/PhysRevX.6.021015}{\emph{Phys. Rev. X}
  {\bfseries 6} (2016) 021015}.

\bibitem{string6}
Y.~Wan, J.C.~Wang and H.~He, \emph{Twisted gauge theory model of topological
  phases in three dimensions},
  \href{https://doi.org/10.1103/PhysRevB.92.045101}{\emph{Phys. Rev. B}
  {\bfseries 92} (2015) 045101}.

\bibitem{ye16a}
P.~Ye, T.L.~Hughes, J.~Maciejko and E.~Fradkin, \emph{Composite particle theory
  of three-dimensional gapped fermionic phases: Fractional topological
  insulators and charge-loop excitation symmetry},
  \href{https://doi.org/10.1103/PhysRevB.94.115104}{\emph{Phys. Rev. B}
  {\bfseries 94} (2016) 115104}.

\bibitem{YeGu2015}
P.~Ye and Z.-C.~Gu, \emph{Topological quantum field theory of three-dimensional
  bosonic abelian-symmetry-protected topological phases},
  \href{https://doi.org/10.1103/PhysRevB.93.205157}{\emph{Phys. Rev. B}
  {\bfseries 93} (2016) 205157}.

\bibitem{corbodism3}
A.~{Kapustin} and R.~{Thorngren}, \emph{{Anomalies of discrete symmetries in
  various dimensions and group cohomology}}, {\emph{ArXiv e-prints} (2014) }
  [\href{https://arxiv.org/abs/1404.3230}{{\ttfamily 1404.3230}}].

\bibitem{YW13a}
P.~Ye and X.-G.~Wen, \emph{Constructing symmetric topological phases of bosons
  in three dimensions via fermionic projective construction and dyon
  condensation}, \href{https://doi.org/10.1103/PhysRevB.89.045127}{\emph{Phys.
  Rev. B} {\bfseries 89} (2014) 045127}.

\bibitem{ye16_set}
S.-Q.~Ning, Z.-X.~Liu and P.~Ye, \emph{Symmetry enrichment in three-dimensional
  topological phases},
  \href{https://doi.org/10.1103/PhysRevB.94.245120}{\emph{Phys. Rev. B}
  {\bfseries 94} (2016) 245120}.

\bibitem{2018arXiv180101638N}
S.-Q.~{Ning}, Z.-X.~{Liu} and P.~{Ye}, \emph{{Topological gauge theory,
  symmetry fractionalization, and classification of symmetry-enriched
  topological phases in three dimensions}}, {\emph{arXiv e-prints} (2018) }
  [\href{https://arxiv.org/abs/1801.01638}{{\ttfamily 1801.01638}}].

\bibitem{2016arXiv161008645Y}
P.~Ye, \emph{Three-dimensional anomalous twisted gauge theories with global
  symmetry: Implications for quantum spin liquids},
  \href{https://doi.org/10.1103/PhysRevB.97.125127}{\emph{Phys. Rev. B}
  {\bfseries 97} (2018) 125127}.

\bibitem{string4}
J.C.~Wang and X.-G.~Wen, \emph{Non-abelian string and particle braiding in
  topological order: Modular $\mathrm{SL}(3,\mathbb{Z})$ representation and
  $(3+1)$-dimensional twisted gauge theory},
  \href{https://doi.org/10.1103/PhysRevB.91.035134}{\emph{Phys. Rev. B}
  {\bfseries 91} (2015) 035134}.

\bibitem{PhysRevLett.114.031601}
J.C.~Wang, Z.-C.~Gu and X.-G.~Wen, \emph{Field-theory representation of
  gauge-gravity symmetry-protected topological invariants, group cohomology,
  and beyond},
  \href{https://doi.org/10.1103/PhysRevLett.114.031601}{\emph{Phys. Rev. Lett.}
  {\bfseries 114} (2015) 031601}.

\bibitem{3loop_ryu}
X.~Chen, A.~Tiwari and S.~Ryu, \emph{Bulk-boundary correspondence in
  (3+1)-dimensional topological phases},
  \href{https://doi.org/10.1103/PhysRevB.94.045113}{\emph{Phys. Rev. B}
  {\bfseries 94} (2016) 045113}.

\bibitem{string10}
J.~{Wang}, X.-G.~{Wen} and S.-T.~{Yau}, \emph{{Quantum Statistics and Spacetime
  Surgery}}, {\emph{ArXiv e-prints} (2016) }
  [\href{https://arxiv.org/abs/1602.05951}{{\ttfamily 1602.05951}}].

\bibitem{2016arXiv161209298P}
P.~Putrov, J.~Wang and S.-T.~Yau, \emph{Braiding statistics and link invariants
  of bosonic/fermionic topological quantum matter in 2+1 and 3+1 dimensions},
  \href{https://doi.org/https://doi.org/10.1016/j.aop.2017.06.019}{\emph{Annals
  of Physics} {\bfseries 384} (2017) 254 }.

\bibitem{Ye:2017aa}
P.~Ye, M.~Cheng and E.~Fradkin, \emph{Fractional $s$-duality, classification of
  fractional topological insulators, and surface topological order},
  \href{https://doi.org/10.1103/PhysRevB.96.085125}{\emph{Phys. Rev. B}
  {\bfseries 96} (2017) 085125}.

\bibitem{Tiwari:2016aa}
A.~Tiwari, X.~Chen and S.~Ryu, \emph{Wilson operator algebras and ground states
  of coupled $\mathit{BF}$ theories},
  \href{https://doi.org/10.1103/PhysRevB.95.245124}{\emph{Phys. Rev. B}
  {\bfseries 95} (2017) 245124}.

\bibitem{2012FrPhy...7..150W}
K.~{Walker} and Z.~{Wang}, \emph{{(3+1)-TQFTs and topological insulators}},
  \href{https://doi.org/10.1007/s11467-011-0194-z}{\emph{Frontiers of Physics}
  {\bfseries 7} (2012) 150} [\href{https://arxiv.org/abs/1104.2632}{{\ttfamily
  1104.2632}}].

\bibitem{zhang2021compatible}
Z.-F.~Zhang and P.~Ye, \emph{Compatible braidings with hopf links, multiloop,
  and borromean rings in $(3+1)$-dimensional spacetime},
  \href{https://doi.org/10.1103/PhysRevResearch.3.023132}{\emph{Phys. Rev.
  Research} {\bfseries 3} (2021) 023132}.

\bibitem{ye19a}
M.-Y.~Li and P.~Ye, \emph{Fracton physics of spatially extended excitations},
  \href{https://doi.org/10.1103/PhysRevB.101.245134}{\emph{Phys. Rev. B}
  {\bfseries 101} (2020) 245134}.

\bibitem{2021LiYeFracton}
M.-Y.~{Li} and P.~{Ye}, \emph{{Fracton physics of spatially extended
  excitations. II. Polynomial ground state degeneracy of exactly solvable
  models}}, {\emph{arXiv e-prints} (2021) arXiv:2104.05735}
  [\href{https://arxiv.org/abs/2104.05735}{{\ttfamily 2104.05735}}].

\bibitem{pretko18string}
S.~Pai and M.~Pretko, \emph{Fractonic line excitations: An inroad from
  three-dimensional elasticity theory},
  \href{https://doi.org/10.1103/PhysRevB.97.235102}{\emph{Phys. Rev. B}
  {\bfseries 97} (2018) 235102}.

\bibitem{hansson2004superconductors}
T.H.~Hansson, V.~Oganesyan and S.L.~Sondhi, \emph{Superconductors are
  topologically ordered},
  \href{https://doi.org/http://dx.doi.org/10.1016/j.aop.2004.05.006}{\emph{Annals
  of Physics} {\bfseries 313} (2004) 497}.

\bibitem{abeffect}
Y.~Aharonov and D.~Bohm, \emph{Significance of electromagnetic potentials in
  the quantum theory},
  \href{https://doi.org/10.1103/PhysRev.115.485}{\emph{Phys. Rev.} {\bfseries
  115} (1959) 485}.

\bibitem{PRESKILL199050}
J.~Preskill and L.M.~Krauss, \emph{Local discrete symmetry and
  quantum-mechanical hair},
  \href{https://doi.org/https://doi.org/10.1016/0550-3213(90)90262-C}{\emph{Nuclear
  Physics B} {\bfseries 341} (1990) 50 }.

\bibitem{PhysRevLett.62.1071}
M.G.~Alford and F.~Wilczek, \emph{Aharonov-bohm interaction of cosmic strings
  with matter}, \href{https://doi.org/10.1103/PhysRevLett.62.1071}{\emph{Phys.
  Rev. Lett.} {\bfseries 62} (1989) 1071}.

\bibitem{PhysRevLett.62.1221}
L.M.~Krauss and F.~Wilczek, \emph{Discrete gauge symmetry in continuum
  theories}, \href{https://doi.org/10.1103/PhysRevLett.62.1221}{\emph{Phys.
  Rev. Lett.} {\bfseries 62} (1989) 1221}.

\bibitem{ALFORD1992251}
M.G.~Alford, K.-M.~Lee, J.~March-Russell and J.~Preskill, \emph{Quantum field
  theory of non-abelian strings and vortices},
  \href{https://doi.org/https://doi.org/10.1016/0550-3213(92)90468-Q}{\emph{Nuclear
  Physics B} {\bfseries 384} (1992) 251 }.

\bibitem{Horowitz1990}
G.T.~Horowitz and M.~Srednicki, \emph{A quantum field theoretic description of
  linking numbers and their generalization},
  \href{https://doi.org/10.1007/BF02099875}{\emph{Communications in
  Mathematical Physics} {\bfseries 130} (1990) 83}.

\bibitem{horowitz89}
G.T.~Horowitz, \emph{Exactly soluble diffeomorphism invariant theories},
  \href{https://doi.org/10.1007/BF01218410}{\emph{Commun. Math. Phys.}
  {\bfseries 125} (1989) 417}.

\bibitem{Baez2011}
J.C.~Baez and J.~Huerta, \emph{An invitation to higher gauge theory},
  \href{https://doi.org/10.1007/s10714-010-1070-9}{\emph{General Relativity and
  Gravitation} {\bfseries 43} (2011) 2335}.

\bibitem{bti6}
P.~Ye and J.~Wang, \emph{Symmetry-protected topological phases with charge and
  spin symmetries: Response theory and dynamical gauge theory in two and three
  dimensions}, \href{https://doi.org/10.1103/PhysRevB.88.235109}{\emph{Phys.
  Rev. B} {\bfseries 88} (2013) 235109}.

\bibitem{RN874}
P.~Cromwell, E.~Beltrami and M.~Rampichini, \emph{The mathematical tourist},
  \href{https://doi.org/10.1007/BF03024401}{\emph{The Mathematical
  Intelligencer} {\bfseries 20} (1998) 53}.

\bibitem{bti1}
A.~Vishwanath and T.~Senthil, \emph{Physics of three-dimensional bosonic
  topological insulators: Surface-deconfined criticality and quantized
  magnetoelectric effect},
  \href{https://doi.org/10.1103/PhysRevX.3.011016}{\emph{Phys. Rev. X}
  {\bfseries 3} (2013) 011016}.

\bibitem{PhysRevB.99.205120}
B.~Han, H.~Wang and P.~Ye, \emph{Generalized wen-zee terms},
  \href{https://doi.org/10.1103/PhysRevB.99.205120}{\emph{Phys. Rev. B}
  {\bfseries 99} (2019) 205120}.

\bibitem{Kapustin2014}
A.~Kapustin and N.~Seiberg, \emph{Coupling a qft to a tqft and duality},
  \href{https://doi.org/10.1007/JHEP04(2014)001}{\emph{Journal of High Energy
  Physics} {\bfseries 2014} (2014) 1}.

\bibitem{PhysRevB.99.235137}
Q.-R.~Wang, M.~Cheng, C.~Wang and Z.-C.~Gu, \emph{Topological quantum field
  theory for abelian topological phases and loop braiding statistics in
  $(3+1)$-dimensions},
  \href{https://doi.org/10.1103/PhysRevB.99.235137}{\emph{Phys. Rev. B}
  {\bfseries 99} (2019) 235137}.

\bibitem{milnor1954link}
J.~Milnor, \emph{Link groups},
  \href{https://doi.org/10.2307/1969685}{\emph{Ann. Math.} {\bfseries 59}
  (1954) 177}.

\bibitem{mellor2003geometric}
B.~Mellor and P.~Melvin, \emph{A geometric interpretation of milnor's triple
  linking numbers}, \href{https://doi.org/10.2140/agt.2003.3.557}{\emph{Algebr.
  Geom. Topol.} {\bfseries 3} (2003) 557}.

\bibitem{lapa17}
M.F.~Lapa, C.-M.~Jian, P.~Ye and T.L.~Hughes, \emph{Topological electromagnetic
  responses of bosonic quantum hall, topological insulator, and chiral
  semimetal phases in all dimensions},
  \href{https://doi.org/10.1103/PhysRevB.95.035149}{\emph{Phys. Rev. B}
  {\bfseries 95} (2017) 035149}.

\bibitem{string7}
D.~Gaiotto, A.~Kapustin, N.~Seiberg and B.~Willett, \emph{Generalized global
  symmetries}, \href{https://doi.org/10.1007/JHEP02(2015)172}{\emph{J. High
  Energy Phys.} {\bfseries 2015} (2015) 1}.

\bibitem{kb2015}
C.W.~von Keyserlingk and F.J.~Burnell, \emph{Walker-wang models and axion
  electrodynamics},
  \href{https://doi.org/10.1103/PhysRevB.91.045134}{\emph{Phys. Rev. B}
  {\bfseries 91} (2015) 045134}.

\bibitem{chen2021loops}
X.~{Chen}, A.~{Dua}, P.-S.~{Hsin}, C.-M.~{Jian}, W.~{Shirley} and C.~{Xu},
  \emph{{Loops in 4+1d Topological Phases}}, {\emph{arXiv e-prints} (2021)
  arXiv:2112.02137} [\href{https://arxiv.org/abs/2112.02137}{{\ttfamily
  2112.02137}}].

\end{thebibliography}

\end{document}